\documentclass[11pt,english]{article}
\PassOptionsToPackage{dvipsnames}{xcolor}
\usepackage{jheppub}
\usepackage{orcidlink}
\usepackage{amsmath, bbold}
\usepackage{cleveref}
\crefname{equation}{eq.}{eqs.}
\Crefname{equation}{Equation}{Equations}
\crefname{section}{section}{sections}
\Crefname{section}{Section}{Sections}
\crefname{subsection}{section}{sections}
\Crefname{subsection}{Section}{Sections}
\crefname{subsubsection}{section}{sections}
\Crefname{subsubsection}{Section}{Sections}
\crefname{figure}{figure}{figures}
\Crefname{figure}{Figure}{Figures}
\crefname{table}{table}{tables}
\Crefname{table}{Table}{Tables}
\crefname{appendix}{appendix}{appendices}
\Crefname{appendix}{Appendix}{Appendices}
\usepackage{appendix}
\usepackage{cancel}
\usepackage{booktabs}
\usepackage{makecell}
\usepackage{adjustbox}
\usepackage{threeparttable}
\usepackage{booktabs}
\usepackage[dvipsnames]{xcolor}
\usepackage[normalem]{ulem}
\usepackage{graphicx, float}%
\usepackage{tikz}
\usetikzlibrary{decorations.pathmorphing, positioning, calc, arrows.meta, math}
\usepackage{pgfplots}
\usepgfplotslibrary{fillbetween}
\pgfplotsset{compat=1.18, set layers}
\graphicspath{{./figures/}}

\usepackage[font=small,labelfont=sc]{caption}

\usepackage{amsmath}
\usepackage{array}
\usepackage{booktabs}
\usepackage{longtable}

\newcommand{\CLFVCurrent}{\textsc{Current}}
\newcommand{\CLFVMid}{\textsc{Mid}}
\newcommand{\CLFVFuture}{\textsc{Future}}
\newcommand{\CLFVNoData}{\textemdash}

\usepackage{array}
\usepackage{booktabs}
\usepackage{makecell}

\newcommand{\me}[1]{\textcolor{blue}{#1}}
\newcommand{\adl}[1]{\textcolor{green!50!black}{#1}}

\newcommand{\SO}[1]{\mathrm{SO}(#1)}
\newcommand{\SU}[1]{\mathrm{SU}(#1)}

\newcommand{\U}[1]{\mathrm{U}(#1)}

\definecolor{darkred}{rgb}{0.6,0,0}
\definecolor{darkpurple}{rgb}{0.5,0,0.5}

\def\Tr{\text{Tr}}
\def\z2{$\mathbb{Z}_2$}
\def\z3{$\mathbb{Z}_3$}
\def\321{$S\U{3}_c \times S\U{2}_L \times \U{1}_\mathrm{Y}$}

\def\555{\ensuremath{SU(5)^3}}

\def\24{\ensuremath{\mathbf{24}}}

\newcommand{\TeV}{~\mathrm{TeV}}

\newcommand{\MPcol}[1]{\textcolor{orange}{#1}}

\newcommand{\MP}[1]{\textsf{\MPcol{[MP: #1]}}}

\usepackage{xcolor}

\newcommand{\commentout}[1]{} %
\author[a]{Andrea Di Lecce\orcidlink{0009-0008-1228-6209}}
\author[a]{, Majid Ekhterachian\orcidlink{0000-0002-1980-2498}}
\author[b,c]{, Marko Pesut\orcidlink{0000-0001-5163-6808}}
\author[d]{and Stefan Stelzl\orcidlink{0000-0001-5964-1054}}

\affiliation[a]{Scuola Normale Superiore and INFN,
Piazza dei Cavalieri 7, 56126 Pisa, Italy}

\affiliation[b]{PSI Center for Neutron and Muon Sciences,
5232 Villigen PSI, Switzerland}

\affiliation[c]{DAMTP, University of Cambridge, Wilberforce Road, Cambridge CB3 0WA}

\affiliation[d]{Institut de F\'isica d'Altes Energies (IFAE)
and Barcelona Institute of Science and Technology (BIST),
Campus UAB, 08193 Bellaterra (Barcelona), Spain}

\emailAdd{andrea.dilecce@sns.it}
\emailAdd{majid.ekhterachian@sns.it}
\emailAdd{marko.pesut@psi.ch} %
\emailAdd{sstelzl@ifae.es}
\title{
The Silence of the Dipoles: Accidental Symmetries in Lepton Flavor 
}

\abstract{
Upcoming experiments are set to dramatically improve the search for charged lepton flavor violation, with the sensitivity to the rates of $\mu\to eee$ decay and $\mu N\to eN$ conversion expected to improve by four orders of magnitude in the near future. We study how these measurements reshape our understanding of fundamental interactions and their implications for the Higgs hierarchy problem. We consider the framework of Strongly Interacting Light Higgs with partial compositeness. In generic models of partial compositeness, the existing constraints from $\mu\to e\gamma$ and the electric dipole moment (EDM) of the electron require the scale of Higgs compositeness to be above $\sim\mathcal{O}(100)$ TeV. These mass scales lie beyond the direct reach of the LHC and envisioned future colliders and are comparable to, or higher than, those accessible to future $\mu\to3e$ and $\mu N\to eN$ searches. We show, however, that models where the strongly coupled sector has an accidental $\U{3}\times\mathrm{CP}$ or $\U{2}\times\U{1}\times\mathrm{CP}$ symmetry, broken by the elementary--composite mixings, can substantially relax these constraints. In the $\U{3}$ case, these symmetries (1) lead to an accidental alignment of the Yukawa and dipole couplings at leading order, suppressing $\mu\to e\gamma$ and the electron EDM, allowing the compositeness scale to lie in the few-TeV range; (2) are probed by the upcoming $\mu\to eee$ and $\mu N\to eN$ experiments; and (3) provide a direct target for current and future collider experiments. In the $\U{2}\times\U{1}$ case, $\mu\to e\gamma$ and the electron EDM remain important constraints, while upcoming searches for $\mu\to eee$, $\mu N\to eN$ and flavor-violating $\tau$ decays provide complementary probes. We discuss how these accidental symmetries can arise in the infrared compatibly with the dynamical generation of flavor hierarchies in the ultraviolet.
}

\begin{document}
\maketitle

\providecommand{\introadd}[1]{\textcolor{brown}{#1}}
\providecommand{\introdel}[1]{\textcolor{gray}{\sout{#1}}}
\providecommand{\introdelnote}[1]{\footnote{\introdel{#1}}}

\newpage

\section{Introduction}
\label{sec:intro}

Upcoming experiments are set to dramatically improve the search
for charged lepton flavor violation, with Mu3e, Mu2e and COMET
aiming to improve the sensitivity to the rates of $\mu\to eee$
decay and coherent $\mu N\to eN$ conversion by four orders of
magnitude~\cite{Mu3e:2020TDR,Mu2e:2019,COMET:2020TDR}.
Together with searches for $\mu\to e\gamma$~\cite{MEGII:2018design},
flavor-violating $\tau$ decays~\cite{banerjee2022snowmass2021whitepaper}
and the electric dipole moment
(EDM) of the electron~\cite{Alarcon:2022ero}, this program will test
lepton interactions with unprecedented sensitivity.
The contributions to charged lepton flavor violation induced solely
by the observed light neutrino masses are far below these
sensitivities, making these processes particularly clean probes
of new physics~\cite{Kuno:1999jp}. \Cref{tab:experimental-inputs} summarizes the principal
muon and electron EDM searches considered in this work.

The impact of this experimental progress depends on the
structure of the underlying interactions. The constraints and reach on the Wilson coefficients of higher dimensional operators involving only Standard Model (SM) fields have recently been studied in \cite{EliasMiro:2021jgu}.
If the dominant
source of charged lepton flavor violation is a dipole
interaction, the present $\mu\to e\gamma$ bound already severely
constrains $\mu\to eee$ and $\mu N\to eN$ conversion. 
Their
projected sensitivities then translate into bounds on the
dipole coefficient comparable to those from $\mu\to e\gamma$.
If instead other interactions dominate, these experiments can
explore a much larger region of presently allowed parameter
space. 

\begin{table}[b]
\centering
\resizebox{\textwidth}{!}{%
\renewcommand{\arraystretch}{1.25}
\setlength{\tabcolsep}{8pt}
\begin{tabular}{|c|c|c|}
\hline
process & current bound & projected sensitivity \\
\hline\hline
$\mu\to e\gamma$
&
$\mathrm{BR}(\mu\to e\gamma)<1.5\times 10^{-13}$
\cite{MEGII:2025new}
&
$\mathrm{BR}(\mu\to e\gamma)<6\times 10^{-14}$
\cite{MEGII:2018design}
\\
\hline
$\mu\to 3e$
&
$\mathrm{BR}(\mu\to 3e)<1.0\times 10^{-12}$
\cite{SINDRUM:1988mueee}
&
$\mathrm{BR}(\mu\to 3e)<1 \times 10^{-16}$
\cite{Mu3e:2013proposal,Mu3e:2020TDR}
\\
\hline
$\mu N\to eN$
&
$\mathrm{CR}(\mu N\to eN)<7.0\times 10^{-13}$
\cite{SINDRUMII:2006mueconv}
&
$\mathrm{CR}(\mu N\to eN)<8 \times 10^{-17}$
(Mu2e) \cite{Mu2e:2019}
\\[-1mm]
&
&
{$\mathrm{CR}(\mu N\to eN)<{\rm few}\times 10^{-17}$}
(COMET) {\cite{COMET:2020TDR}}
\\
\hline
Electron EDM
&
$|d_e|<4.1\times 10^{-30}\,e\,{\rm cm}$
\cite{Roussy:2022cmp}
&
$|d_e|\lesssim 10^{-32}\,e\,{\rm cm}$
\cite{Alarcon:2022ero}
\\
\hline
\end{tabular}
}
\caption{
Current bounds and projected sensitivities for charged
lepton flavor violation and the electron EDM.
Here $\mathrm{CR}(\mu N\to eN)\equiv
\Gamma(\mu^- N\to e^- N)/\Gamma_{\rm capt}(\mu^- N)$
denotes the coherent conversion rate normalized to the ordinary
muon capture rate. Below we use the projection of Mu2e to
represent the future reach for $\mu N\to eN$ processes in the
different scenarios.
}
\label{tab:experimental-inputs}
\end{table}

The expectation of new physics within this reach is deeply tied to the nature of the Higgs sector. Although the discovery of the Higgs boson completed the Standard Model as a renormalizable description of particle physics at the electroweak scale, the presence of a light, apparently elementary scalar unavoidably implies that the electroweak scale is unstable against quantum corrections caused by the presence of heavy degrees of freedom beyond the Standard Model (BSM). Natural solutions to this hierarchy problem therefore point to new dynamics not far above the TeV scale in order to stabilize the Higgs potential. However, the absence of direct signals at the LHC and of indirect deviations in precision observables has put this naturalness paradigm under increasing experimental pressure: viable models of natural electroweak symmetry breaking tend to either tolerate a larger separation between the electroweak and new-physics scales, or feature a non-generic structure that suppresses the most sensitive collider and precision observables.

In particular, flavor provides one of the most stringent constraints on natural solutions to the Higgs hierarchy problem. In the SM, all quark and charged-lepton flavor breakings are encoded in the Yukawa matrices. This highly restricted structure leads to the GIM suppression of flavor-changing neutral currents and to charged-lepton flavor conservation in the limit of vanishing neutrino masses. At the same time, flavor remains unexplained within the SM: the Yukawa couplings are free parameters, and their non-generic hierarchical pattern constitutes one of the open questions in particle physics, the so-called flavor puzzle. The issue with flavor is therefore twofold: one would like to understand the origin of the flavor hierarchies, while at the same time avoiding the generically unsuppressed BSM sources of flavor transitions that would spoil the accidental flavor symmetries and the highly restricted CP-violating structure of the SM. In particular, the absence of deviations in rare processes typically pushes generic flavor- and CP-violating new physics far above the scales directly probed at colliders and motivated by the Higgs hierarchy problem. Any realistic TeV-scale solution to the hierarchy problem must therefore also specify how flavor and CP violation are controlled and protected.

We study these questions in the broad framework of a Strongly
Interacting Light Higgs (SILH)~\cite{Giudice:2007fh}, with
fermion masses generated through partial
compositeness~\cite{Kaplan:1991dc}. A particularly motivated
realization is a composite pseudo-Nambu--Goldstone Higgs, whose
lightness is protected by an approximate global symmetry of
the strong sector~\cite{Agashe:2004rs,Panico:2015jxa}. In this scenario, elementary fermions mix linearly with fermionic operators of
this sector, and acquire their Yukawa couplings through these
mixings. In an approximately conformal regime, differences in
the operator dimensions can generate large hierarchies among
the mixings through renormalization-group (RG) evolution, even
starting from non-hierarchical ultraviolet (UV) couplings.
Partial compositeness can thus relate the fermion mass
hierarchies to the dynamics responsible for Higgs compositeness.
The same mixings also
control the new contributions to flavor
observables mediated by other composite resonances.

The suppression associated with the small mixings of light
fermions is, however, insufficient in the generic, anarchic
scenario. 
In particular, the constraints from
$\mu\to e\gamma$ and the electron EDM, both mediated by dipole operators, require compositeness
scales of order $100$~TeV or higher for sizable strong-sector
couplings~\cite{Keren-Zur:2012buf,Frigerio:2018uwx, Glioti:2024hye}.
Such scales are far above those motivated by the Higgs
hierarchy problem and outside the direct reach of the LHC
and envisioned future colliders. 
A symmetry that suppresses
these processes could therefore change both the allowed
compositeness scale and the significance of the experimental
program.

The role of flavor symmetries in partial compositeness in the lepton sector has
been explored in several directions. Lepton minimal flavor
violation (MFV)~\cite{Redi:2013pga} provides strong protection,
but also suppresses charged lepton flavor violation to
negligible levels. Less restrictive possibilities studied in the literature include partial lepton universality \cite{Redi:2013pga,Glioti:2024hye} and a
$\mathrm{U}(1)^3\times\mathrm{CP}$ symmetry of the strong
sector~\cite{Frigerio:2018uwx}. 
A setup with universal composite-sector masses and Yukawa couplings and generic elementary--composite mixings was
studied in ref.~\cite{Feruglio2015LFV} under the name intermediate flavor violation. This leads to alignment of the leading dipole with the charged-lepton Yukawa matrix,
while flavor-changing currents remain.
Another way suggested to weaken
the constraints is by separating the dynamical
scales associated with flavor~\cite{Vecchi:2012fv,Panico:2016ull}.
Various symmetry assumptions have also been explored to weaken the strong flavor constraints in the quark sector \cite{Fitzpatrick:2007sa,Cacciapaglia:2007fw,Santiago:2008vq,Csaki:2008eh,Csaki:2009wc,Delaunay:2010dw,Redi:2011zi,Barbieri:2012bh,Barbieri2012,Barbieri:2012uh,Redi:2012uj,Covone:2024elw,Davighi:2025cqx, Pesut:2025ook, Davighi:2023iks}.
A systematic survey of the symmetries in the quark sector has recently been done
in ref.~\cite{Glioti:2024hye}. 

In this work, we study symmetry
assumptions in the lepton sector that suppress dipole operators and thus permit a low compositeness scale while
retaining observable lepton flavor violation.
We study strong sectors with an accidental
$\mathrm{U}(3)\times\mathrm{CP}$ or
$\mathrm{U}(2)\times\mathrm{U}(1)
\times\mathrm{CP}$ symmetry, broken by the
elementary--composite mixings. Neutrino masses admit several realizations in the
framework of Composite Higgs, see e.g.~\cite{Frigerio:2018uwx,Agashe:2024uvp}, and, due to their smallness, typically lead to negligible additional sources of flavor and CP violation. We thus focus on the charged-lepton sector.

In the
$\mathrm{U}(3)$ case, the composite operators that mix with left- and
right-handed charged leptons transform as triplets of $\mathrm{U}(3)$. This leads to an alignment of the dipole and the Yukawa couplings at leading order, which suppresses the rate of $\mu \to e \gamma$.
The diagonal CP-odd coefficient controlling the electron EDM starts at sixth order in the mixings. 
This alignment is due to the common flavor quantum numbers under the SM flavor symmetry group of the dipole operator and the Yukawa; in contrast, operators with other flavor quantum numbers are in general not aligned and lead to flavor violation at leading order. For example, flavor-changing
$Z$ couplings therefore generally remain unsuppressed and mediate
$\mu\to eee$ and $\mu N\to eN$ conversion. Within this scenario, these
processes provide the leading tests of the lepton sector and allow for the
compositeness scale to lie in the few-TeV range.
The forthcoming searches can then probe compositeness scales an order of magnitude higher.

We then study a smaller flavor symmetry of the strongly coupled sector, $\mathrm{U}(2)\times \mathrm{U}(1)$.
The $\mathrm{U}(2)$ can provide sufficient protection to the first two generations, while the processes involving $\tau$ can be enhanced compared to the $\mathrm{U}(3)$ case. The $\mu\to e\gamma$ and electron EDM bounds remain important, while upcoming searches for $\mu\to eee$, $\mu N\to eN$ conversion and flavor-violating
$\tau$ decays provide complementary probes.

The flavor symmetries considered in this work may seem to be in tension with the dynamical origin of the charged-lepton hierarchies: a flavor symmetry enforces equal dimensions for operators in the same multiplet, whereas the usual RG explanation of the hierarchies relies on their differences. We show how this can be reconciled if the symmetry is accidental in the infrared, rather than imposed throughout the whole RG flow. Above a flavor scale $\Lambda_f$, non-universal dimensions can generate hierarchical mixings, while below $\Lambda_f$, an approximate flavor symmetry can emerge. The symmetry-breaking deformations within the strong sector are then irrelevant, so that their effects are suppressed at the compositeness scale. This is similar to the construction recently developed in~\cite{Agashe:2025tge}. We discuss the conditions under which this picture gives sufficient protection.

The paper is organized as follows. In
\cref{sec:partial_compositeness}, we introduce the
partial-compositeness framework and review the anarchic and
MFV scenarios. In \cref{sec:flavorprotection}, we study the
accidental $\mathrm{U}(3)\times\mathrm{CP}$ and
$\mathrm{U}(2)\times\mathrm{U}(1)
\times\mathrm{CP}$ symmetries, derive their consequences for different observables and, in \cref{sec:symandhier}, discuss the compatibility of flavor symmetries with the
dynamical generation of flavor hierarchies. In \cref{sec:comparison}, we compare
the bounds and projections with those obtained under other
flavor symmetry assumptions considered in the literature, before concluding in
\cref{sec:conclusion}. %

\section{Partial compositeness}
\label{sec:partial_compositeness}
We begin by reviewing partial compositeness in the lepton sector and the power counting for the effective operators relevant to flavor and CP violation. We then discuss two reference scenarios: anarchy, where generic flavor and CP violation impose stringent bounds on the compositeness scale, and minimal flavor violation, where charged-lepton flavor violation is suppressed to negligible levels. These two limits provide the context for the symmetry assumptions studied in the following section, which suppress the most constraining dipole contributions while retaining potentially observable flavor violation.

\subsection{Framework and power counting}
We briefly review the mechanism of partial compositeness (PC)
to generate Yukawa couplings in composite Higgs models \cite{Kaplan:1991dc, Panico:2015jxa}. We denote by $m_*$ the typical mass of the composite resonances of the strongly coupled sector and by $g_*$ their typical coupling. We assume that $g_*$ can still be treated perturbatively, in analogy to the couplings among mesons, $\sim 4\pi/\sqrt{N_c}$ or glueballs $\sim 4\pi/N_c$ in large-$N_c$ QCD.
Within  partial compositeness, 
 elementary fermions with SM quantum numbers  linearly mix with fermionic operators of the strong dynamics \cite{Kaplan:1991dc,Giudice:2007fh,Panico:2015jxa}. For the lepton sector, we have
\begin{equation}
  \mathcal L_{\rm mix}(\Lambda_{\rm UV}) =
  \sum_{i,a}
  \left[
    \lambda_{\ell}^{ai}(\Lambda_{\rm UV})
    \Lambda_{\rm UV}^{5/2-\Delta_{\ell}^a}\,
    \left(\bar{\mathcal O}_\ell^a \ell_L^i \right)
    +
    \lambda_{e}^{ai}(\Lambda_{\rm UV})
    \Lambda_{\rm UV}^{5/2-\Delta_{e}^a}\,
    \left(\bar{\mathcal O}_e^a e_R^i \right)
  \right]
  + {\rm h.c.}\, ,
  \label{eq:pc_uv_mixing}
\end{equation}
where $i=1,2,3$ labels the  flavor index of the elementary lepton fields $e_R^i$ and $\ell^i_L$, and $a$ labels composite-sector operators $\mathcal O_{\psi }^a$, which have scaling dimension $\Delta_{\psi}^a$. In the regime where the mixings $\lambda_\psi^{ai}$ are perturbative, their RG evolution starting from the UV scale $\Lambda_{\rm UV}$ is given by
\begin{equation}
  \lambda_{\psi}^{ai}(m_*) \simeq
  \lambda_{\psi}^{ai}(\Lambda_{\rm UV})
  \left( \frac{m_*}{\Lambda_{\rm UV}} \right)^{\Delta_{\psi}^a-5/2},
  \qquad
  \psi=\ell,e\, .
  \label{eq:lambda_running}
\end{equation}
Thus irrelevant mixings with $\Delta_{\psi }^a>5/2$ are suppressed in the infrared (IR), whereas relevant mixings can grow until they become $\mathcal{O}(g_*)$. 
Note that starting from anarchic mixings in the UV and with a mild variation in $\Delta_\psi^a$, one obtains a hierarchical structure in the mixings at the scale $m_*$.
It is useful to define the dimensionless mixings
\begin{equation}
  \epsilon_{\psi}^{ai} \equiv \frac{\lambda_{\psi}^{ai}(m_*)}{g_*}\, ,
  \qquad
  |\epsilon_\psi^{ai}|\lesssim 1\, .
  \label{eq:epsilon_def}
\end{equation} 
As the strong sector develops a gap at the scale $m_*$, the operators $\mathcal O_\psi^a$ interpolate fermionic
resonances $\Psi_{\psi,a}$ with the same SM quantum numbers as $\psi$. The light
mass eigenstate is schematically
\begin{equation}
  \psi_{\rm SM}^i
  \simeq
  \psi_{\rm el}^i
  +
  \epsilon_{\psi}^{ai}\,\Psi_{\psi,a} ,
\end{equation}
where $\epsilon_{\psi}^{ai}$ measures the degree of composite admixture in a light SM fermion.
One can then estimate the charged-lepton Yukawa matrix, noting that they interact with the composite Higgs through their composite admixture,
\begin{equation}
  (Y_e)_{ij} \sim
  g_*\, (\epsilon_\ell^\dagger c_Y \epsilon_e)_{ij}\, ,
  \label{eq:pc_yukawa_matrix}
\end{equation}
where $c_Y^{ab}$ is a dimensionless matrix %
which, in the absence of flavor symmetries, has generic complex  $O(1)$ entries.
The singular values $y_i$ of the Yukawa matrix are parametrically given in terms of the eigenvalues $\epsilon_i^{\psi}$ of the dimensionless mixing matrices,
\begin{equation}
  y_i \equiv \frac{\sqrt{2}m_i}{v}
  \sim g_*\,\epsilon_i^\ell\epsilon_i^e\,,
  \qquad
  i=e,\mu,\tau\, .
  \label{eq:epsilon_product}
\end{equation}
Thus the charged-lepton mass hierarchy originates from the hierarchy in the degree of compositeness: lighter leptons are more elementary, while heavier leptons have
a larger overlap with the strong sector. The mass fixes only the product
$\epsilon_i^\ell\epsilon_i^e$, leaving free how compositeness is shared
between the two chiralities. We parametrize this freedom as
\begin{equation}
 r_i^2
        \equiv
        \frac{\epsilon_i^\ell}{\epsilon_i^e},
        \qquad
        \epsilon_i^\ell
        \sim
        \sqrt{\frac{y_i}{g_\ast}}\,r_i,
        \qquad
        \epsilon_i^e
        \sim
        \sqrt{\frac{y_i}{g_\ast}}\,\frac1{r_i},
        \qquad
        i=1,2,3\,,
  \label{eq:r_param}
\end{equation}
where $r_i=1$ corresponds to comparable left- and right-handed compositeness, $r_i\gg1$ to a more composite lepton doublet, and $r_i\ll1$ to a more composite right-handed charged lepton.

Neutrino masses can be accommodated in several ways.
With conserved
lepton number, small Dirac masses can be generated in a way similar to that for charged leptons and quarks with partially
composite right-handed neutrinos. Majorana masses can instead
be generated in a different manner. 
One possibility is that the strong sector preserves lepton number and the lepton-number breaking arises from bilinear mixing of the left-handed lepton doublet with an electroweak triplet operator of the strong sector~\cite{Keren-Zur:2012buf}. Alternatively, the strong sector can violate lepton number by a small amount which through the usual linear mixing of the left-handed doublet with operators of the strong sector leads to neutrino masses. 
The source of lepton-number violation in the strong sector could arise from a heavy elementary
Majorana singlet which also mixes linearly with the strong sector~\cite{Huber:2003sf,Agashe:2015izu}.

More generally, at energies below $m_*$, the resonances can be integrated out. The resulting effective field theory (EFT) is organized by the power counting \cite{Giudice:2007fh}
\begin{equation}
  \mathcal L_{\rm EFT}
  =
  \frac{m_*^4}{g_*^2}
  \left[
  \mathcal{L}_0
  \left(
    \frac{D_\mu}{m_*},
    \frac{g_* H}{m_*},
    \frac{g_{\rm SM} A_\mu}{m_*},
    \frac{g_* \epsilon_\psi \psi}{m_*^{3/2}}
  \right)
  +
  \frac{g_*^2}{16\pi^2}\,
  \mathcal{L}_1
  \left(
    \frac{D_\mu}{m_*},
    \frac{g_* H}{m_*},
    \frac{g_{\rm SM} A_\mu}{m_*},
    \frac{g_* \epsilon_\psi \psi}{m_*^{3/2}}
  \right)
  + \cdots
  \right] .
  \label{eq:pc_nda}
\end{equation}
The first term denotes tree-level matching from the strong sector, while
$\mathcal{L}_1$ denotes loop-generated structures. The SM field strength appears in operators with the combination $ g_X X_{\mu\nu}/m_*^2$.
Thus an operator $\mathcal O\sim H^{n_H}D^{n_D}X^{n_X}\psi^{n_\psi}$, of dimension $d = n_H+n_D+2n_X+3n_\psi/2$, has the parametric Wilson coefficient
\begin{equation}
  \tilde C_\mathcal{O} \sim 
  c_{\mathcal O}
  \left(\frac{g_*^2}{16\pi^2}\right)^{n_l}
  \frac{
    g_*^{\,n_H+n_\psi-2}
  }{
    m_*^{d-4}
  }\prod_{X} g_X
  \prod_{\psi}\epsilon_{\psi}\, ,
  \label{eq:pc_general_rule}
\end{equation}
with $c_{\mathcal O} \sim O(1)$. Throughout this work, tilded Wilson coefficients denote quantities defined %
prior to the rotations that diagonalize the charged-lepton Yukawa matrix. The integer $n_l$ counts whether the operator is generated at tree level ($n_l=0$), or first appears at one loop ($n_l=1$).

The most important operators for our discussion are dipole operators,
\begin{equation}
     \mathcal{O}_{eX}^{ij} = (\bar\ell_L^i\sigma^{\mu\nu}e_R^j)H X_{\mu\nu},
\end{equation}
where $X_{\mu \nu}=B_{\mu \nu},W_{\mu \nu}$,
and 
Higgs-lepton current operators, 
\begin{align}
    \mathcal{O}_{H\ell}^{(1)ij} &= (H^\dagger i\overleftrightarrow D_\mu H) (\bar\ell_L^i\gamma^\mu\ell_L^j)\,,
  \qquad&
  \mathcal{O}_{H\ell}^{(3)ij} &= (H^\dagger i\overleftrightarrow D_\mu^A H) (\bar\ell_L^i\tau^A\gamma^\mu\ell_L^j)\,,
  \\[2mm]
  \mathcal{O}_{He}^{ij} &= (H^\dagger i\overleftrightarrow D_\mu H) (\bar e_R^i\gamma^\mu e_R^j)\,. \label{eq: Higgs current operators}
\end{align}
Lepton flavor violation can also come from 
dimension-six Yukawa-like operators 
\begin{equation}
    \mathcal{O}_{eH}^{ij} = (H^\dagger H)(\bar\ell_L^i H e_R^j),
\end{equation}
and 
four-fermion operators, e.g.
\begin{equation}
    \mathcal{O}_{\ell e}^{ijkl} = (\bar\ell_L^i\gamma_\mu\ell_L^j) (\bar e_R^k\gamma^\mu e_R^l)\,, \qquad
  \mathcal{O}_{\ell q}^{ijkl} = (\bar\ell_L^i\gamma_\mu\ell_L^j) (\bar q_L^k\gamma^\mu q_L^l), \quad \dots\,.
\end{equation}
The corresponding Wilson coefficients scale as
\begin{align}
\begin{alignedat}{4}
  [\tilde C_{eX}]_{ij}
  &\sim{}
  \frac{g_*^2}{16\pi^2}\frac{g_X g_*}{m_*^2}
  (\epsilon_\ell^\dagger c_{eX}\epsilon_e)_{ij}\,,
  \qquad&
  [\tilde C_{H\ell}^{(1,3)}]_{ij}
  &\sim{}
  \frac{g_*^2}{m_*^2}
  (\epsilon_\ell^\dagger c_{H\ell}^{(1,3)}\epsilon_\ell)_{ij}\,,
  \\[2mm]
   [\tilde C_{He}]_{ij}
  &\sim{}
  \frac{g_*^2}{m_*^2}
  (\epsilon_e^\dagger c_{He}\epsilon_e)_{ij}\,,
  \qquad&
  [\tilde C_{eH}]_{ij}
  &\sim{}
  \frac{g_*^3}{m_*^2}
  (\epsilon_\ell^\dagger c_{eH}\epsilon_e)_{ij}\,,
  \\[2mm]
  [\tilde C_{\ell e}]_{ijkl}
  &\sim{}
  \frac{g_*^2}{m_*^2}
  (\epsilon_\ell^{ai})^\ast \epsilon_\ell^{bj}
  (\epsilon_e^{ck})^\ast \epsilon_e^{dl}
  c_{\ell e}^{abcd}\,,
  \qquad&
  [\tilde C_{\ell q}]_{ijkl}
  &\sim{}
  \frac{g_*^2}{m_*^2}
  (\epsilon_\ell^{ai})^\ast \epsilon_\ell^{bj}
  (\epsilon_q^{ck})^\ast \epsilon_q^{dl}
  c_{\ell q}^{abcd}\,, \quad \dots \,,
\end{alignedat}
\label{eq:pc_scaling_ops}
\end{align}
where $g_X=g,g'$ are electroweak gauge couplings, and where we assume that the dipole operators are generated at one-loop order in the coupling $g_*$.\footnote{These operators could also arise at tree level, and thus be less suppressed. In this work, however, we assume the one-loop suppression as it is typically found in holographic constructions, see e.g.~\cite{Liu:2016idz}.}
We denote the Wilson coefficient of the electromagnetic dipole operator by $
\tilde{C}_{e \gamma}^{i j}=c_W \tilde{C}_{e B}^{i j}-s_W \tilde{C}_{e W}^{i j}$ where $c_W$ and $s_W$ are the cosine and sine of the Weinberg angle.
The coefficients $c_{\mathcal O}$ are tensors in strong-sector flavor space and their indices are contracted with the mixing matrices $\epsilon_\psi^{ai}$.
Generically the operators involving four elementary fermions come with four mixing insertions, and thus their effects are more suppressed. The effect of the operator $\mathcal{O}_{eH}^{ij}$ on the most sensitive flavor observables is suppressed either by a loop or by additional small mixing insertions.

This power counting %
indicates how many powers of $g_*$, $m_*$, SM gauge couplings, loop factors, Higgs fields and compositeness insertions accompany each %
operator, while all flavor assumptions considered in the next sections act by restricting the order-one tensors $c_{\mathcal O}$ and the spurions $\epsilon_\psi$.

\subsection{Anarchic scenario}
\label{subsec:anarchic-bounds}
 We first review the standard anarchic realization of partial compositeness in the lepton sector. In this scenario, the hierarchies in the mixings arise through RG running as in eq.~\eqref{eq:lambda_running}.
The strong dynamics is not assumed to have a lepton-flavor symmetry. The charged-lepton hierarchies originate from hierarchical elementary--composite mixings, while the strong-sector flavor tensors are anarchic, with generic order-one complex entries i.e. $c_\mathcal{O}^{ab}\sim\mathcal O(1)$.\footnote{The tensors associated with non-Hermitian operators, such as $c_Y$, $c_{e\gamma}$ and $c_{eH}$, are generic complex matrices, while the tensors $c_{H\ell}^{(1,3)}$ and $c_{He}$ multiplying Hermitian current operators are Hermitian matrices.} 

We order the strong-sector operators according to their scaling dimension (and thus the hierarchy induced on the mixings), and apply a unitary rotation of the elementary fields to bring the mixings to triangular form with real positive diagonal elements,
\begin{equation}
        \epsilon_\ell =
        \begin{pmatrix}
        \epsilon^\ell_1
        &
        a^\ell_{12}\epsilon^\ell_1
        &
        a^\ell_{13}\epsilon^\ell_1
        \\
        0
        &
        \epsilon^\ell_2
        &
        a^\ell_{23}\epsilon^\ell_2
        \\
        0
        &
        0
        &
        \epsilon^\ell_3
        \end{pmatrix},
        \qquad
        \epsilon_e =
        \begin{pmatrix}
        \epsilon^e_1
        &
        a^e_{12}\epsilon^e_1
        &
        a^e_{13}\epsilon^e_1
        \\
        0
        &
        \epsilon^e_2
        &
        a^e_{23}\epsilon^e_2
        \\
        0
        &
        0
        &
        \epsilon^e_3
        \end{pmatrix}.
\label{eq:triangular-eps-anarchic}
\end{equation}
Here the %
$a^{\ell,e}_{ij}$ are complex coefficients of order unity and $\epsilon^\psi_i$ can be parametrically small with hierarchies among them, ordered as
\begin{equation}
\epsilon^\ell_1<\epsilon^\ell_2<\epsilon^\ell_3, \qquad\epsilon^e_1<\epsilon^e_2<\epsilon^e_3\,.
\end{equation}
The three singular values of each matrix scale as $\epsilon^i_\psi$. In the anarchic scenario this ordering coincides with the charged-lepton mass ordering.
In this basis, the charged-lepton Yukawa matrix has the parametric structure
\begin{equation}
        (Y_e)_{ij}
        \sim
        g_\ast\,\epsilon^\ell_i\epsilon^e_j.
\label{eq:Yukawa-anarchic-parametric}
\end{equation}
 The Yukawa matrix can be diagonalized by unitary rotations, which scale as
\begin{equation}
        (U_L)_{ij}
        \sim
        \min\left(
        \frac{\epsilon_i^\ell}{\epsilon_j^\ell},
        \frac{\epsilon_j^\ell}{\epsilon_i^\ell}
        \right),
        \qquad
        (U_R)_{ij}
        \sim
        \min\left(
        \frac{\epsilon_i^e}{\epsilon_j^e},
        \frac{\epsilon_j^e}{\epsilon_i^e}
        \right),
\label{eq:ULUR-anarchic}
\end{equation}
leading to diagonal entries given by eq.~\eqref{eq:epsilon_product}.
Crucially, after rotating to the charged-lepton mass basis, the dipole and current operators are generically not diagonal, thus leading to lepton flavor violation. %
We now discuss the lepton flavor violation (LFV) induced by various classes of operators.
\paragraph{Dipole operators:}

Before diagonalizing the Yukawa matrix, the coefficient of the electromagnetic dipole operator has the form
\begin{equation}
        \tilde C_{e\gamma}^{ij}
        \sim
        \frac{e}{16\pi^2}
        \frac{g_\ast^3}{m_\ast^2}
        (\epsilon_\ell^\dagger c_{e\gamma}\epsilon_e)_{ij},
\label{eq:dipole-anarchic-interaction}
\end{equation}
where $c_{e\gamma}$ is a generic complex order-one matrix independent of $c_Y$. %
In the mass basis, one then finds for the coefficients of operators relevant for $\mu\to e$ transitions 
\begin{equation}
        C_{e\gamma}^{\,e\mu}
        \sim
        \frac{e}{16\pi^2}
        \frac{g_\ast^2}{m_\ast^2}
        \sqrt{y_e y_\mu}\,
        \frac{r_e}{r_\mu},
        \qquad
        C_{e\gamma}^{\,\mu e}
        \sim
        \frac{e}{16\pi^2}
        \frac{g_\ast^2}{m_\ast^2}
        \sqrt{y_e y_\mu}\,
        \frac{r_\mu}{r_e}.
\label{eq:dipole-anarchic-general-r}
\end{equation}
\paragraph{Electron EDM:}

The anarchic scenario also contains generic CP-violating phases. The flavor-diagonal dipole therefore induces an electron EDM of order
\begin{equation}
        \mathrm{Im}\,C_{e\gamma}^{\,ee}
        \sim
        \frac{e}{16\pi^2}
        \frac{g_\ast^2}{m_\ast^2}
        y_e \Delta_{\rm CP}^e
        \Rightarrow
        d_e \equiv v\sqrt{2} \mathrm{Im}\,C_{e\gamma}^{\,ee}
        \sim
        \frac{g_\ast^2}{8\pi^2}
        \frac{e\,m_e}{m_\ast^2}\Delta_{\rm CP}^e \,,
\label{eq:edm-anarchic-scaling}
\end{equation}
where $\Delta_{\rm CP}^e$ is, generically, a dimensionless order-one CP-odd factor. If, instead, CP is an accidental symmetry of the strong sector, all strong-sector tensors can be chosen real in a common CP basis. In particular, $c_Y$ and $c_{e\gamma}$ are simultaneously real in this basis. This suppresses but does not eliminate CP violation entirely, as %
the elementary--composite mixings generically still carry phases. 
Parametrically, in the mass basis, we find
\begin{equation}
d_e
\sim
\frac{g_\ast^2}{8\pi^2}
\frac{e m_e}{m_\ast^2}
\sqrt{\frac{m_e}{m_\mu}}
\left[\frac{r_e}{r_\mu}
+\frac{r_\mu}{r_e}
\right],
\label{eq:edm-anarchic-CP}
\end{equation}
where each of the two terms is implicitly understood to carry an independent order-one coefficient.\footnote{We note that this result is more suppressed than the estimate given in \cite{Frigerio2018} in the same scenario. We find that the terms with the parametrics estimated in \cite{Frigerio2018} cancel.} %
CP invariance of the strong sector suppresses the electron-EDM coefficient by a factor $\sqrt{m_e/m_\mu}$ and hence relaxes the corresponding bound on $m_\ast$ only by a factor \hbox{$(m_\mu/m_e)^{1/4}\simeq 4$}. %

\paragraph{Higgs-current operators:}

Similarly, we can estimate the coefficients of the Higgs-current operators %
in the mass basis to scale as
\begin{equation}
        C_{H\ell}^{(1)e\mu} \sim  C_{H\ell}^{(3)e\mu}
        \sim
        \frac{g_\ast}{m_\ast^2}
        y_\mu
        \sqrt{\frac{y_e}{y_\mu}}\,
        r_e r_\mu,
        \qquad
        C_{He}^{e\mu}
        \sim
        \frac{g_\ast}{m_\ast^2}
        y_\mu
        \sqrt{\frac{y_e}{y_\mu}}\,
        \frac{1}{r_e r_\mu}.
\label{eq:current-anarchic-general-r}
\end{equation}
These operators induce lepton-flavor-violating $Z$ couplings after electroweak symmetry breaking and contribute to both $\mu\to 3e$ and $\mu N \to e N$ processes at tree level.

\paragraph{Summary and constraints:}

We summarize in \cref{tab:anarchic-summary} the estimates of the size of the Wilson coefficients and the main constraints on the anarchic scenario from the lepton sector. All constraints are understood up to coefficients of order unity.
\begin{table}[H]
\centering
\setlength{\tabcolsep}{3pt}
\renewcommand{\arraystretch}{1.25}
\resizebox{\textwidth}{!}{%
\begin{tabular}{|c|c|c|c|}
\hline
Anarchic PC
&
$C_{e\gamma}^{e\mu},\,C_{e\gamma}^{\mu e}$
&
$C_{H\ell}^{e\mu}$, \, $C_{He}^{e\mu}$
&
$\frac{16\pi^2}{e} \frac{m_\ast^2}{y_eg_\ast^2}\times\mathrm{Im}\,C_{e\gamma}^{ee}$
\\
\hline
scaling
&
\begin{tabular}{@{}c@{}}
$\displaystyle
C_{e\gamma}^{e\mu}
\sim
\frac{e}{16\pi^2}
\frac{g_\ast^2}{m_\ast^2}
\sqrt{y_e y_\mu}\,
\frac{r_e}{r_\mu}$\\[2mm]
$\displaystyle
C_{e\gamma}^{\mu e}
\sim
\frac{e}{16\pi^2}
\frac{g_\ast^2}{m_\ast^2}
\sqrt{y_e y_\mu}\,
\frac{r_\mu}{r_e}$
\end{tabular}
&
\begin{tabular}{@{}c@{}}
$\displaystyle
C_{H\ell}^{e\mu}
\sim
\frac{g_\ast}{m_\ast^2}
\sqrt{y_e y_\mu}\,
r_e r_\mu$\\[2mm]
$\displaystyle
C_{He}^{e\mu}
\sim
\frac{g_\ast}{m_\ast^2}
\sqrt{y_e y_\mu}\,
\frac{1}{r_e r_\mu}$
\end{tabular}
&
\begin{tabular}{@{}c@{}}
$\displaystyle
\Delta_{\rm CP}^e \sim 1
\quad (\cancel{\rm CP})$\\[2mm]
$\displaystyle
\Delta_{\rm CP}^e \sim \sqrt{\frac{m_e}{m_\mu}} \; \max\left[\frac{r_e}{r_\mu}
, \frac{r_\mu}{r_e} \right]
\quad (\rm CP)$
\end{tabular}
\\
\hline
current bounds on $m_\ast[\text{TeV}]$
&
\begin{tabular}{@{}c@{}}
$\displaystyle
26g_\ast
\quad(\mu\to e\gamma)$\\[1mm]
$\displaystyle
4.6g_\ast
\quad(\mu\to 3e)$\\[1mm]
$\displaystyle
4.4g_\ast
\quad(\mu N\to eN)$
\end{tabular}
&
\begin{tabular}{@{}c@{}}
$\displaystyle
2.9\sqrt{g_\ast}
\quad(\mu N\to eN)$\\[1mm]
$\displaystyle
1.1\sqrt{g_\ast}
\quad(\mu\to 3e)$
\end{tabular}
&
\begin{tabular}{@{}c@{}}
$\displaystyle
180g_\ast
\quad(\cancel{\rm CP})
\quad\text{(eEDM)}$\\[1mm]
$\displaystyle
47g_\ast
\quad(\rm CP)
\quad\text{(eEDM)}$
\end{tabular}
\\
\hline
future projections on $m_\ast[\text{TeV}]$
&
\begin{tabular}{@{}c@{}}
$\displaystyle
32g_\ast
\quad(\mu\to e\gamma)$\\[1mm]
$\displaystyle
46g_\ast
\quad(\mu\to 3e)$\\[1mm]
$\displaystyle
39g_\ast
\quad(\mu N\to eN)$
\end{tabular}
&
\begin{tabular}{@{}c@{}}
$\displaystyle
20\sqrt{g_\ast}
\quad(\mu N\to eN)$\\[1mm]
$\displaystyle
11\sqrt{g_\ast}
\quad(\mu\to 3e)$
\end{tabular}
&
\begin{tabular}{@{}c@{}}
$\displaystyle
3600g_\ast
\quad(\cancel{\rm CP})
\quad \text{(eEDM)}$\\[1mm]
$\displaystyle
940g_\ast
\quad(\rm CP)
\quad\text{(eEDM)}$
\end{tabular}
\\
\hline
\end{tabular}%
}
\caption{
Parametric scaling of the Wilson coefficients, current bounds and future reach on the compositeness scale $m_*$ in anarchic partial compositeness. 
The numbers shown  correspond to the left--right symmetric case, $r_i=1$. We show the corresponding experiment in parentheses with the sensitivity given in \cref{tab:experimental-inputs}.
The EDM constraint is shown for two scenarios: one in which the strong sector preserves CP and one in which it does not.
All constraints should be understood as estimates with $\mathcal{O}(1)$ uncertainty.}
\label{tab:anarchic-summary}
\end{table}

\Cref{tab:anarchic-summary} highlights the severe flavor and CP constraints on anarchic partial compositeness in the lepton sector. 
The strongest bound comes from the electron EDM and points to a scale $m_*$ in the $\gtrsim 100$~TeV
range, even if the strong sector is assumed to be CP invariant.

Likewise, dipole operators dominate the CP-conserving lepton-flavor-violating constraints on $m_*$. Currently the strongest
bounds come from $\mu \to e \gamma$. While future $\mu \to 3e$ and $\mu N\to eN$ sensitivities will significantly increase, the resulting limits on the scale $m_*$ will remain comparable to the current ones. On the other hand, Higgs-current constraints are relatively weak, but they are projected to become an order of magnitude more stringent with future data.

In summary, even though anarchic partial compositeness gives an elegant dynamical explanation of charged-lepton hierarchies, the current flavor and CP constraints in this scenario push the scale of
compositeness to scales $\gtrsim 100$~TeV, far from the natural expectations and out of the reach
of any foreseeable future collider. While future electron EDM experiments will probe higher compositeness scales, the remarkable progress in lepton-flavor-violating searches will not open up much new parameter space in this scenario.

\subsection{Minimal flavor violation}
\label{subsec:MFV}

We now discuss the maximally flavor-protected charged-lepton limit in
which the lepton sector satisfies minimal flavor violation
(MFV)~\cite{DAmbrosio:2002vsn,Cirigliano:2005ck,Redi:2013pga,Glioti:2024hye}. In the lepton sector, the approximate flavor symmetry is
\begin{equation}
  G_\ell^{\rm MFV}
  =
  \U{3}_\ell\times\U{3}_e\,.
\end{equation}
Up to small effects responsible for neutrino masses, all the breaking of this symmetry is controlled by the SM Yukawa spurion, which transforms as
\begin{equation}
  Y_e
  \sim
  (\mathbf 3_\ell,\overline{\mathbf 3}_e)\,.
\end{equation}
MFV by itself is a flavor hypothesis and does not exclude flavor-singlet CP phases~\cite{Isidori:2012ts}. In the following we additionally impose CP invariance of the relevant strong-sector matching coefficients and consider the charged-lepton MFV$\times$CP limit.\footnote{Without the CP assumption, the MFV scenario inherits the $m_\ast /g_\ast \gtrsim 180$ TeV bound from the EDM estimate in the anarchic scenario.} 

In partial compositeness, MFV follows from a flavor-universal strong sector together with the additional requirement that one elementary–composite mixing is flavor universal. More precisely, it can be realized as follows: the lepton sector of the strong dynamics is assumed to respect an accidental flavor symmetry
\begin{equation}
        G_{\rm strong}
        =
        \mathrm{U}(3)_{\mathcal O} \times \mathrm{CP}.
\label{eq:MFV-strong-symmetry}
\end{equation}
The composite operators transform as fundamentals under the strong-sector flavor group,
\begin{equation}
        \mathcal O_\ell^a \sim \mathbf 3_{\mathcal O},
        \qquad
        \mathcal O_e^a \sim \mathbf 3_{\mathcal O},
        \qquad
        a=1,2,3.
\end{equation}
The only sources of lepton-flavor breaking are the elementary--composite mixings,
\begin{equation}
        \mathcal L_{\rm mix}
        =
        g_\ast\,(\epsilon_\ell)_{a i} m_\ast^{5/2-\Delta_\ell^a}\,
        \bar{\mathcal O}_\ell^a \ell_L^i 
        +
        g_\ast\,(\epsilon_e)_{a i}m_\ast^{5/2-\Delta_e^a}\,
        \bar{\mathcal O}_e^a e_R^i 
        +
        \mathrm{h.c.}\, .
\label{eq:MFV-mixing}
\end{equation}
Under $\mathrm{U}(3)_\ell \times \mathrm{U}(3)_e \times \mathrm{U}(3)_{\mathcal O}$, the mixings $\epsilon_{\ell,e}$ transform as
\begin{equation}
        \epsilon_\ell \sim (\overline{\mathbf 3}_\ell,\mathbf 1_e,\mathbf 3_{\mathcal O}),
        \qquad
        \epsilon_e \sim (\mathbf 1_\ell,\overline{\mathbf 3}_e,\mathbf 3_{\mathcal O}).
\label{eq:MFV-mixing-spurions}
\end{equation}
The two standard PC realizations of MFV are 
\begin{align}
  {\rm LU}: \mathrm{U}(3)_{\ell} \times \mathrm{U}(3)_e \times \mathrm{U}(3)_{\mathcal{O}} \xrightarrow{\epsilon_{\ell}=\epsilon_L \mathbf{1}} \mathrm{U}(3)_{\ell+\mathcal{O}} \times \mathrm{U}(3)_e\qquad&
  \epsilon_e
  \simeq
  \frac{Y_e}{g_\ast c_Y\epsilon_L},
  \label{eq:MFV-LLU}
  \\[1mm]
  {\rm RU}:\mathrm{U}(3)_{\ell} \times \mathrm{U}(3)_e \times \mathrm{U}(3)_{\mathcal{O}} \xrightarrow{\epsilon_e=\epsilon_R \mathbf{1}} \mathrm{U}(3)_{\ell} \times \mathrm{U}(3)_{e+\mathcal{O}}\qquad&
  \epsilon_\ell^\dagger
  \simeq
  \frac{Y_e}{g_\ast c_Y\epsilon_R}\,,
  \label{eq:MFV-RLU}
\end{align}
where LU and RU denote left and right universality,
respectively, and $Y_e \simeq g_\ast c_Y\,\epsilon_\ell^\dagger\epsilon_e$ with $c_Y \in \mathbb R$. In either case the residual symmetry in the charged-lepton
mass basis is $\U{1}_e\times\U{1}_\mu\times\U{1}_\tau$ and there is no charged-lepton-flavor violation. Under the additional CP assumption, the diagonal dipole coefficients are real, so the direct charged-lepton contribution to the electron EDM vanishes.\footnote{Apart from the lepton sector, there is also a non-zero contribution to the electron EDM leaking from 
the quark sector.
In scenarios where flavor symmetries help to reduce the quark-sector bounds on $m_*$ to the few-TeV range (see e.g. \cite{Glioti:2024hye,Redi:2011zi}), such contributions to the electron EDM are negligible. 
}

The dominant constraints therefore come from lepton-flavor-preserving processes.
In the left-universality realization, the leading Higgs-current coefficients scale as:\footnote{Additional suppression can be implemented via custodial protection (see the discussion in \cref{sec:U3CP}).}
\begin{equation}
  (C_{H\ell}^{(1,3)})_{ii}
  \sim
  \frac{g_\ast^2\epsilon_L^2}{m_\ast^2},
  \qquad 
  (C_{He})_{ii}
  \sim
  \frac{y_i^2}{\epsilon_L^2m_\ast^2}\,,
\end{equation}
and analogously for right universality %
\cite{Giudice:2007fh,Redi:2013pga,Glioti:2024hye}.
The leading constraints arise from the LEP/SLC $Z$-pole measurements limiting the %
deviations of $Z$ couplings, $|\delta g_Z|\sim C_{H\ell,He}v^2\lesssim10^{-3}$. In particular, the left-universality scenario is dominantly constrained by the universal left-handed $Z\ell\ell$ coupling, and analogously the right-universality scenario by the universal right-handed coupling \cite{Contino:2013kra,Glioti:2024hye}, leading to
\begin{equation}
m_\ast \gtrsim 8 \,g_\ast\epsilon_{L,R}\ {\rm TeV}\,.
\label{eq:Zcouplingbound}
\end{equation}
The opposite chirality is fixed by the charged-lepton mass. The lowest scale allowed occurs for $r_\tau \sim 1$ and gives
\begin{equation}
  \epsilon_{L,R}^2
  \simeq
  \frac{y_\tau}{g_\ast},
  \qquad
  m_\ast
  \gtrsim
  8\sqrt{g_\ast y_\tau}\ {\rm TeV}
  \simeq
  0.8\sqrt{g_\ast}\ {\rm TeV}\,.
  \label{eq:MFV-Z-bound-opt}
\end{equation}
Finally, the leading flavor-universal four-lepton interaction scales
as
\begin{equation}
  C_{\ell\ell\ell\ell}\sim
  \frac{g_\ast^2\epsilon_L^4}{m_\ast^2}\,,
\end{equation}
in left universality and analogously for right universality. The leading bounds come from electron scattering and lead to
\cite{deBlas:2022ofj,Glioti:2024hye}
\begin{equation}
  m_\ast
  \gtrsim
  4\,g_\ast\epsilon_{L,R}^2\ {\rm TeV},
\end{equation}
which is weaker than the bound of eq.~\eqref{eq:Zcouplingbound}. 

\section{Lepton flavor protection with accidental symmetries} 
\label{sec:flavorprotection}
The discussion of \cref{subsec:anarchic-bounds} shows that partial-compositeness models with anarchic lepton flavor are severely constrained %
by $\mu\to e\gamma$ and by the electron EDM. %
As a result, the characteristic scale $m_*$ of the composite sector is pushed well above the TeV scale. These constraints can be relaxed with the assumption of MFV in the lepton sector  (see \cref{subsec:MFV}). In this scenario, lepton-flavor violation is essentially absent at the level relevant for the next generation of experiments.

The main motivation of our work is therefore to study an intermediate possibility: a flavor symmetry strong enough to make a compositeness scale of order a few TeV compatible with the $\mu\to e\gamma$ and electron EDM constraints, while weak enough to leave observable lepton-flavor-violating effects in future $\mu\to 3e$ and $\mu N\to eN$ searches. 

We assume that the strong sector respects a flavor symmetry $G_{\text{strong}}$, which is broken by the elementary-composite mixings. For $G_{\text{strong}}$, we consider two different cases. First, in \cref{sec:U3CP}, we study $G_{\text{strong}}\equiv\U{3}\times \rm CP$, which, as we will see, leads to alignment of dipole operators with the Yukawa matrix at leading order, thereby strongly relaxing the bounds from $\mu \to e \gamma$ and the electron EDM. Then, in \cref{sec:U2U1CP}, we consider $G_{\text{strong}}\equiv\U{2}\times \U{1} \times \rm CP$, which treats the third-generation leptons separately. In this scenario, processes involving $\mu$ and $e$ can be more protected than transitions involving the $\tau$. Finally, in \cref{sec:symandhier}, we discuss how the hierarchical structure of the lepton Yukawas can arise consistently with an accidental strong-sector flavor symmetry which provides protection from flavor and CP constraints. We also estimate the effects of strong-sector flavor-symmetry breaking captured by higher-dimensional operators.

\subsection{$\U{3}\times \rm CP$}
\label{sec:U3CP}
We assume that the strong sector respects an accidental flavor symmetry
\begin{equation}
        G_{\rm strong}
        \equiv
        \mathrm{U}(3)_{\mathcal O}
        \times \mathrm{CP}\,.
\label{eq:U3CP-symmetry}
\end{equation}
The composite operators mixing with the elementary leptons transform as fundamentals under $\U{3}_{\mathcal{O}}$,
\begin{equation}
        \mathcal O_\ell^a \sim \mathbf 3_{\mathcal O},
        \qquad
        \mathcal O_e^a \sim \mathbf 3_{\mathcal O},
        \qquad
        a=1,2,3,
\end{equation}
and the only sources of lepton-flavor and CP breaking are the elementary--composite mixings.
Under $\mathrm{U}(3)_\ell \times \mathrm{U}(3)_e \times \mathrm{U}(3)_{\mathcal O}$, the mixings $\epsilon_{\ell,e}$ transform as
\begin{equation}
        \epsilon_\ell \sim (\overline{\mathbf 3}_\ell,\mathbf 1_e,\mathbf 3_{\mathcal O}),
        \qquad
        \epsilon_e \sim (\mathbf 1_\ell,\overline{\mathbf 3}_e,\mathbf 3_{\mathcal O}).
\label{eq:spurions-U3}
\end{equation}
Unlike the MFV scenario, we assume that the mixings are generic and not flavor universal. The crucial point is that both the dipole coefficients and the Yukawa transform under the flavor group as $ (\mathbf 3_\ell,\overline{\mathbf{3}}_e,\mathbf 1_{\mathcal O})$, which implies that there is a unique contraction generating them at the second order in the mixings and, consequently, they are aligned (see also ~\cite{Feruglio2015LFV}). This effect suppresses the bounds on the scale of new physics from both the electron EDM and $\mu \to e\gamma$. As we will show, the electron EDM is further suppressed and does not even arise at fourth order in the mixings.

Before estimating the size of the Wilson coefficients relevant for lepton-flavor and CP violation, it is important to identify two different scenarios. To discuss them, we must distinguish the basis in which $\U{3}$ is manifest from the basis in which the mixings are row-hierarchical.
These two bases can coincide or, instead, be related by a generic unitary rotation. We denote these two cases by \emph{aligned} and \emph{generic non-aligned} scenarios, respectively.

To make this distinction more precise, we start in a basis where each mixing, denoted here by $\hat{\epsilon}
_\ell$ and $\hat{\epsilon}_e$, is a row-hierarchical upper-triangular matrix, as given in \cref{eq:triangular-eps-anarchic}. 
It is convenient, however, to consider a basis where $\mathcal{O}^a_e$ and $\mathcal{O}^a_\ell$ are aligned in the strong-sector $\U{3}_\mathcal{O}$ space.
This can be achieved by a unitary rotation $R$ acting only on $\mathcal{O}_e$.
In such a basis, the mixings $\epsilon_\ell$ and $\epsilon_e$
are related to the row-hierarchical mixings by
\begin{equation}
        \epsilon_\ell
        = \hat{\epsilon}
_\ell,
        \qquad
        \epsilon_e
        =
        R \hat{\epsilon}_e\,,
\label{eq:U3-mixings-symbasis}
\end{equation}
where $\epsilon_e$ is not generally row-hierarchical. Note that in this basis the $\U{3}_\mathcal{O}$ symmetry is manifest in the sense that every two-index $\U{3}_\mathcal{O}$-invariant tensor is proportional to the identity. In particular
\begin{equation}
        c_{\mathcal O}^{ab}
        =
        c_{\mathcal O}\,\delta^a{}_{b} \,,
        \label{eq:cabid}
\end{equation}
with $c_{\mathcal O}$ being an $\mathcal{O}(1)$ constant. Additionally, due to the CP invariance of the strong sector, $c_{\mathcal O}$ can be chosen real.

The two scenarios are specified as follows: the aligned scenario corresponds to the case where $R= \mathbb{1}$,  while the generic non-aligned case corresponds to $R$ being a generic unitary matrix.\footnote{If CP is a symmetry of the strong sector in the UV where the hierarchies are generated, there is a basis where the mixings $\hat{\epsilon}_{\ell,e}$ are row-hierarchical and $c^{ab}_{\mathcal{O}}$ real. This basis is then related to the one where $\U{3}_\mathcal{O}$ is manifest and $c_\mathcal{O}$ real by an orthogonal $O(3) \subset \U{3}$ rotation $R$ rather than a generic $\U{3}$ rotation. We will discuss this case briefly when estimating the electron EDM. \label{footnote:Rorthogonal}} We first consider the generic non-aligned case in
\cref{subsec:nonalignedU3}, before focusing in \cref{subsec:alignedU3} on the aligned case.

\subsubsection{Generic non-aligned $\U{3}$}
\label{subsec:nonalignedU3}
In the generic non-aligned case, the only assumption we make about the mixings is that their singular values are hierarchical. In the basis defined in \cref{eq:U3-mixings-symbasis}, they can therefore be written as
\begin{equation}
        \epsilon_\ell
        =
        V_\ell D_\ell W_\ell^\dagger,
        \qquad
        \epsilon_e
        =
       V_e D_e W_e^\dagger,
        \qquad
        D_\psi=\mathrm{diag}(\epsilon_1^\psi,\epsilon_2^\psi,\epsilon_3^\psi),
\label{eq:triangular-eps}
\end{equation}
where $\psi=\ell,e$, the matrices $V_\psi$ and $W_\psi$ are unitary, and the singular values $\epsilon_i^\psi$ are real, positive and hierarchically ordered. With independent $\U{3}_\ell\times \U{3}_e$ rotations of the elementary fields one can set both of $W_\psi$ matrices to identity. The Yukawa matrix is then given by
\begin{equation}
        (Y_e)^{ij}
        \simeq  g_\ast c_Y (\epsilon_{\ell}^*)^{\,\, i}_a \epsilon_e^{a j} =
        g_\ast c_Y\,
        (D_\ell
        R^\prime
        D_e)^{ij} ,
        \qquad
        R^\prime\equiv V_\ell^\dagger  V_e,
\label{eq:Yukawa-U3-SVD}
\end{equation}
and the charged-lepton Yukawa  singular values scale as:\footnote{The unitary matrices diagonalizing the charged-lepton Yukawa matrix have the parametric form 
$(U_L)_{ij}\sim\min\left[\varepsilon_\ell^i/\varepsilon_\ell^j,\varepsilon_\ell^j/\varepsilon_\ell^i\right]$,
and similarly for $U_R$ with $\varepsilon_\ell\to\varepsilon_e$, with the phases depending on the assumption on CP, see discussion around eq.~\eqref{eq:DeltaCPe-U3-2}.}
\begin{equation}
        y_i
        \sim
        g_\ast\,\epsilon_i^\ell\epsilon_i^e,
        \qquad
        i=e,\mu,\tau \,.
\label{eq:yukawa-scaling}
\end{equation}
As mentioned above, the dipole and the Yukawa matrices are aligned at the second order in mixings. The dominant lepton-flavor-violating effects then arise at the fourth order for the dipole operators, or at the leading (second) order from the Higgs-current operators. This is because Higgs-current operators involve the same chiralities and transform differently from the Yukawas under the flavor symmetry group.

The suppression of the dipole strongly weakens the present $\mu\to e\gamma$ and electron EDM constraints, while the current operators are probed by  $\mu\to 3e$ and $\mu N\to eN$ experiments which are expected to remarkably improve their sensitivity.\footnote{Higgs-current operators also mediate lepton-flavor-violating tau decays such as $\tau\to 3\mu$ or $\tau\to \mu \rho$, discussed in appendix~\ref{app:taudecays}. For the scenario considered here and for $r_i=1$, the strongest current bound on $m_\ast$ from tau decays is about a factor of six weaker than that from $\mu N\to eN$ conversion, placing the two within an order of magnitude. The projected improvements in muon searches will further widen this gap.} We will discuss these  operators in turn, starting with the dipole.

\paragraph{Dipole operators:}

With two insertions of the elementary--composite mixings, the $\mathrm{U}(3)_{\mathcal{O}}$  symmetry allows only the contraction
\begin{equation}
        \tilde C_{e\gamma}^{ij}\big|_{\epsilon^2}
        =
        \frac{e}{16\pi^2}\,
        \frac{g_\ast^3}{m_\ast^2}\,
        c_{e\gamma}^{(0)}
        (\epsilon_\ell^\dagger\epsilon_e)_{ij} = \frac{e}{16\pi^2}\,
        \frac{g_\ast^2}{m_\ast^2}\,
        \frac{c_{e\gamma}^{(0)}}{c_Y}\,
        (Y_e)_{ij}\,,
\label{eq:dipole-two-spurion-coefficient}
\end{equation}
where $c_{e\gamma}^{(0)}$ is an order-one coefficient and real due to the CP symmetry of the strong sector. %
Consequently, at order $\epsilon^2$, the dipole operator does not mediate charged-lepton flavor violation. Moreover, after diagonalizing the Yukawa couplings, there are no phases in $C_{e\gamma}$, and thus no electron EDM is generated at this order.

 The first lepton-flavor-violating dipole operators arise at order $\mathcal{O}(\epsilon^4)$. $\tilde{C}_{e\gamma}$ transforms under $\mathrm{U}(3)_\ell \times \mathrm{U}(3)_e \times \mathrm{U}(3)_{\mathcal O}$ as $ (\mathbf 3_\ell,\overline{\mathbf{3}}_e,\mathbf 1_{\mathcal O})$, and including all contractions up to fourth order in $\epsilon$, we have
\begin{align}
        \tilde C_{e\gamma}
        =
        \frac{e}{16\pi^2}\,
        \frac{g_\ast^3}{m_\ast^2}
        &\bigg[
      c_{e\gamma}^{(0)}\,\epsilon_\ell^\dagger\epsilon_e
        +
a_{e\gamma}^{(R)}\,\epsilon_\ell^\dagger\epsilon_e\Tr(\epsilon_e^\dagger\epsilon_e)+a_{e\gamma}^{(L)}\,\epsilon_\ell^\dagger\epsilon_e \Tr(\epsilon_\ell^\dagger\epsilon_\ell) \nonumber\\ 
&+
        c_{e\gamma}^{(R)}\,
        (\epsilon_\ell^\dagger\epsilon_e)
        (\epsilon_e^\dagger\epsilon_e)
        +
        c_{e\gamma}^{(L)}\,
        (\epsilon_\ell^\dagger\epsilon_\ell)
        (\epsilon_\ell^\dagger\epsilon_e)
        \bigg],
\label{eq:dipole-up-to-four-spurions}
\end{align}
where $a_{e\gamma}^{(L)},a_{e\gamma}^{(R)},c_{e\gamma}^{(L)},c_{e\gamma}^{(R)}$ are real and order one.
The terms in the first line are proportional to the Yukawa matrix and remain aligned.\footnote{There are also corrections to the Yukawa matrix at fourth order in $\epsilon$. Including them modifies the rotations defining the charged-lepton mass basis, so that the leading two-spurion dipole is no longer exactly diagonalized by these rotations. However, we have assumed that the leading dipole arises at one-loop order, while the Yukawa coupling has a leading contribution at tree level. The fourth-order terms in $\epsilon$ for both the Yukawa and dipole couplings arise at one loop. They are therefore relatively more important for the dipole coupling and can be neglected for the Yukawa coupling at the order considered here.}
The first term is aligned with $Y_e$, while the last two terms are the leading sources of charged-lepton flavor violation. Going to the mass basis, one finds
\begin{equation}
        C_{e\gamma} =
        \frac{e}{16\pi^2}\,
        \frac{g_\ast^2}{m_\ast^2}
        \frac{1}{c_Y}
        \bigg[
        c_{e\gamma}\,Y_e^{\rm diag}
        +
       c_{e\gamma}^{(R)}\,
        Y_e^{\rm diag}\,
        X_R
        +
        c_{e\gamma}^{(L)}\,
        X_L\,
        Y_e^{\rm diag}
        \bigg],
\label{eq:dipole-mass-basis-four-spurion}
\end{equation}
where $c_{e \gamma}$ includes the subleading diagonal contributions from $a_{e\gamma}^{(L/R)}$, and where we have defined
\begin{equation}
        X_R
        =
        U_R^\dagger
        (\epsilon_e^\dagger\epsilon_e)
        U_R = U_R^\dagger
        D_e^2
        U_R ,
        \qquad
        X_L
        =
        U_L^\dagger
        (\epsilon_\ell^\dagger\epsilon_\ell)
        U_L= U_L^\dagger
        D_\ell^2
        U_L.
\label{eq:XL-XR-def}
\end{equation}
Here $U_{L,R}$ are the unitary rotations on the elementary fields that diagonalize the Yukawa matrix.
The matrices $X_L$ and $X_R$ are Hermitian matrices, and their $\mu-e$ off-diagonal entries scale as, %
\begin{equation}
        (X_L)_{e\mu}
        \sim
        \epsilon^{\ell}_{e}\epsilon^{\ell}_{\mu}
        \sim
        \frac{\sqrt{y_e y_\mu}}{g_\ast}\,r_e r_\mu,
        \qquad
        (X_R)_{e\mu}
        \sim
        \epsilon^{e}_{e}\epsilon^{e}_{\mu}
        \sim
        \frac{\sqrt{y_e y_\mu}}{g_\ast}\,\frac1{r_e r_\mu}\,.
\label{eq:XL-XR-generic-U3-scaling}
\end{equation}
Substituting these relations into \cref{eq:dipole-mass-basis-four-spurion} yields the corresponding off-diagonal dipole operator coefficients:
\begin{align}
        C_{e\gamma}^{e\mu}
        &\sim
        \frac{e}{16\pi^2}\,
        \frac{g_\ast^2}{m_\ast^2}
        \left[
        y_\mu\,
        \frac{\sqrt{y_e y_\mu}}{g_\ast}\,r_e r_\mu
        +
        y_e\,
        \frac{\sqrt{y_e y_\mu}}{g_\ast}\,\frac1{r_e r_\mu}
        \right],
        \\
        C_{e\gamma}^{\mu e}
        &\sim
        \frac{e}{16\pi^2}\,
        \frac{g_\ast^2}{m_\ast^2}
        \left[
        y_\mu\,
        \frac{\sqrt{y_e y_\mu}}{g_\ast}\,\frac1{r_e r_\mu}
        +
        y_e\,
        \frac{\sqrt{y_e y_\mu}}{g_\ast}\,r_e r_\mu
        \right].
\label{eq:mue-dipole-r}
\end{align}
Thus the contribution to the $\mu\to e\gamma$ rate is controlled by
\begin{equation}
         \max\left\{
        |C_{e\gamma}^{e\mu}|,
        |C_{e\gamma}^{\mu e}|
        \right\}\sim
        \frac{e}{16\pi^2}\,
        \frac{g_\ast^2}{m_\ast^2}
        y_\mu\,
        \frac{\sqrt{y_e y_\mu}}{g_\ast}
        \max \left\{
        r_e r_\mu,
        \frac1{r_e r_\mu}
        \right\}.
\label{eq:mue-dipole-generic-U3-envelope}
\end{equation}
Comparing %
with the anarchic scenario, there is an additional parametric suppression,
\begin{equation}
        \frac{
        C_{e\gamma}^{\U{3}}
        }{
        C_{e\gamma}^{\rm anarchic}
        }
        \sim
        \frac{y_\mu}{g_\ast} \frac{\max \left\{
        r_e r_\mu,
        \frac1{r_e r_\mu}
        \right\}}{\max \left\{\frac{r_\mu}{r_e},\frac{r_e}{r_\mu}\right\}}\,,
\label{eq:dipole-coefficient-suppression-U3}
\end{equation}
which brings the bound on the scale of compositeness $m_*$ from $\mu \to e \gamma$ down by about two orders of magnitude, making it subdominant.

We also note that $\mathcal{O}_{eH}$ shares the same flavor quantum numbers as the Yukawa and dipole operators. Similar to the dipole case, it is automatically aligned with Yukawa at leading order. Its effects are always subdominant for the processes we are considering.

\paragraph{Electron EDM:}
Up to the order discussed in eq.~\eqref{eq:dipole-mass-basis-four-spurion}, there is no contribution to the electron EDM. More precisely, $c_{e\gamma}$ is real and, since $X_L$ and $X_R$ are both Hermitian and $c_{e\gamma}^{(L/R)}$ are real, the diagonal entries of $C_{e\gamma}$ are real up to this order, and do not contribute to the electron EDM.

The first non-vanishing leptonic EDM arises at sixth order in the mixing. We parametrize it as
\begin{equation}
        \mathrm{Im}\,C_{e\gamma}^{ee}
        \sim
        \frac{e}{16\pi^2}
        \frac{g_\ast^2}{m_\ast^2}
        y_e\,\Delta_{\rm CP}^{e},
\label{eq:EDM-suppression-U3}
\end{equation}
where $\Delta_{\rm CP}^e$ is the phase of $C^{ee}_{e\gamma}$ and captures the additional suppression compared to the anarchic scenario. One such contribution in the mass basis is given by
\begin{equation}
        C_{e\gamma}^{(6)}
        \supset
        \frac{e}{16\pi^2}
        \frac{g_\ast^2}{m_\ast^2}
        c_{LR}\,
        X_LY_e^{\rm diag}X_R,
        \qquad
        c_{LR}\in\mathbb R.
\label{eq:six-spurion-edm-U3}
\end{equation}
This leads to
\begin{equation}
        \Delta_{\rm CP}^{e}\big|_{\rm gen}
        \sim
        \frac1{y_e}\,
        \mathrm{Im}\left(X_LY_e^{\rm diag}X_R\right)_{ee}
        \sim
        \sum_{k=\mu,\tau}
        \frac{y_k}{y_e}\,
       \mathrm{Im}\left[ (X_L)_{ek}(X_R)_{ke} \right]\,.
\label{eq:DeltaCPe-U3}
\end{equation}
The size of $ \Delta_{\rm CP}^{e}\big|_{\rm gen}$ depends on our assumption about CP in the strong sector in the UV.
As we will discuss below in \cref{sec:symandhier}, one can imagine that CP is a symmetry of the strong sector in the UV, where the hierarchical structure of the mixings is generated. However, one can also imagine a case where CP is a symmetry of the strong sector only in the IR, where the $\U{3}_{\mathcal{O}}$ symmetry is present. These two cases lead to a different scaling of $\Delta_{\rm CP}^{e}\big|_{\rm gen}$: in the first case, as discussed in footnote~\ref{footnote:Rorthogonal}, the strong-sector matrices $c$ are real in the same basis where the mixings are row-hierarchical. In this case, $R$ in eq.~\eqref{eq:U3-mixings-symbasis} is an orthogonal matrix and $R'$ in eq.~\eqref{eq:Yukawa-U3-SVD} differs from $R$ only by a parametrically small unitary rotation.
$R'$ is thus predominantly real, with parametrically small phases.
On the other hand, in the second case, the $c$'s are real in a basis with generic mixings of eq.~\eqref{eq:triangular-eps}, leading to a generic unitary $R'$.  
We then find
\begin{align}
        \Delta_{\rm CP}^{e}\big|_{\rm gen}
        \sim
       \begin{cases}
                  \frac{y_\tau^2}{g_\ast^2}  &\text{CP in the IR,} \\
           \frac{y^2_\tau}{g_*^2} \sqrt{\frac{y_{\mu}}{y_\tau}}(\frac{r_\mu}{r_\tau}+\frac{r_\tau}{r_\mu}) &\text{CP in the UV} 
       \end{cases}.
\label{eq:DeltaCPe-U3-2}
\end{align}
Both estimates are parametrically much smaller than the anarchic CP-violating dipole estimate in \cref{eq:edm-anarchic-scaling} as well as the anarchic CP-conserving estimate in \cref{eq:edm-anarchic-CP}. %

\paragraph{Current operators:}

At two-spurion order, the Higgs-current operators have the parametric scaling
\begin{align}
        \mathcal L_{H\ell}
        &\supset
        \frac{g_\ast^2}{m_\ast^2}\,
        c_{H\ell}\,
        (\epsilon_\ell^\dagger\epsilon_\ell)_{ij}
        \left(H^\dagger i\overleftrightarrow D_\mu H\right)
        \bar \ell_L^i\gamma^\mu \ell_L^j ,
        \\
        \mathcal L_{He}
        &\supset
        \frac{g_\ast^2}{m_\ast^2}\,
        c_{He}\,
        (\epsilon_e^\dagger\epsilon_e)_{ij}
        \left(H^\dagger i\overleftrightarrow D_\mu H\right)
        \bar e_R^i\gamma^\mu e_R^j ,
\end{align}
with real order-one coefficients $c_{H\ell}$ and $c_{He}$. Unlike the dipole operator, these structures are already not aligned with the charged-lepton Yukawa matrix at second order in the mixings. After rotating to the charged-lepton mass basis, the leading $\mu e$ components are given by %
\begin{equation}
        C_{H\ell}^{e\mu}
        \sim
        \frac{g_\ast^2}{m_\ast^2}
        (X_L)_{e\mu}
        \sim
        \frac{g_\ast}{m_\ast^2}
        \sqrt{y_e y_\mu}\,r_e r_\mu,
        \qquad
        C_{He}^{e\mu}
        \sim
        \frac{g_\ast^2}{m_\ast^2}
        (X_R)_{e\mu}
        \sim
        \frac{g_\ast}{m_\ast^2}
        \sqrt{y_e y_\mu}\,\frac1{r_e r_\mu}.
\label{eq:current-U3-scaling}
\end{equation}
Therefore the %
off-diagonal components of the
current operators have the same scaling as in the anarchic scenario. 

While $\U{3}_{\mathcal O}$ flavor symmetry does not by itself protect the flavor-changing $Z$ couplings, additional protection can arise from a custodial $P_{LR}$ symmetry, provided the elementary--composite mixings
respect it. As discussed in appendix~\ref{sec:PLR}, the embeddings considered here allow the zero-momentum charged-lepton coupling of either chirality to be protected, but not both simultaneously while generating the charged-lepton Yukawa interaction. The current of the unprotected
chirality therefore remains. Momentum-dependent interactions also survive for the protected chirality. Derivative $Z$ exchange carries an additional low-energy suppression $|q^2|/m_Z^2$, with the momentum transfer set by the mass of the decaying lepton, $|q^2|\sim m_{\ell_j}^2$. Derivative photon exchange generates a contact interaction without this momentum suppression; its amplitude is suppressed parametrically by a factor $\sim e^2/g_*^2$ relative to the Higgs-current contribution for the same chirality without $P_{LR}$.
We remind the reader that modifications to flavor-preserving $Z$ couplings lead to the constraint in eq.~\eqref{eq:MFV-Z-bound-opt}, which applies to this scenario and the other scenarios we discuss below.
The $P_{LR}$ symmetry provides a protection from such flavor preserving deviations as well, but can only protect a single charged-lepton chirality at a time.\footnote{As discussed in appendix~\ref{sec:PLR}, custodial symmetry can also protect the $Z$ couplings to right-handed leptons from corrections induced by Higgs-current operators, provided their embeddings satisfy $T_R^3=0$.}

\begin{figure}[t]
    \centering
\includegraphics[width=0.99\linewidth]{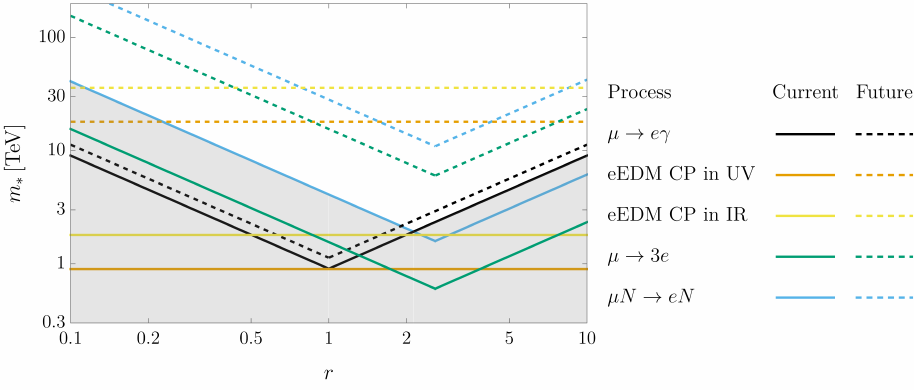}
    \caption{
  Current bound (solid) and future  sensitivity (dashed) from lepton-flavor-violating observables for the generic non-aligned $\U{3}$ scenario with additional $P_{LR}$ symmetry protecting left-handed charged leptons for $r_e=r_\mu=r_\tau=r$ and fixed $g_*=2$. The electron EDM constraint is shown for two scenarios: one in which CP is preserved in the UV and one in which CP and $\U{3}$ arise accidentally in the IR at the same scale. The gray shaded region is excluded by current constraints. 
    \label{fig:U3nonalignedPLR} }
\end{figure}

\paragraph{Discussion:}
\begin{figure}[t]
    \centering
\includegraphics[width=0.99\linewidth]{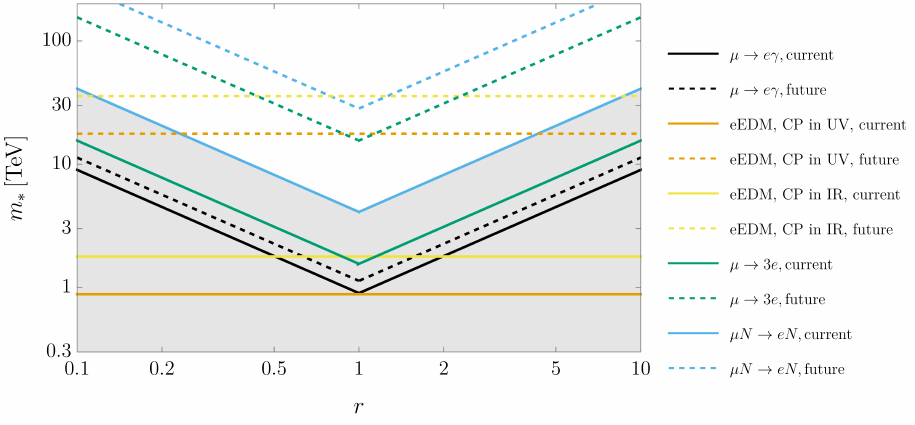}
    \caption{Current bound (solid) and future  sensitivity (dashed) from lepton-flavor-violating observables for the generic non-aligned $\U{3}$ scenario without assuming additional $P_{LR}$  symmetry for $r_e=r_\mu=r_\tau=r$ and fixed $g_*=2$. The electron EDM constraint is shown for two scenarios: one in which CP is preserved in the UV and one in which CP and $\U{3}$ arise accidentally in the IR at the same scale. The gray shaded region is excluded by current constraints. 
    \label{fig:U3nonaligned} }
\end{figure}
We summarize in \cref{tab:U3CP-summary} the estimates of the size of the Wilson coefficients and the main constraints on the generic non-aligned $\mathrm{U}(3)\times\mathrm{CP}$ scenario. We also show current constraints and future reach in figures~\ref{fig:U3nonaligned} and~\ref{fig:U3nonalignedPLR} without and with the additional assumption of $P_{LR}$ protection.

The phenomenological picture is qualitatively different from anarchic partial compositeness. In this scenario, off-diagonal dipole operators arise only at fourth order in the mixings and thus the current bound from $\mu \to e \gamma$ is around the TeV scale. Moreover, the imaginary part of the dipole operators arises only at sixth order in the mixing, and the bound on $m_*$ from the electron EDM is also around the TeV scale. Already now the strongest constraints in the lepton sector arise from $\mu N \to e N$ and $\mu \to 3e$ and are in the few-TeV range. The future improvements of the rates by four orders of magnitude will therefore probe this scenario up to tens of TeV. 
Another way to see this is by noting that in the anarchic scenario 
\begin{equation}
        \frac{\mathrm{BR}(\mu\to3e)}
             {\mathrm{BR}(\mu\to e\gamma)}
        \simeq
        \frac{\alpha}{3\pi}
        \left[
        \log\left(\frac{m_\mu^2}{m_e^2}\right)
        -\frac{11}{4}
        \right]
        \simeq
        6\times10^{-3}.
\end{equation}
On the other hand, for the scenario considered here, this ratio is enhanced by four orders of magnitude,
\begin{equation}
\frac{\text{BR}(\mu \to 3e)}{\text{BR}(\mu \to e \gamma)} \sim \frac{\pi}{6 \, \alpha}\sim 70.
\label{eq:brscaling}
\end{equation}
The expected future improvement in electron EDM complements these probes in a similar mass range.

\begin{table}[t]
\centering
\setlength{\tabcolsep}{3pt}
\renewcommand{\arraystretch}{1.25}
\resizebox{\textwidth}{!}{%
\begin{tabular}{|c|c|c|c|}
\hline
Generic $\mathrm{U}(3)\times\mathrm{CP}$
&
$C_{e\gamma}^{e\mu},\,C_{e\gamma}^{\mu e}$
&
$C_{H\ell}^{e\mu},\,C_{He}^{e\mu}$
&
$\frac{16\pi^2}{e} \frac{m_\ast^2}{y_eg_\ast^2}\times\mathrm{Im}\,C_{e\gamma}^{ee}$
\\
\hline
scaling
&
\begin{tabular}{@{}c@{}}
$\displaystyle
C_{e\gamma}^{e\mu}
\sim
\frac{e}{16\pi^2}
\frac{g_\ast^2}{m_\ast^2}
y_\mu
\frac{\sqrt{y_e y_\mu}}{g_\ast}
r_e r_\mu$
\\[2mm]
$\displaystyle
C_{e\gamma}^{\mu e}
\sim
\frac{e}{16\pi^2}
\frac{g_\ast^2}{m_\ast^2}
y_\mu
\frac{\sqrt{y_e y_\mu}}{g_\ast}
\frac1{r_e r_\mu}$
\end{tabular}
&
\begin{tabular}{@{}c@{}}
$\displaystyle
C_{H\ell}^{e\mu}
\sim
\frac{g_\ast}{m_\ast^2}
\sqrt{y_e y_\mu}\,
r_e r_\mu$
\\[2mm]
$\displaystyle
C_{He}^{e\mu}
\sim
\frac{g_\ast}{m_\ast^2}
\sqrt{y_e y_\mu}\,
\frac1{r_e r_\mu}$
\end{tabular}
&
\begin{tabular}{@{}c@{}}
$\displaystyle
\Delta_{\rm CP}^{e}\big|_{\rm UV}
\sim
\frac{y_\tau^2}{g_\ast^2}
\sqrt{\frac{y_\mu}{y_\tau}} \; \max
\left[
\frac{r_\mu}{r_\tau}
,
\frac{r_\tau}{r_\mu}
\right]$
\\[2mm]
$\displaystyle
\Delta_{\rm CP}^{e}\big|_{\rm IR}
\sim
\frac{y_\tau^2}{g_\ast^2}$
\end{tabular}
\\
\hline
current bounds on $m_\ast$ [TeV]
&
\begin{tabular}{@{}c@{}}
$\displaystyle
0.64\sqrt{g_\ast}
\quad(\mu\to e\gamma)$
\\[1mm]
$\displaystyle
0.11\sqrt{g_\ast}
\quad(\mu\to3e)$
\\[1mm]
$\displaystyle
0.11\sqrt{g_\ast}
\quad(\mu N\to eN)$
\end{tabular}
&
\begin{tabular}{@{}c@{}}
$\displaystyle
2.9\sqrt{g_\ast}
\quad(\mu N\to eN)$
\\[1mm]
$\displaystyle
1.1\sqrt{g_\ast}
\quad(\mu\to3e)$
\end{tabular}
&
\begin{tabular}{@{}c@{}}
$\displaystyle
0.9
\quad(\text{CP in the UV})$
\\[1mm]
$\displaystyle
1.8
\quad(\text{CP in the IR})$
\end{tabular}
\\
\hline
future projections on $m_\ast$ [TeV]
&
\begin{tabular}{@{}c@{}}
$\displaystyle
0.8 \sqrt{g_\ast}
\quad(\mu\to e\gamma)$
\\[1mm]
$\displaystyle
1.1 \sqrt{g_\ast}
\quad(\mu\to3e)$
\\[1mm]
$\displaystyle
0.95 \sqrt{g_\ast}
\quad(\mu N\to eN)$
\end{tabular}
&
\begin{tabular}{@{}c@{}}
$\displaystyle
20 \sqrt{g_\ast}
\quad(\mu N\to eN)$
\\[1mm]
$\displaystyle
11\sqrt{g_\ast}
\quad(\mu\to3e)$
\end{tabular}
&
\begin{tabular}{@{}c@{}}
$\displaystyle
18 
\quad(\text{CP in the UV})$
\\[1mm]
$\displaystyle
36
\quad(\text{CP in the IR})$
\end{tabular}
\\
\hline
\end{tabular}%
}
\caption{
Parametric scaling of the Wilson coefficients, current bounds and future reach on the compositeness scale $m_*$ in the generic non-aligned $\mathrm{U}(3)_{\mathcal O}\times\mathrm{CP}$ scenario. 
The numbers shown correspond to the left--right symmetric case, $r_i=1$. We show the corresponding experiment in parentheses with the sensitivity given in \cref{tab:experimental-inputs}.
The EDM constraint is shown for two scenarios: one in which CP is a symmetry in the UV and one in which it arises accidentally at the same scale as $\mathrm{U}(3)_{\mathcal O}$.
All constraints should be understood as estimates with $\mathcal{O}(1)$ uncertainty.  }

\label{tab:U3CP-summary}
\end{table}
\noindent

\subsubsection{Aligned $\U{3}_{\mathcal{O}}$}\label{subsec:alignedU3}
The other realization of the $\U{3}$ scenario corresponds to mixings that are simultaneously row-hierarchical and upper-triangular in the same basis i.e. $R = \mathbb 1$. As we will see in \cref{sec:symandhier}, this can be realized in a scenario where a $\U{1}^3$ subgroup of the strong-sector $\U{3}$ symmetry group persists to the UV.

\begin{figure}[t]
    \centering
\includegraphics[width=0.99\linewidth]{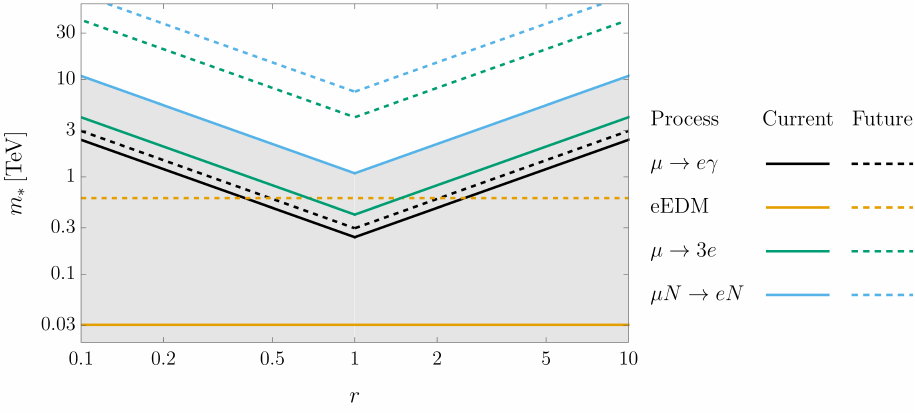}
    \caption{
   Current bound (solid) and future  sensitivity (dashed) from lepton-flavor-violating observables for the aligned $\U{3}$ scenario without assuming additional $P_{LR}$ symmetry for $r_e=r_\mu=r_\tau=r$ and fixed $g_*=2$. The gray shaded region is excluded by current constraints.  
    \label{fig:U3aligned} }
\end{figure}

\begin{figure}[t]
    \centering
\includegraphics[width=0.99\linewidth]{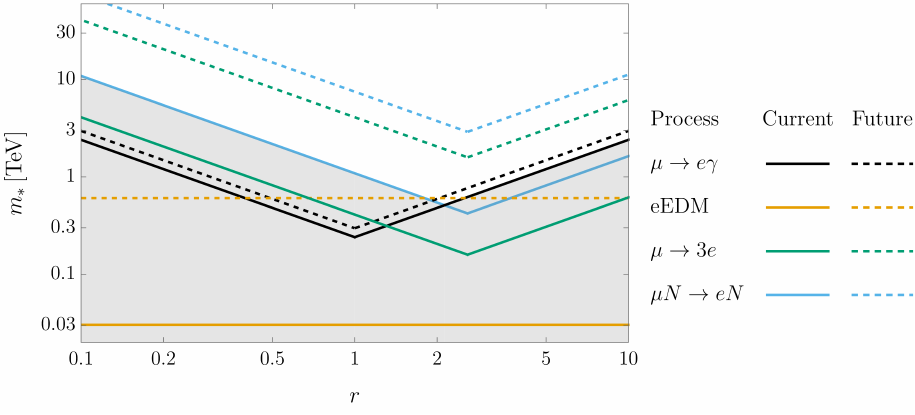}
    \caption{ Current bound (solid) and future  sensitivity (dashed) from lepton-flavor-violating observables for the aligned $\U{3}$ scenario with an additional $P_{LR}$ symmetry protecting left-handed charged leptons for $r_e=r_\mu=r_\tau=r$ and fixed $g_*=2$. The gray shaded region is excluded by current constraints.  
    \label{fig:U3alignedPLR} }
\end{figure}

In this aligned case, we can start in a basis where the mixings are row-hierarchical and the $\U{3}_\mathcal{O}$ symmetry is manifest. The off-diagonal components of the Yukawa matrix scale as $Y_{ij} \sim \min(y_i, y_j)$. Going to the mass basis requires a rotation \begin{equation}
    (U_L)_{ij} \sim (U_R)_{ij} \sim y_i/y_j \quad \text{for} \quad i<j.
\end{equation}
As in the generic non-aligned case, the dipole coefficients are given by the expression in \cref{eq:dipole-up-to-four-spurions}. However, in this case, for $X_L$ and $X_R$ in the mass basis, we find
\begin{equation}
        (X_L)_{e\mu}^{\rm al.}
        \sim
        \frac{y_e}{g_\ast}
        \left[
        r_e^2+r_\mu^2+
        \frac{y_\mu}{y_\tau}r_\tau^2
        \right] \,,
        \qquad
        (X_R)_{e\mu}^{\rm al.}
        \sim
        \frac{y_e}{g_\ast}
        \left[
        \frac{1}{r_e^2}+\frac{1}{r_\mu^2}
        +\frac{y_\mu}{y_\tau}\frac{1}{r_\tau^2}
        \right]\,,
\end{equation}
which are suppressed compared to the non-aligned case \eqref{eq:XL-XR-generic-U3-scaling}.
This leads to
\begin{equation}
       \max \left\{\left|C_{e \gamma}^{e \mu}\right|,\left|C_{e \gamma}^{\mu e}\right|\right\}_{\rm{al.}}
        \sim
        \frac{e}{16\pi^2}
        \frac{g_\ast^2}{m_\ast^2}
        y_\mu\,
        \frac{y_e}{g_\ast}
        \max \left(
        r_e^2,
        \frac{1}{r_e^2},r_\mu^2,\frac{1}{r_\mu^2},
        \frac{y_\mu}{y_\tau}r_\tau^2
        ,\frac{y_\mu}{y_\tau}\frac{1}{r_\tau^2}
        \right).
\end{equation}
Regarding the Higgs-current operators, they scale as
\begin{equation}
        C_{H\ell}^{e\mu}\big|_{\rm al.}
        \sim
        \frac{g_\ast}{m_\ast^2}
        y_e
        \left[
        r_e^2+r_\mu^2+
        \frac{y_\mu}{y_\tau}r_\tau^2
        \right],
        \qquad
        C_{He}^{e\mu}\big|_{\rm al.}
        \sim
        \frac{g_\ast}{m_\ast^2}
        y_e
        \left[
        \frac{1}{r_e^2}+\frac{1}{r_\mu^2}
        +\frac{y_\mu}{y_\tau}\frac{1}{r_\tau^2}
        \right].
\end{equation}
Finally, the electron EDM again arises only at sixth order in the mixings, as given by eq.~\eqref{eq:DeltaCPe-U3},
\begin{equation}
        \mathrm{Im}\,C_{e\gamma}^{ee}\big|_{\rm al.}
        \sim
        \frac{e}{16\pi^2}
        \frac{g_\ast^2}{m_\ast^2}
        y_e\,
        \frac{y_e}{g_\ast^2}y_\tau \left[ \frac{\max_i(r_i)}{\min_j(r_j)}
         \right]^2\,.
\end{equation}
We summarize in \cref{tab:U3CP-aligned-summary} the estimates of the size of the Wilson coefficients and the main constraints on the aligned $\mathrm{U}(3)\times\mathrm{CP}$ scenario.\footnote{For the scenario considered here, tau LFV  is dominantly mediated by the Higgs-current operators. For $r_i=1$, they yield slightly weaker bounds on $m_\ast$ than muon LFV. For $r_\tau$ deviating from order one, these bounds can become important, similar to the scenario discussed below in section~\ref{sec:U2U1CP}.} We again show current constraints and future reach in figures~\ref{fig:U3aligned} and~\ref{fig:U3alignedPLR} without and with the additional assumption of $P_{LR}$ protection.

For $r_i=1$, the aligned limit suppresses the $\mu-e$ coefficients of both the dipole operators and Higgs-current operators relative to generic $\mathrm{U}(3)$ by $\sim \sqrt{y_e/y_\mu}$. 
Additionally, setting again $r_i=1$, the imaginary part of the $e-e$ component of the dipole operator contributing to the electron EDM is suppressed compared to the generic non-aligned $\U{3}$ case with CP in the UV by
$\sim y_e/\sqrt{y_\mu y_\tau}$.

Finally, we note that the ratio of the branching fractions for $\mu \to e e e$ and $\mu \to e \gamma$ is again enhanced by four orders of magnitude compared with the anarchic case and is given by eq.~\eqref{eq:brscaling}. %

\begin{table}[t]
\centering
\setlength{\tabcolsep}{3pt}
\renewcommand{\arraystretch}{1.25}
\resizebox{\textwidth}{!}{%
\begin{tabular}{|c|c|c|c|}
\hline
Aligned $\mathrm{U}(3)\times\mathrm{CP}$
&
$\max\{|C_{e\gamma}^{e\mu}|,|C_{e\gamma}^{\mu e}|\}$
&
$\max\{|C_{H\ell}^{e\mu}|,\,|C_{He}^{e\mu}|\}$
&
$\frac{16\pi^2}{e} \frac{m_\ast^2}{y_eg_\ast^2}\times\mathrm{Im}\,C_{e\gamma}^{ee}$
\\
\hline
scaling
&
\begin{tabular}{@{}c@{}}
$\displaystyle
\frac{e}{16\pi^2}
\frac{g_\ast^2}{m_\ast^2}
y_\mu
\frac{y_e}{g_\ast} \times$
\\[2mm] $ \max \left(
        r_e^2,
        \frac{1}{r_e^2},r_\mu^2,\frac{1}{r_\mu^2},
        \frac{y_\mu}{y_\tau}r_\tau^2
        ,\frac{y_\mu}{y_\tau}\frac{1}{r_\tau^2}
        \right)$
        \end{tabular}
&
\begin{tabular}{@{}c@{}}
$\displaystyle \frac{g_\ast}{m_\ast^2}
y_e \times $ \\[2mm] $ \max \left(
        r_e^2,
        \frac{1}{r_e^2},r_\mu^2,\frac{1}{r_\mu^2},
        \frac{y_\mu}{y_\tau}r_\tau^2
        ,\frac{y_\mu}{y_\tau}\frac{1}{r_\tau^2}
        \right)$
 \end{tabular}
&
\begin{tabular}{@{}c@{}}
$\displaystyle
\Delta_{\rm CP}^{e}\big|_{\rm al}
\sim
\frac{y_e}{g_\ast^2}y_\tau \left[ \frac{\max_i(r_i)}{\min_j(r_j)}
         \right]^2$
\end{tabular}
\\
\hline
current bounds on $m_\ast$ [TeV]
&
\begin{tabular}{@{}c@{}}
$\displaystyle
0.17\sqrt{g_\ast}
\quad(\mu\to e\gamma)$
\\[1mm]
$\displaystyle
0.03\sqrt{g_\ast}
\quad(\mu\to3e)$
\\[1mm]
$\displaystyle
0.03\sqrt{g_\ast}
\quad(\mu N\to eN)$
\end{tabular}
&
\begin{tabular}{@{}c@{}}
$\displaystyle
0.77\sqrt{g_\ast}
\quad(\mu N\to eN)$
\\[1mm]
$\displaystyle
0.29\sqrt{g_\ast}
\quad(\mu\to3e)$
\end{tabular}
&
$\displaystyle
0.03$ 
\\
\hline
future projections on $m_\ast$ [TeV]
&
\begin{tabular}{@{}c@{}}
$\displaystyle
0.21\sqrt{g_\ast}
\quad(\mu\to e\gamma)$
\\[1mm]
$\displaystyle
0.30\sqrt{g_\ast}
\quad(\mu\to3e)$
\\[1mm]
$\displaystyle
0.25\sqrt{g_\ast}
\quad(\mu N\to eN)$
\end{tabular}
&
\begin{tabular}{@{}c@{}}
$\displaystyle
5.3\sqrt{g_\ast}
\quad(\mu N\to eN)$
\\[1mm]
$\displaystyle
2.9\sqrt{g_\ast}
\quad(\mu\to3e)$
\end{tabular}
&
$\displaystyle
0.6$
\\
\hline
\end{tabular}%
}
\caption{Parametric scaling of the Wilson coefficients, current bounds and future reach on the compositeness scale $m_*$ in the aligned $\mathrm{U}(3)_{\mathcal O}\times\mathrm{CP}$ scenario. 
The numbers shown correspond to the left--right symmetric case, $r_i=1$. We show the corresponding experiment in parentheses with the sensitivity given in \cref{tab:experimental-inputs}.
All constraints should be understood as estimates with $\mathcal{O}(1)$ uncertainty.}
\label{tab:U3CP-aligned-summary}
\end{table}

\subsection{$\mathrm{U}(2)\times\mathrm{U}(1)\times\mathrm{CP}$}
\label{sec:U2U1CP}
One may ask whether the full $\U{3}_\mathcal{O}$ symmetry is necessary to suppress $\mu\to e$ transitions and the electron EDM sufficiently to allow TeV-scale compositeness. At the same time, an ongoing experimental program is searching for lepton-flavor violation in $\tau$ decays, see e.g.~\cite{banerjee2022snowmass2021whitepaper}. Together, these considerations motivate scenarios in which transitions between the first two lepton families are more strongly suppressed than those involving the third.
Following this line, we consider a scenario where the strong sector has a symmetry
\begin{equation}
    G_{\text{strong}}= \mathrm{U}(2)_{\mathcal{O}}\times\mathrm{U}(1)_\mathcal{O}\times\mathrm{CP}\,.
\end{equation} 
We assume the composite operators that mix with the elementary leptons to transform as
    \begin{equation}
        \mathcal O_\ell^a \sim \left(\mathbf 2_0 \oplus \mathbf{1_1}\right)_{\mathcal O},
        \qquad
        \mathcal O_e^a \sim \left(\mathbf 2_0 \oplus \mathbf{1_1}\right)_{\mathcal O}\,,
       \label{transf U{2}}
\end{equation}
so that the mixings have the following transformation properties under $\U{3}_\ell \times \U{3}_e \times G_{\text{strong}}$,
\begin{equation}
        \epsilon_\ell \sim (\overline{\mathbf 3}_\ell,\mathbf 1_e,\mathbf {2_0}\oplus\mathbf {1_1}),
        \qquad
        \epsilon_e \sim (\mathbf 1_\ell,\overline{\mathbf 3}_e,\mathbf {2_0} \oplus \mathbf{1_1}).
\label{eq:spurions-U2 X U1}
\end{equation}
It is worth emphasizing an important difference with respect to the $\U{3}_\mathcal{O}$ scenario in forming invariants of $G_{\text{strong}}$ that transform as $(\mathbf 3_\ell,\overline{\mathbf 3}_e)$. In particular, at leading order in the mixings, there are now two such invariants, obtained from contractions of $\U{2}_{\mathcal{O}}$ doublets and singlets, and therefore the dipole and Yukawa matrices are not automatically aligned at this order. 

There are three different scenarios to consider. First, if both $\U{2}_\mathcal{O}$ and $\U{1}_\mathcal{O}$ are not manifest in the basis where the mixings are row-hierarchical, all observables scale as in the anarchic case, and we do not pursue this case further. Next, we consider the case in which the $\U{2}_\mathcal{O}$-singlet operators have smaller anomalous dimensions than the doublet operators. Their associated mixings therefore become larger in the IR, so that the dominant row of each mixing matrix corresponds to the singlet direction. In this scenario, $\U{1}_\mathcal{O}$ is manifest in the row-hierarchical basis. Finally, there is a scenario in which the full $\U{2}_\mathcal{O}\times \U{1}_\mathcal{O}$ is manifest while the mixings are row-hierarchical. We now discuss the last two cases.

\subsubsection{Non-aligned $\U{2}_\mathcal{O}$}
We first discuss the case where the $\U{2}_\mathcal{O}$ is not manifest in the basis where the mixings are row-hierarchical.
Let us work in a basis where $\U{2}_\mathcal{O}$ is manifest. This can again be achieved by a $\U{2}$ unitary rotation $R$ acting only on $\mathcal{O}_e$ with $R=\text{diag}(R_{2\times2},\mathbb{1})$.
In such a basis, the mixings $\epsilon_\ell$ and $\epsilon_e$ are given by
\begin{equation}
        \epsilon_\ell
        = \hat{\epsilon}
_\ell,
        \qquad
        \epsilon_e
        =
        R \hat{\epsilon}_e\,,
\label{eq:U2-mixings-symbasis}
\end{equation}
where $\hat{\epsilon}$ denotes the mixing in the row-hierarchical basis.\footnote{Similar to above, if CP is a symmetry of the strong sector in the UV where the hierarchies are generated, all the phases can be put in the mixings in the row-hierarchical basis. Thus an orthogonal rotation $R$ suffices to go to the basis where $\U{2}_\mathcal{O}$ is manifest. We find that this does not change our results for the electron EDM.}
The charged-lepton Yukawa matrix is
\begin{equation}
        Y_e
        \sim
        g_\ast
        \left[
        \hat{\epsilon}_\ell^\dagger c_Y^{(2)} \hat{\epsilon}_e
        +
        \hat{\epsilon}_\ell^\dagger c_Y^{(3)} \hat{\epsilon}_e
        \right],
        \qquad
        c_Y^{(3)} = 
        \begin{pmatrix}
       0
        &
        0
        &
        0
        \\
        0
        &
        0
        &
        0
        \\
        0
        &
        0
        &
        c_Y
        \end{pmatrix},
          \qquad 
        c_Y^{(2)} = \begin{pmatrix} R_{2\times 2} &
        0 \\
        0
        &
        0
         \end{pmatrix}\,,
\label{eq:U2U1-generic-Yukawa}
\end{equation}
where $R_{2\times 2}$ represents the $\U{2}$ rotation above.
The coefficients of the dipole operators have the same structure with the replacement $c_Y\to c_{e\gamma}$.
After going to the mass basis, the coefficients of the dipole operators scale as
\begin{equation}
    C^{ij}_{e\gamma} \propto Y^{\rm diag}_{ij}+(c_{e\gamma}-c_Y)(U_L)_{3i}^\ast y_{\tau}(U_R)_{3j} \label{dipole U2},\qquad
        {c_{e\gamma},c_Y}\in\mathbb R.
\end{equation} 
The matrices that diagonalize the Yukawa scale as
\begin{equation}
       (U_L)_{ij} \sim \begin{pmatrix}
      1 
        &
        \frac{\epsilon_\ell^1}{\epsilon_\ell^2}
        &
        \frac{\epsilon^1_\ell \epsilon^2_e}{y_\tau/g_*}
        \\
        \frac{\epsilon_\ell^1}{\epsilon_\ell^2}
        &
        1
        &
        \frac{y_\mu}{y_\tau}
        \\
        \frac{y_e}{y_\tau}
        &
         \frac{y_\mu}{y_\tau}
        &
       1
        \end{pmatrix}, \qquad  (U_R)_{ij}\sim \begin{pmatrix}
      1 
        &
        \frac{\epsilon_e^1}{\epsilon_e^2}
        &
        \frac{\epsilon^1_e \epsilon^2_\ell}{y_\tau/g_*}
        \\
        \frac{\epsilon_e^1}{\epsilon_e^2}
        &
        1
        &
        \frac{y_\mu}{y_\tau}
        \\
        \frac{y_e}{y_\tau}
        &
         \frac{y_\mu}{y_\tau}
        &
       1
        \end{pmatrix}.
\end{equation}
Consequently, the off-diagonal $\tau-\mu$ entry of the coefficient of the dipole operators scales as
\begin{equation}\label{eq:U2taumudip}
   C_{e\gamma}^{\tau \mu} \sim C_{e\gamma}^{\mu \tau} \sim \frac{e}{16\pi^2}  \frac{g_*^2}{m_\ast^2}\, y_\mu, 
\end{equation}
while the $\mu - e$ entry scales as
\begin{equation}
    C_{e\gamma}^{e\mu}\sim C_{e\gamma}^{\mu e} \sim 
    \frac{e}{16\pi^2} \frac{g_*^2}{m_\ast^2}\,y_\mu\,\dfrac{y_e}{y_\tau}. 
\label{eq:eq:U2muedip}
\end{equation}
The imaginary part of $C^{ee}_{e\gamma}$ responsible for the electron EDM is given by:\footnote{Similarly, for the muon EDM, we have $\text{Im} \left(C^{\mu \mu}_{e\gamma}\right) \sim \frac{e}{16\pi^2} \frac{g_*^2}{m_\ast^2}\,y_\mu\,\dfrac{y_\mu}{y_\tau}$. The enhancement of the muon EDM relative to the electron EDM is thus $m_\mu^2/m_e^2$, in contrast to the usual $m_\mu/m_e$. Nevertheless, the planned muon EDM measurements \cite{2collaboration2026improveddirectlimitmuon} will still not provide competitive probes.} 
\begin{equation}
   \text{Im} \left(C^{ee}_{e\gamma}\right) \sim \frac{e}{16\pi^2} \frac{g_*^2}{m_\ast^2}\,y_e\,\dfrac{y_e}{y_\tau}.
\label{eq:U2eedm}
\end{equation}
\begin{figure}[t]
    \centering
\includegraphics[width=0.99\linewidth]{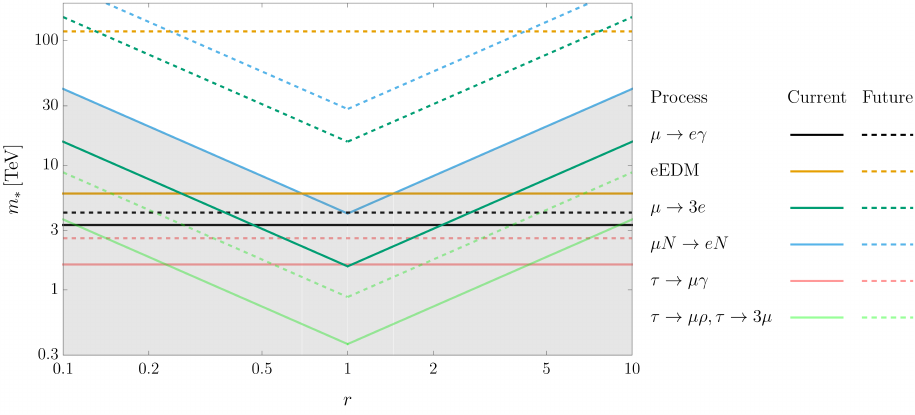}
    \caption{
    Current bound (solid) and future  sensitivity (dashed) from lepton-flavor-violating observables for the generic non-aligned $\U{2}$ scenario without assuming additional $P_{LR}$ symmetry for $r_e=r_\mu=r_\tau=r$ and fixed $g_*=2$. The gray shaded region is excluded by current constraints.
    \label{fig:U2nonaligned} }
\end{figure}
Finally, for the Higgs-current operators, the $\mu -e$ component scales as in the anarchic case,
\begin{equation}
        C_{H\ell}^{(1)e\mu} \sim  C_{H\ell}^{(3)e\mu}
        \sim
        \frac{g_\ast}{m_\ast^2}
        y_\mu
        \sqrt{\frac{y_e}{y_\mu}}\,
        r_e r_\mu,
        \qquad
        C_{He}^{e\mu}
        \sim
        \frac{g_\ast}{m_\ast^2}
        y_\mu
        \sqrt{\frac{y_e}{y_\mu}}\,
        \frac{1}{r_e r_\mu},
\label{eq:current-U2-general}
\end{equation}
while the $\tau - \mu$ component is given by
\begin{equation}
        C_{H\ell}^{(1)\tau \mu} \sim  C_{H\ell}^{(3)\tau \mu}
        \sim
        \frac{g_\ast}{m_\ast^2}
        y_\mu
         \left(r^2_\mu + r^2_\tau\right),
        \qquad
        C_{He}^{\tau \mu}
        \sim
        \frac{g_\ast}{m_\ast^2}
        y_\mu
        \,
         \left( \frac{1}{r^2_\mu} + \frac{1}{r^2_\tau} \right).
\label{eq:current-U2mutau-general}
\end{equation}
We summarize the scaling of the different coefficients in \cref{tab:dipole_constraints}. We show current constraints and future reach in \cref{fig:U2nonaligned}.
We discuss the main constraints and future projections from $\tau$ decays in appendix~\ref{app:taudecays}.
The dominant constraints on this scenario come from the electron EDM, with $\mu \to e \gamma$ and $\tau \to \mu \gamma$ being comparable. To compare the relative sensitivity of $\mu \to e \gamma$ and $\tau \to \mu \gamma$, note that the ratio of these branching fractions is given by
\begin{equation}
    \frac{\text{BR}(\tau \to \mu \gamma)}{\text{BR}(\mu \to e \gamma)} \sim \frac{y_\mu^2}{y_e^2}.
\end{equation}
In particular, this ratio is enhanced compared to the anarchic scenario by four orders of magnitude. With the future improvement of $\mu \to eee$ and $\mu N \to e N$, these will be the dominant probes of this scenario, along with the improving electron EDM.

\begin{table}[H]
\centering
\setlength{\tabcolsep}{3pt}
\renewcommand{\arraystretch}{1.25}
\resizebox{\textwidth}{!}{%
\begin{tabular}{|c|c|c|c|}
\hline
Observable & Operator coefficient
& bound $m_\ast$ [TeV]
& projection $m_\ast$ [TeV] \\
\hline

$\tau \to \mu \gamma$
&
$\displaystyle
C^{\tau\mu}_{e\gamma}
\sim \frac{e}{16\pi^2}\frac{g_\ast^2}{m_\ast^2}
\,y_\tau\,\frac{y_\mu}{y_\tau}$
&
$0.67\,g_\ast$
&
$1.05\,g_\ast$
\\
\hline

$\mu \to e \gamma$
&
$\displaystyle
C^{\mu e}_{e\gamma}
\sim \frac{e}{16\pi^2}\frac{g_\ast^2}{m_\ast^2}
\,y_\mu\,\frac{y_e}{y_\tau}$
&
$1.7\,g_\ast$
&
$2.1\,g_\ast$
\\
\hline

EDM
&
$\displaystyle
\mathrm{Im}\left(C^{ee}_{e\gamma}\right)
\sim \frac{e}{16\pi^2}
\frac{g_\ast^2}{m_\ast^2}
\,y_e\,\frac{y_e}{y_\tau}$
&
$3.0\,g_\ast$
&
$60\,g_\ast$
\\
\hline

$\tau \to \mu(\rho^0,\pi^0,\pi^+\pi^-)$
&
\begin{tabular}{@{}c@{}}
$\displaystyle
C^{ \tau \mu}_{H\ell} 
\sim \frac{g_\ast}{m_\ast^2}\,y_\mu\,
\max\!\left(r_\mu^2,r_\tau^2\right)$
\\[1mm]
$\displaystyle
C^{\tau\mu}_{He}
\sim \frac{g_\ast}{m_\ast^2}\,y_\mu\,
\max\!\left(r_\mu^{-2},r_\tau^{-2}\right)$
\end{tabular}
&
$0.23\,\sqrt{g_\ast}$
&
$ 0.57 \,\sqrt{g_\ast}$
\\
\hline

$\tau \to \mu\phi $
&
\begin{tabular}{@{}c@{}}
$\displaystyle
C^{ \tau \mu}_{H\ell} 
\sim \frac{g_\ast}{m_\ast^2}\,y_\mu\,
\max\!\left(r_\mu^2,r_\tau^2\right)$
\\[1mm]
$\displaystyle
C^{\tau\mu}_{He}
\sim \frac{g_\ast}{m_\ast^2}\,y_\mu\,
\max\!\left(r_\mu^{-2},r_\tau^{-2}\right)$
\end{tabular}
&
$0.20\,\sqrt{g_\ast}$
&
$0.47\,\sqrt{g_\ast}$
\\
\hline

$\tau\to3\mu$
&
\begin{tabular}{@{}c@{}}
$\displaystyle
C^{ \tau \mu}_{H\ell} 
\sim \frac{g_\ast}{m_\ast^2}\,y_\mu\,
\max\!\left(r_\mu^2,r_\tau^2\right)
$\\[1mm]
$\displaystyle
C_{He}^{\tau\mu}
\sim \frac{g_\ast}{m_\ast^2}\,y_\mu\,
\max\!\left(r_\mu^{-2},r_\tau^{-2}\right)
$
\end{tabular}
&
\begin{tabular}{@{}c@{}}
$0.22\,\sqrt{g_\ast}$\\[1mm]

\end{tabular}
&
\begin{tabular}{@{}c@{}}
$(0.41\text{--}0.70)\,\sqrt{g_\ast}$\\[1mm]

\end{tabular}
\\
\hline
\hline
\multicolumn{4}{|c|}{Non-aligned $\U{2}$}
\\
\hline

$\mu \to eee$, $\mu N \to e N$
&
\begin{tabular}{@{}c@{}}
$\displaystyle
C^{\mu e}_{H\ell}
\sim \frac{g_\ast}{m_\ast^2}\,
y_\mu\sqrt{\frac{y_e}{y_\mu}}\,
r_e r_\mu$
\\[1mm]
$\displaystyle
C^{\mu e}_{He}
\sim \frac{g_\ast}{m_\ast^2}\,
y_\mu\sqrt{\frac{y_e}{y_\mu}}\,
\frac{1}{r_e r_\mu}$
\end{tabular}
&
$2.9\,\sqrt{g_\ast}$
&
$20\,\sqrt{g_\ast}$
\\
\hline
\hline
\multicolumn{4}{|c|}{Aligned $\U{2}$}
\\
\hline

$\mu \to eee$, $\mu N \to e N$
&
\begin{tabular}{@{}c@{}}
$\displaystyle
C^{e\mu}_{H\ell}
\sim \frac{g_\ast}{m_\ast^2}\,y_e
\; \max\left[
r_e^2,\;r_\mu^2\;,\frac{y_\mu}{y_\tau}r_\tau^2
\right]$
\\[1mm]
$\displaystyle
C^{e\mu}_{He}
\sim \frac{g_\ast}{m_\ast^2}\,y_e\; \max
\left[
\frac{1}{r_e^2},\;\frac{1}{r_\mu^2}
\;,\frac{y_\mu}{y_\tau}\frac{1}{r_\tau^2}
\right]$
\end{tabular}
&
$0.77\,\sqrt{g_\ast}$
&
$5.3\,\sqrt{g_\ast}$
\\
\hline
\end{tabular}%
}
\caption{
Parametric scaling of the Wilson coefficients, current bounds
and future reach on the compositeness scale $m_*$ in both the non-aligned and aligned
$\mathrm{U}(2)_{\mathcal O}\times
\mathrm{U}(1)_{\mathcal O}\times\mathrm{CP}$ scenarios. The numbers shown correspond to the left--right symmetric case, $r_i=1$. 
All constraints should be understood as estimates with $\mathcal{O}(1)$ uncertainty.
 The current $\tau\to\mu\gamma$ bound uses the Belle limit
$\mathrm{BR}<4.2\times10^{-8}$~\cite{2021}, and its Belle~II projection is $7.0\times10^{-9}$
\cite{theatlascollaboration2025projectionskeymeasurementsheavy}. The strongest current hadronic bound comes from $\tau\to\mu\rho^0$, with $\mathrm{BR}<1.7\times10^{-8}$~\cite{2023}, while the strongest projected hadronic reach comes from
$\tau\to\mu\pi^0$, with sensitivity $7.1\times10^{-10}$ \cite{banerjee2022snowmass2021whitepaper}. 
For $\tau\to3\mu$, we use the current limit
$1.9\times10^{-8}$ \cite{BelleII:Tau3Mu2024,LHCb:Tau3Mu2026}
and the Belle~II projected sensitivity $(0.2\text{--}1.7)\times10^{-9}$
\cite{theatlascollaboration2025projectionskeymeasurementsheavy}.
}
\label{tab:dipole_constraints}
\end{table}

\subsubsection{Aligned $\U{2}_\mathcal{O}$}
\begin{figure}[t]
    \centering
\includegraphics[width=0.99\linewidth]{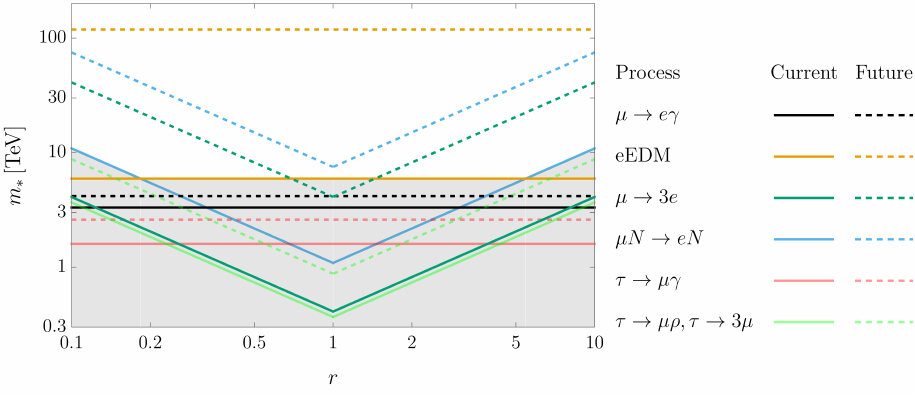}
    \caption{ Current bound (solid) and future  sensitivity (dashed) from lepton-flavor-violating observables for the aligned $\U{2}$ scenario without assuming additional $P_{LR}$ symmetry for $r_e=r_\mu=r_\tau=r$ and fixed $g_*=2$. The gray shaded region is excluded by current constraints.
    \label{fig:U2aligned} }
\end{figure}

\begin{figure}[t]
    \centering
\includegraphics[width=0.99\linewidth]{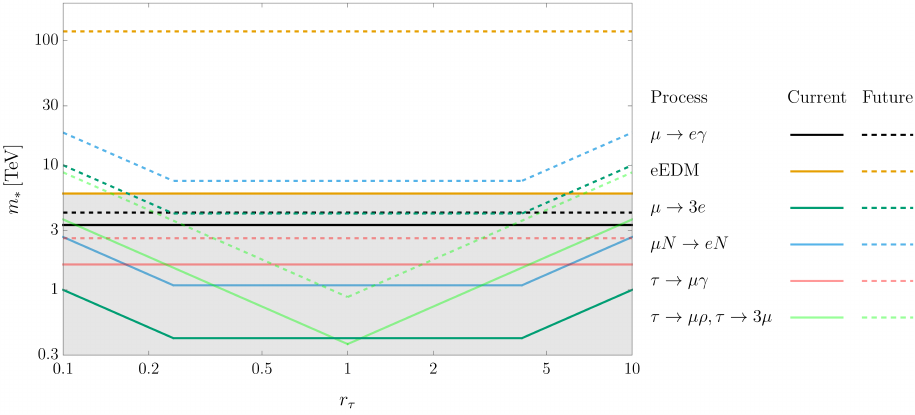}
    \caption{
     Current bound (solid) and future  sensitivity (dashed) from lepton-flavor-violating observables for the aligned $\U{2}$ scenario without assuming additional $P_{LR}$ symmetry for fixed $r_e=r_\mu=1$ and fixed $g_*=2$ as a function of $r_\tau$. The gray shaded region is excluded by current constraints.
    \label{fig:U2alignedtau} }
\end{figure}

Finally, we discuss the scenario of aligned $\U{2}_\mathcal{O}$. In this case, the $\U{2}_\mathcal{O}$ can be made manifest in the row-hierarchical basis for the mixings. As we see below, this can be realized, for example, if a $\U{1}^3$ subgroup of $\U{2}_\mathcal{O}\times \U{1}_\mathcal{O}$ survives in the UV when the hierarchies between the first two families are generated.
Also in this case, the leading-order lepton-flavor-violating dipoles and EDM coefficients are given by eqs.~\eqref{eq:U2taumudip}, \eqref{eq:eq:U2muedip}, and \eqref{eq:U2eedm}.
This comes from the fact that they are generated through mixing with the third generation, and are therefore not affected by the relative orientation of $\U{2}_\mathcal{O}$. On the other hand, the Higgs-current operators scale differently. In particular,
the $\mu -e$ component scales as 
\begin{equation}
    C_{H\ell}^{e\mu}\sim
\frac{g_\ast}{m_\ast^2}
y_e
\left[
 r_e^2+r_\mu^2+
 \frac{y_\mu}{y_\tau}r_\tau^2
\right], \qquad 
C_{He}^{e\mu}\sim \frac{g_\ast}{m_\ast^2}
y_e
\left[
 \frac1{r_e^2}+\frac1{r_\mu^2}+
 \frac{y_\mu}{y_\tau}\frac1{r_\tau^2}
\right]\,,
\label{eq:current-U2-aligned}
\end{equation}
while the $\tau - \mu$ component is still given by
\begin{equation}
        C_{H\ell}^{\tau \mu} \sim  
        \frac{g_\ast}{m_\ast^2}
        y_\mu
        \left(r^2_\mu + r^2_\tau\right),
        \qquad
        C_{He}^{\tau \mu}
        \sim
        \frac{g_\ast}{m_\ast^2}
        y_\mu
        \,
        \left( \frac{1}{r^2_\mu} + \frac{1}{r^2_\tau} \right).
\label{eq:current-U2mutau-aligned}
\end{equation}
 We show current constraints and future reach in \cref{fig:U2aligned} for $r_e=r_\mu=r_\tau$ and in \cref{fig:U2alignedtau} for $r_e=r_\mu=1$. 
This scenario is dominantly constrained by $\mu \to e\gamma$, $\tau \to \mu \gamma$ and the electron EDM, all coming from dipole operators. The expected improvement of $\mu \to eee $ and $\mu N \to e N$ will probe comparable energy scales, while for $r_e=r_\mu=r_\tau$, $\tau$ decays remain subdominant. However, for fully composite left- or right-handed $\tau$ corresponding to $r_\tau \sim 10 \sqrt{g_\ast}$ or $r_\tau \sim 0.1 /\sqrt{g_\ast}$, $\tau$ decays can probe unconstrained scales.
Finally the improvement in the electron EDM will dominate the future reach.

\subsection{Symmetries and hierarchies}
\label{sec:symandhier}

The scenarios discussed in \cref{sec:U3CP} assume that the strong sector is invariant under $\U{3}_{\mathcal O}\times {\rm CP}$ near the scale $m_\ast$.\footnote{We focus here for concreteness on the $\U{3}$ case, but a similar discussion applies to the $\U{2}_{\mathcal O}\times\U{1}_{\mathcal O}\times {\rm CP}$ scenario of \cref{sec:U2U1CP}.} At first sight, %
this symmetry seems to be in tension with the usual partial-compositeness explanation for the origin of flavor hierarchies. In anarchic partial compositeness, the hierarchy of the elementary--composite mixings is generated by RG evolution. Starting from order-one mixings at $\Lambda_{\rm UV}$, one obtains as in eq.~\eqref{eq:lambda_running},
\begin{equation}
  \lambda_{\psi}^{ai}(m_\ast)
  \simeq
  \lambda_{\psi}^{ai}(\Lambda_{\rm UV})
  \left(\frac{m_\ast}{\Lambda_{\rm UV}}\right)^{\Delta_\psi^a-5/2},
  \qquad
  \psi=\ell,e\, .
  \label{eq:symhier-usual-running}
\end{equation}
The hierarchies then come from non-degenerate scaling dimensions $\Delta_\psi^a$. If, instead, an exact $\U{3}_{\mathcal O}$ symmetry acted on the three operators $\mathcal O_\psi^a$ during the whole RG flow, their dimensions would be equal, $\Delta_\psi^1=\Delta_\psi^2=\Delta_\psi^3$. In that case, no hierarchies would be dynamically generated among the lepton mixings. The hierarchies would therefore have to originate from the UV values of the mixings, defeating part of the motivation for partial compositeness.

This tension is avoided if $\U{3}_{\mathcal O}\times {\rm CP}$ arises as an accidental symmetry of the strong sector in the IR. 
A simple way to parametrize this possibility is to introduce an intermediate scale $\Lambda_f$, with
\begin{equation}
        \Lambda_{\rm UV}\gg \Lambda_f \gg m_\ast\, .
        \label{eq:symhier-scales}
\end{equation}
Above $\Lambda_f$, the strong dynamics is not assumed to respect the full $\U{3}_{\mathcal O}$ symmetry, while below $\Lambda_f$, the relevant and marginal interactions satisfy the $\U{3}_{\mathcal{O}}$ symmetry, with symmetry-violating effects encoded in irrelevant deformations. 
In this way, the flavor hierarchies are generated in the UV, while the accidental IR symmetry provides the protection from the most constraining LFV and CP-violating effects.
To see this more explicitly, note that above $\Lambda_f$, the dimensions of the operators that mix with the elementary leptons are generally different,
\begin{equation}
        \Delta_\psi^a=\frac52+\gamma_\psi^a,
        \qquad
        \gamma_\psi^1,\gamma_\psi^2,\gamma_\psi^3
        \ \hbox{not necessarily equal}\, .
        \label{eq:symhier-dimensions-above}
\end{equation}
Starting from anarchic mixings at $\Lambda_{\rm UV}$, the RG running down to $\Lambda_f$ generates hierarchical mixings,
\begin{equation}
        \lambda_{\psi}^{ai}(\Lambda_f)
        \sim
        \lambda_{\psi}^{ai}(\Lambda_{\rm UV})
        \left(\frac{\Lambda_f}{\Lambda_{\rm UV}}\right)^{\gamma_\psi^a},
        \qquad
        \psi=\ell,e\, .
        \label{eq:symhier-stage-one}
\end{equation}

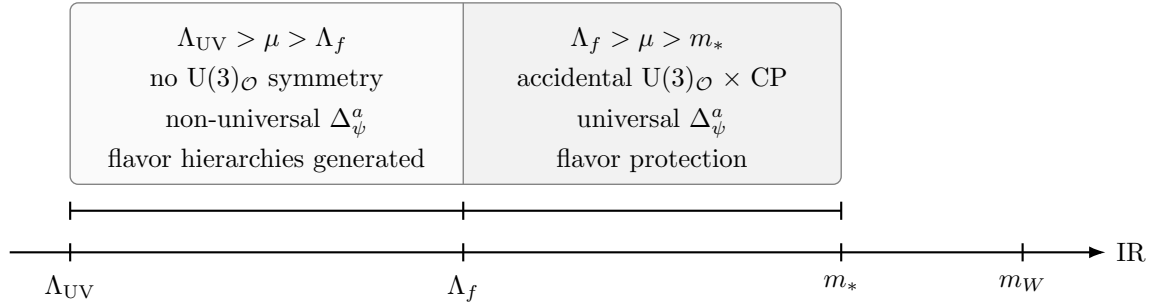
\begin{figure}[t]
\centering
\begin{tikzpicture}[
    font=\small,
    x=1cm,
    y=1cm,
    axis/.style={
        thick,
        -{Latex[length=2.2mm,width=1.6mm]}
    },
    tick/.style={thick},
    frame/.style={
        draw=black!50,
        line width=0.5pt,
        rounded corners=3pt
    },
    regime/.style={
        align=center,
        inner sep=0pt,
        outer sep=0pt
    }
]

\coordinate (UV) at (0,0);
\coordinate (Lf) at (5.2,0);
\coordinate (ms) at (10.2,0);
\coordinate (mW) at (12.6,0);

\begin{scope}
    \clip[rounded corners=3pt]
        ($(UV)+(0,0.90)$) rectangle ($(ms)+(0,3.30)$);

    \fill[black!2]
        ($(UV)+(0,0.90)$) rectangle ($(Lf)+(0,3.30)$);

    \fill[black!5]
        ($(Lf)+(0,0.90)$) rectangle ($(ms)+(0,3.30)$);
\end{scope}

\draw[frame]
    ($(UV)+(0,0.90)$) rectangle ($(ms)+(0,3.30)$);

\draw[black!50,line width=0.5pt]
    ($(Lf)+(0,0.90)$) -- ($(Lf)+(0,3.30)$);

\node[regime]
    at ($(UV)!0.5!(Lf)+(0,2.80)$) {
    $\Lambda_{\rm UV}>\mu>\Lambda_f$
};

\node[regime]
    at ($(Lf)!0.5!(ms)+(0,2.80)$) {
    $\Lambda_f>\mu>m_\ast$
};

\node[regime,text width=4.7cm]
    at ($(UV)!0.5!(Lf)+(0,1.76)$) {
    no $\U{3}_{\mathcal O}$ symmetry\\[2pt]
    non-universal $\Delta_\psi^a$\\[2pt]
    flavor hierarchies generated
};

\node[regime,text width=4.5cm]
    at ($(Lf)!0.5!(ms)+(0,1.76)$) {
    accidental $\U{3}_{\mathcal O}\times\mathrm{CP}$\\[2pt]
    universal $\Delta_\psi^a$\\[2pt]
    flavor protection
};

\draw[thick]
    ($(UV)+(0,0.55)$) -- ($(ms)+(0,0.55)$);

\foreach \p in {UV,Lf,ms}{
    \draw[tick]
        ($(\p)+(0,0.43)$) -- ($(\p)+(0,0.67)$);
}

\draw[axis]
    (-0.8,0) -- (13.7,0) node[right] {IR};

\foreach \p/\lab in {
    UV/{$\Lambda_{\rm UV}$},
    Lf/{$\Lambda_f$},
    ms/{$m_\ast$},
    mW/{$m_W$}
}{
    \draw[tick]
        ($(\p)+(0,-0.12)$) -- ($(\p)+(0,0.12)$);
    \node[below=5pt] at (\p) {\lab};
}

\end{tikzpicture}
\caption{
Schematic two-stage RG flow generating charged-lepton hierarchies
with accidental infrared $\U{3}_{\mathcal O}\times\mathrm{CP}$
protection. Above $\Lambda_f$, non-universal operator dimensions
can generate hierarchical elementary--composite mixings.
Below $\Lambda_f$, the strong sector is approximately
$\U{3}_{\mathcal O}\times\mathrm{CP}$ invariant, with irrelevant
symmetry-breaking deformations suppressed at $m_\ast$ by powers
of $m_\ast/\Lambda_f$.
}
\label{fig:symhier-scales}
\end{figure}

The $\U{3}_{\mathcal O}$ symmetry can emerge below $\Lambda_f$, for example, if the interactions that distinguish the three strong-sector flavor directions are mediated by states with masses of order $\Lambda_f$ or above.
Once these states are integrated out, the accidental symmetry emerges if the relevant and marginal interactions of the strong sector below $\Lambda_f$  respect $\U{3}_{\mathcal O}$, while the residual breaking appears only through irrelevant operators suppressed by powers of $\Lambda_f$.

In the second stage of the RG evolution, the operators in a $\U{3}_{\mathcal O}$ irreducible representation have a common dimension $\Delta_\psi$, so that
\begin{equation}
        \lambda_{\psi}^{ai}(m_\ast)
        \simeq
        \left(\frac{m_\ast}{\Lambda_f}\right)^{\Delta_\psi-5/2}
        \lambda_{\psi}^{ai}(\Lambda_f)\,,
        \label{eq:symhier-stage-two}
\end{equation}
and the running is flavor universal. It preserves the relative hierarchies generated above $\Lambda_f$, and the strong-sector tensors that enter the matching at $m_\ast$ are approximately $\U{3}_{\mathcal O}$ invariant.

 Below the scale $\Lambda_f$, $\U{3}_{\mathcal O}$-breaking effects within the strong sector are captured by irrelevant operators in the strong-sector Lagrangian, 
\begin{equation}
        \mathcal L_{\rm strong}
        =
        \mathcal L_{\rm strong}^{\rm sym}
        +
        \sum_{n}
        \frac{c_n}{\Lambda_f^{\Delta_{n}-4}}\,
        \mathcal O^{(n)}_{\rm br},
        \qquad
        \Delta_{n}>4\, .
        \label{eq:symhier-breaking-operators}
\end{equation}
Here $\mathcal L_{\rm strong}^{\rm sym}$ denotes the $\U{3}_{\mathcal O}$-invariant part of the strong-sector Lagrangian, while $\mathcal O^{(n)}_{\rm br}$ denotes  operators that transform non-trivially under $\U{3}_{\mathcal O}$. The index $n$ labels the different symmetry-breaking deformations.
Generically, the operator with the lowest dimension gives the leading symmetry-breaking effect. Therefore, for simplicity, we denote the lowest dimension of these operators by $\Delta_{\rm br}$ and estimate its effect below. 
After running down to the scale $m_\ast$, this induces $\U{3}_{\mathcal O}$-breaking corrections to the strong-sector tensors,
\begin{equation}
        (c_\mathcal{O})^{a}_b
        =
        c_\mathcal{O}^{(0)}\delta^{a}_b
        +
        \delta_{\rm br}\,(\widetilde c_\mathcal{O})^{a}_b,
        \qquad
        \delta_{\rm br}
        \sim
        \left(\frac{m_\ast}{\Lambda_f}\right)^{\Delta_{\rm br}-4},
        \label{eq:symhier-breaking-delta}
\end{equation}
where $\mathcal{O}=Y,e\gamma,H\ell,He,\ldots$ and $(\widetilde c_\mathcal{O})^{a}_b$ is an order-one $\U{3}_{\mathcal O}$-non-invariant tensor. %
Its specific structure depends on the underlying UV flavor symmetry: the maximal symmetry compatible with the dynamical hierarchy generation is $\mathrm{U}(1)^3$;\footnote{We discuss this option in the aligned scenario.} for simplicity we assume here a fully anarchic matrix, corresponding to the least symmetric case.
These corrections generically reintroduce the usual anarchic scaling of LFV operators, but multiplied by the small symmetry-breaking parameter $\delta_{\rm br}$.\footnote{In general there can be another source of LFV generated at $\Lambda_f$, from the mixing of elementary-fermion bilinears with $\U{3}$-singlet operators of the strong sector with scaling dimension $\Delta_{s}$. These effects are suppressed similar to eq.~\eqref{eq:symhier-breaking-delta}, replacing $\Delta_{\rm br}$ with $3+\Delta_{s}$. }

For example, the correction to the electromagnetic dipole is estimated as: 
\begin{align}
        \left(C_{e\gamma}^{e\mu}\right)_{\rm br} \sim 
        \delta_{\rm br}\,
        \frac{e}{16\pi^2}
        \frac{g_\ast^2}{m_\ast^2}
        \sqrt{y_e y_\mu}\,
        \frac{r_e}{r_\mu}, \qquad  \left(C_{e\gamma}^{\mu e}\right)_{\rm br} \sim
        \delta_{\rm br}\,
        \frac{e}{16\pi^2}
        \frac{g_\ast^2}{m_\ast^2}
        \sqrt{y_e y_\mu}\,
        \frac{r_\mu}{r_e}
        .
        \label{eq:symhier-dipole-breaking}
\end{align}
In particular, consider the left--right symmetric case $r_i \sim 1$. The current $\mu\to e\gamma$ bound on an anarchic dipole gives $m_\ast\gtrsim 26\,g_\ast\TeV$, as summarized in \cref{tab:anarchic-summary}. Therefore, the present bound is avoided provided 
\begin{equation}
        \delta_{\rm br}
        \lesssim
        \left(
        \frac{m_\ast}{26\,g_\ast\TeV}
        \right)^2 .
        \label{eq:symhier-delta-bound}
\end{equation}
This translates into
\begin{equation} 
        \Lambda_f
        \gtrsim
        m_\ast
        \left(
        \frac{26\,g_\ast\TeV}{m_\ast}
        \right)^{\frac{2}{\Delta_{\rm br}-4}},
        \label{eq:symhier-Lambdaf-general}
\end{equation}
which, for example for $\Delta_{\rm br}=6$, gives $\Lambda_f \gtrsim 80\TeV  \left(\frac{g_\ast}{3}\right)$. Thus an accidental $\U{3}_{\mathcal O}$ symmetry emerging below a scale of order $100\TeV$ is already sufficient to protect a few-TeV composite sector from the present $\mu\to e\gamma$ constraint. 

So far, we have only discussed the emergence of the flavor symmetry $\U{3}_{\mathcal O}$ in the IR. Similarly, CP may be an accidental symmetry of the strong dynamics. However, the scale above which CP is broken, which we denote by $\Lambda_{\rm CP}$, can be independent of $\Lambda_f$. We therefore consider two distinct scenarios. First, one can consider a case where $\Lambda_{\rm CP} \gg \Lambda_f$, in particular such that the hierarchies in the mixings are generated while CP is a symmetry of the strong sector. In this case, at $\Lambda_f$, there exists a basis where all the CP-violating phases are in the mixings, while the mixings are row-hierarchical. In the second scenario, CP and the $\U{3}_{\mathcal O}$ can emerge around the same scale,  $\Lambda_{\rm CP}\sim \Lambda_f$. In this case, while below $\Lambda_{\rm CP}$ all the CP phases can still be carried by the mixings, the mixings are not row-hierarchical in this basis. We considered both of these cases in \cref{sec:U3CP}. 
Similarly to the above, the effects of $\rm CP$ breaking can be captured by higher-dimensional CP-odd operators, suppressed by powers of the scale $\Lambda_{\rm CP}$. Therefore, for $\Lambda_{\rm CP}\sim \Lambda_f$, the EDM constraints can dominate over the $\mu \to e \gamma$ constraints and point to scales higher than those in Eq.~\eqref{eq:symhier-Lambdaf-general}.

As discussed in \cref{sec:U3CP}, the basis where the $\U{3}_{\mathcal O}$ is manifest may be the same as or different from the basis where the mixings have the row-hierarchical structure. 
The non-aligned case can generically arise from the discussion above around eq.~\eqref{eq:symhier-stage-one} without further symmetry assumptions in the UV.

We now provide a picture of how the aligned case can be realized: consider, for example, that a $\U{1}^3$ subgroup of the $\U{3}_{\mathcal O}$ symmetry is preserved all the way to the UV. Then, when the hierarchies are generated, the CFT operators can be simultaneously assigned scaling dimensions and $\U{1}^3$ charges. Assume, for example, that the three operators $\mathcal{O}_\ell^a$ with the lowest dimensions are charged under different $\U{1}_a$ factors, with the $\mathcal{O}_e^a$ operators carrying the same corresponding $\U{1}^3$ charges. This charge assignment is the same as the one discussed in the $\U{1}^3$ scenario in \cref{subsubsec:U13CP}. At the scale $\Lambda_f$ the  hierarchical mixings $\lambda_\psi^{ai} \propto (\Lambda_f/\Lambda_{\rm UV})^{\gamma_\psi^a}$ carry a spurious $\U{1}_a$ charge. This charge assignment guarantees that, below the scale $\Lambda_f$, the mixings are row-hierarchical in the same basis where the $\U{3}_\mathcal{O}$ symmetry is manifest.\footnote{In this case where the $\U{1}^3$ symmetry is present in the UV, the  LFV dipole operators are more suppressed compared to eq.~\eqref{eq:symhier-dipole-breaking} and therefore $\Lambda_f$ can be smaller.}

\section{Comparison with other flavor-protected scenarios}
\label{sec:comparison}
We now compare the scenarios we discussed in \cref{sec:flavorprotection} with other scenarios within the partial compositeness framework discussed in the literature. In \cref{sec:partial_compositeness} we have already reviewed the limiting cases of anarchic partial compositeness and MFV. In this section we review two intermediate scenarios that give rise to LFV, but have more protection than the anarchic case. In \cref{subsubsec:U13CP}, we consider a strong sector with a $\U{1}^3\times \text{CP}$ symmetry, as studied in Ref.~\cite{Frigerio2018}. In \cref{sec:pluandpru}, we consider a strong sector with a $\U{2} \times \U{1} \times \text{CP}$ symmetry and partial left or right universality of the mixings, as discussed in Refs.~\cite{Redi:2013pga,Glioti:2024hye}.

\subsection{$\mathrm{U}(1)^3\times\mathrm{CP}$}
\label{subsubsec:U13CP}

\begin{table}[t]
\centering
\setlength{\tabcolsep}{3pt}
\renewcommand{\arraystretch}{1.25}
\resizebox{\textwidth}{!}{%
\begin{tabular}{|c|c|c|c|c|}
\hline
$\U{1}^3$
&
$C_{e\gamma}^{e\mu, \mu e}$
&
$C_{H\ell}^{e\mu}$
&
$C_{He}^{e\mu}$
&
$\mathrm{Im}\,C_{e\gamma}^{ee}$
\\
\hline
scaling
&
$\displaystyle
\frac{e}{16\pi^2}
\frac{g_\ast^2}{m_\ast^2}
y_e$
&
$\displaystyle
\frac{g_\ast}{m_\ast^2}
y_e \; \max
\left[
r_e^2,r_\mu^2,\frac{y_\mu}{y_\tau}r_\tau^2
\right]$
&
$\displaystyle
\frac{g_\ast}{m_\ast^2}
y_e \; \max
\left[
\frac1{r_e^2}, \frac1{r_\mu^2},
 \frac{y_\mu}{y_\tau}\frac1{r_\tau^2}
\right]$
&
$\displaystyle
\frac{e}{16\pi^2}
\frac{g_\ast^2}{m_\ast^2}
y_e\frac{y_e}{y_\mu}$
\\
\hline
bound on $m_\ast$ [TeV]
&
\begin{tabular}{@{}c@{}}
$\displaystyle
6.8 g_\ast
\quad(\mu\to e\gamma)$\\[1mm]
$\displaystyle
1.2g_\ast
\quad(\mu\to 3e )$\\[1mm]
$\displaystyle
1.2g_\ast
\quad(\mu N\to eN)$
\end{tabular}
&
\begin{tabular}{@{}c@{}}
$\displaystyle
0.78\sqrt{g_\ast}
\quad(\mu N\to eN)$\\[1mm]
$\displaystyle
0.3\sqrt{g_\ast}
\quad(\mu\to 3e)$
\end{tabular}
&
\begin{tabular}{@{}c@{}}
$\displaystyle
0.78\sqrt{g_\ast}
\quad(\mu N\to eN)$\\[1mm]
$\displaystyle
0.3\sqrt{g_\ast}
\quad(\mu\to 3e )$
\end{tabular}
&
$\displaystyle
12g_\ast
\quad(\text{eEDM})$
\\
\hline
projection on $m_\ast$ [TeV]
&
\begin{tabular}{@{}c@{}}
$\displaystyle
8.6g_\ast
\quad(\mu\to e\gamma)$\\[1mm]
$\displaystyle
12g_\ast
\quad(\mu\to 3e)$\\[1mm]
$\displaystyle
10g_\ast
\quad(\mu N\to eN)$
\end{tabular}
&
\begin{tabular}{@{}c@{}}
$\displaystyle
5.3\sqrt{g_\ast}
\quad(\mu N\to eN)$\\[1mm]
$\displaystyle
2.8\sqrt{g_\ast}
\quad(\mu\to 3e)$
\end{tabular}
&
\begin{tabular}{@{}c@{}}
$\displaystyle
5.3\sqrt{g_\ast}
\quad(\mu N\to eN)$\\[1mm]
$\displaystyle
2.8\sqrt{g_\ast}
\quad(\mu\to 3e)$
\end{tabular}
&
$\displaystyle
250 g_\ast
\quad(\text{eEDM})$
\\
\hline
\end{tabular}%
}
\caption{
Parametric scaling of the Wilson coefficients, current bounds and future reach on the compositeness scale $m_*$ in the $\mathrm{U}(1)^3\times\mathrm{CP}$ scenario. 
The numbers shown correspond to the left--right symmetric case, $r_i=1$. 
All constraints should be understood as estimates with $\mathcal{O}(1)$ uncertainty.
}
\label{tab:U13CP-summary}
\end{table}

We now consider the scenario of $\mathrm{U}(1)^3\times\mathrm{CP}$, following the construction of ref.~\cite{Frigerio:2018uwx}. In particular, this scenario can suppress LFV more strongly than the anarchic scenario, while allowing the fermion mass hierarchies to be generated dynamically in the presence of the symmetry.  
This scenario is characterized by the assumption that the strong sector has a
\begin{equation}
  G_{\text{strong}}= \U{1}_{1} \times \U{1}_{2} \times \U{1}_{3} \times \mathrm{CP}
\label{eq:U1symstrong}  
\end{equation} symmetry, with the operators $\mathcal{O}_{\ell,e}^a$ having the same charges under
$\U{1}_{a}$ for each $a = 1,2,3$. More precisely, the mixings $\epsilon_{\ell, e}$ transform under $\U{1}_{1} \times \U{1}_{2} \times \U{1}_{3}$ as
\begin{align}
\epsilon_{\ell,e}^{1} &\sim (+1,\,0,\,0), \qquad
\epsilon_{\ell,e}^{2} \sim (0,\,+1,\,0), \qquad
\epsilon_{\ell,e}^{3} \sim (0,\,0,\,+1)\,.
\end{align}
In a row-hierarchical basis for the mixings, the charged-lepton Yukawa matrix takes the form
\begin{equation}
        (Y_e)_{ij}
        =
        g_\ast\,
        \left(
        \epsilon_\ell^\dagger c_Y \epsilon_e
        \right)_{ij},
        \qquad
        c_Y=\mathrm{diag}(c_Y^1,c_Y^2,c_Y^3),
        \qquad
        c_Y^a\in\mathbb R ,
\label{eq:Yukawa-U13}
\end{equation}
where $c_Y$ is forced to be diagonal by the $\U{1}^3$ symmetry and is real following the assumption of CP invariance. In \cref{tab:U13CP-summary}, we summarize the results of \cite{Frigerio2018} for the most relevant coefficients for lepton-flavor and CP violation, update the bounds and provide future projections for this scenario.

Compared to the anarchic scenario, the off-diagonal coefficients of the dipole operator are suppressed by a factor of $\sqrt{y_e/y_\mu}$, weakening the bound on $m_*$ by a factor of $\left(y_\mu/y_e\right)^{1/4} \approx 4$. The electron EDM is also suppressed by $\sqrt{y_e/y_\mu}$ compared to the anarchic scenario with CP invariance of the strong sector. %

Overall, the scenario is significantly constrained, and the electron EDM points to new physics in the tens of TeV range.
Future improvements in $\mu \to eee$ and $\mu N \to e N$ will probe the theory only close to the current electron EDM bound.

\subsection{$\mathrm{U}(2)\times\mathrm{U}(1)\times\mathrm{CP}$ with partial universality}
\label{sec:pluandpru}

\begin{table}[t]
\centering
\setlength{\tabcolsep}{3pt}
\renewcommand{\arraystretch}{1.25}
\resizebox{\textwidth}{!}{%
\begin{tabular}{|c|c|c|c|}
\hline
pRU $\mathrm{U}(2)\times\mathrm{U}(1)\times\mathrm{CP}$
&
dominant $C_{e\gamma}$
&
$C_{H\ell}^{\mu e}$
&
$C_{He}^{\mu e}$
\\
\hline
scaling
&
$\displaystyle
\frac{e}{16\pi^2}
\frac{g_\ast^2}{m_\ast^2}
a_\ell 
y_\mu
\left(\frac{y_\mu}{y_\tau}\right)^2$
&
$\displaystyle
\frac{g_\ast}{m_\ast^2}
a_\ell 
\frac{y_\mu^3}{y_\tau^2}r_e^2$
&
$\displaystyle
\frac{g_\ast}{m_\ast^2}
a_\ell 
y_\tau
\left(\frac{y_\mu}{y_\tau}\right)^2
\frac1{r_\tau^2}$
\\
\hline
bound on $m_\ast$ [TeV]
&
\begin{tabular}{@{}c@{}}
$\displaystyle
5.8 \sqrt{a_\ell}\, g_\ast
\quad(\mu\to e\gamma)$\\[1mm]
$\displaystyle
1.0 \sqrt{a_\ell} \, g_\ast
\quad(\mu\to 3e)$\\[1mm]
$\displaystyle
0.99\sqrt{a_\ell}\,g_\ast
\quad(\mu N\to eN)$
\end{tabular}
&
\begin{tabular}{@{}c@{}}
$\displaystyle
0.66 \sqrt{a_\ell}\, \sqrt{g_\ast}
\quad(\mu N\to eN)$\\[1mm]
$\displaystyle
0.24 \sqrt{a_\ell}\, \sqrt{g_\ast}
\quad(\mu\to 3e)$
\end{tabular}
&
\begin{tabular}{@{}c@{}}
$\displaystyle
2.7\sqrt{a_\ell}\,\sqrt{g_\ast}
\quad(\mu N\to eN)$\\[1mm]
$\displaystyle
1 \sqrt{a_\ell}\, \sqrt{g_\ast}
\quad(\mu \to 3e )$
\end{tabular}
\\
\hline
projection on $m_\ast$ [TeV]
&
\begin{tabular}{@{}c@{}}
$\displaystyle
7.4\sqrt{a_\ell}\, g_\ast
\quad(\mu\to e\gamma)$\\[1mm]
$\displaystyle
10 \sqrt{a_\ell}\, g_\ast
\quad(\mu\to 3e)$\\[1mm]
$\displaystyle
8.7\sqrt{a_\ell}\,g_\ast
\quad(\mu N\to eN)$
\end{tabular}
&
\begin{tabular}{@{}c@{}}
$\displaystyle
4.5 \sqrt{a_\ell}\, \sqrt{g_\ast}
\quad(\mu N\to eN)$\\[1mm]
$\displaystyle
2.4 \sqrt{a_\ell}\, \sqrt{g_\ast}
\quad(\mu\to 3e)$
\end{tabular}
&
\begin{tabular}{@{}c@{}}
$\displaystyle
19 \sqrt{a_\ell} \,\sqrt{g_\ast}
\quad(\mu N\to eN)$\\[1mm]
$\displaystyle
10 \sqrt{a_\ell} \,\sqrt{g_\ast}
\quad(\mu\to 3e)$
\end{tabular}
\\
\hline
\hline
pLU $\mathrm{U}(2)\times\mathrm{U}(1)\times\mathrm{CP}$
&
dominant $C_{e\gamma}$
&
$C_{H\ell}^{e\mu}$
&
$C_{He}^{e\mu}$
\\
\hline
scaling
&
$\displaystyle
\frac{e}{16\pi^2}
\frac{g_\ast^2}{m_\ast^2}
a_e
y_\mu
\left(\frac{y_\mu}{y_\tau}\right)^2$
&
$\displaystyle
\frac{g_\ast}{m_\ast^2}
a_e 
y_\tau
\left(\frac{y_\mu}{y_\tau}\right)^2
r_\tau^2$
&
$\displaystyle
\frac{g_\ast}{m_\ast^2}
a_e
\frac{y_\mu^3}{y_\tau^2}\frac1{r_e^2}$
\\
\hline
bound on $m_\ast$ [TeV]
&
\begin{tabular}{@{}c@{}}
$\displaystyle
5.8\sqrt{a_e} \, g_\ast
\quad(\mu\to e\gamma)$\\[1mm]
$\displaystyle
1.0\sqrt{a_e} \,g_\ast
\quad(\mu\to 3e)$\\[1mm]
$\displaystyle
0.99\sqrt{a_e}\,g_\ast
\quad(\mu N\to eN)$
\end{tabular}
&
\begin{tabular}{@{}c@{}}
$\displaystyle
2.7\sqrt{a_e} \,\sqrt{g_\ast}
\quad(\mu N\to eN)$\\[1mm]
$ \displaystyle 
1 \sqrt{a_e} \,\sqrt{g_\ast}
\quad(\mu\to 3e)$
\end{tabular}

&
\begin{tabular}{@{}c@{}}
$\displaystyle
0.66\sqrt{a_e} \,\sqrt{g_\ast}
\quad(\mu N\to eN)$\\[1mm]
$\displaystyle
0.23\sqrt{a_e} \,\sqrt{g_\ast}
\quad(\mu\to 3e)$
\end{tabular}
\\
\hline
projection on $m_\ast$ [TeV]
&
\begin{tabular}{@{}c@{}}
$\displaystyle
7.4\sqrt{a_e} \, g_\ast
\quad(\mu\to e\gamma)$\\[1mm]
$\displaystyle
10\sqrt{a_e} \,g_\ast
\quad(\mu\to 3e)$\\[1mm]
$\displaystyle
8.7\sqrt{a_e}\,g_\ast
\quad(\mu N\to eN)$
\end{tabular}
&
\begin{tabular}{@{}c@{}}
$\displaystyle
19 \sqrt{a_e} \,\sqrt{g_\ast}
\quad(\mu N\to eN)$\\[1mm]
$\displaystyle
10 \sqrt{a_e} \,\sqrt{g_\ast}
\quad(\mu\to 3e)$
\end{tabular}
&
\begin{tabular}{@{}c@{}}
$\displaystyle
4.5\sqrt{a_e} \,\sqrt{g_\ast}
\quad(\mu N\to eN)$\\[1mm]
$\displaystyle
2.3\sqrt{a_e} \,\sqrt{g_\ast}
\quad(\mu\to 3e)$
\end{tabular}
\\
\hline
\end{tabular}}
\caption{
Parametric scaling of the Wilson coefficients, current bounds and future reach on the compositeness scale $m_*$ in the pRU/pLU scenario. 
The numbers shown correspond to the left--right symmetric case, $r_\tau = r_e =1$. 
All constraints should be understood as estimates with $\mathcal{O}(1)$ uncertainty.
Note that in these scenarios there is no electron EDM and that $a_\ell, a_e$ can be $O(1)$ or as small as $y_e/y_\mu$ as discussed in the main text. 
}
\label{tab:pRU-summary}
\end{table}

Finally, we consider, as in \cref{sec:U2U1CP}, a strong sector with a $\mathrm{U}(2)\times\mathrm{U}(1)\times\mathrm{CP}$ symmetry. We now impose the additional assumption of partial universality on the mixings \cite{Glioti:2024hye}. As for the case of MFV, there are two partial-universality scenarios: partial left universality (pLU) and partial right universality (pRU). For concreteness, we focus on partial right universality, where the partially universal right-handed mixings break
\begin{equation}
  \U{2}_\mathcal{O} \times \U{1}_\mathcal{O} \times \U{3}_e \to   \U{2}_{\mathcal{O}+e} \times \U{1}_{\mathcal{O}+e}.
\end{equation}
Finally, the left-handed mixings break it to $\U{1}_Y \times \U{1}_{e+\ell+\mathcal{O}}$. The assumption of partial universality of the mixing distinguishes this case from the realization of $\mathrm{U}(2)\times\mathrm{U}(1)\times\mathrm{CP}$ discussed above in \cref{sec:U2U1CP}. Explicitly, in \cite{Glioti:2024hye}, the mixings were brought to the form
 \begin{equation}
        \epsilon_e
        \simeq
        \begin{pmatrix}
        \epsilon_{e} & 0 & 0 \\
        0 & \epsilon_{e} & 0 \\
        0 & 0 & \epsilon_{e3}
        \end{pmatrix},
        \qquad
        \epsilon_\ell^\dagger
        \simeq
        \frac1{g_\ast}
        \begin{pmatrix}
        y_e/\epsilon_{e} & 0 & 0 \\
        0 & y_\mu/\epsilon_{e} & 0 \\
        a_\ell y_\mu/\epsilon_{e} &
        b_\ell y_\mu/\epsilon_{e} &
        y_\tau/\epsilon_{e3}
        \end{pmatrix},
\label{eq:pRU-epsilon-texture}
\end{equation}
where $b_\ell\sim\mathcal O(1)$. Ref.~\cite{Glioti:2024hye} considered $a_\ell \sim \mathcal O(1)$ and pointed out that $a_\ell$ can be as small as $O(y_e/y_\mu)$. In the UV picture described in \cref{sec:symandhier}, we find the latter case.

One can show that all CP phases in the mixings can be removed and therefore no lepton EDMs are generated.
The results for the lepton-flavor-violating coefficients for both  partial right universality and partial left universality are summarized in \cref{tab:pRU-summary}. 

Compared to the anarchic scenario, the bound from the electron EDM disappears, and the $\mu-e$ component of the dipole operator is suppressed by
\begin{equation}
        \frac{C_{e\gamma}^{\rm pRU}}{C_{e\gamma}^{\rm anarchic}}
        \sim
        \frac{a_{\ell} (y_\mu/y_\tau)^2}{\sqrt{y_e/y_\mu}}
        \sim 0.05 \, a_{\ell}.
\label{eq:pRU-pLU-vs-anarchy}
\end{equation}
Consequently, for $a_\ell\sim\mathcal{O}(1)$, the strongest current bound comes from $\mu\to e\gamma$ and requires $m_*\gtrsim 10~\mathrm{TeV}$. Future searches for $\mu\to eee$ and $\mu N\to eN$ will improve
upon this bound, probing compositeness scales of about $30~\mathrm{TeV}$. On the other hand, for $a_\ell\sim\mathcal{O}(y_e/y_\mu)$, the constraints are substantially weaker, and the scenario remains viable at the TeV scale.

\begin{figure}[t]
    \centering
\includegraphics[width=0.99\linewidth]{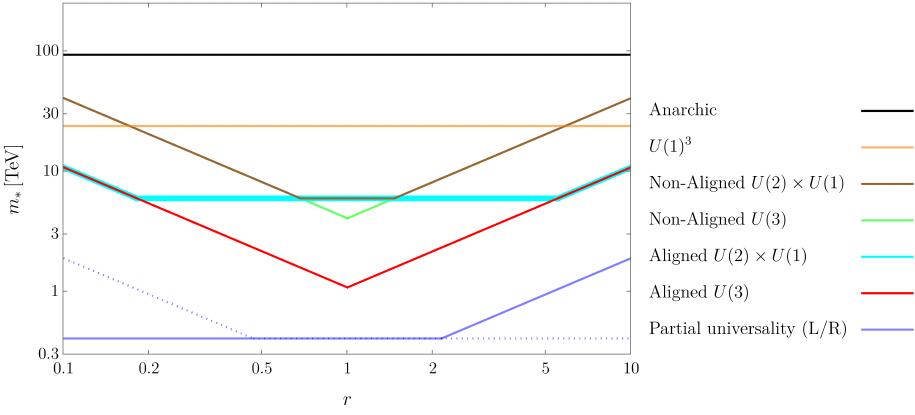}
\caption{Comparison of current bounds on the compositeness scale
$m_*$ from lepton-flavor-violating observables and the electron EDM
in the different flavor scenarios considered.
For the scenarios of partial universality we set $r_\tau=r_e=r$, while in other scenarios we show the bounds as a function of $r=r_e=r_\mu=r_\tau$. 
All bounds should
be understood as estimates with $\mathcal{O}(1)$ uncertainty.}
\label{fig:comparison_models}
\end{figure}

\begin{table}[t]
\centering
\setlength{\tabcolsep}{4pt}
\renewcommand{\arraystretch}{1.2}
\resizebox{\textwidth}{!}{%
\begin{tabular}{|l|c|c|c|c|}
\hline
\textbf{Scenario} &
\textbf{Bounds} &
\textbf{Dipole} &
\textbf{Higgs-current} &
\textbf{eEDM}
\\
\hline\hline

Anarchic &
Current &
$26g_\ast$ &
$2.9\sqrt{g_\ast}$ &
\begin{tabular}{@{}l@{}}
$180g_\ast$ ($\cancel{\rm CP}$)\\
$47g_\ast$ (CP)
\end{tabular}
\\
\cline{2-5}
&
Future &
$46g_\ast$ &
$20\sqrt{g_\ast}$ &
\begin{tabular}{@{}l@{}}
$3600g_\ast$ ($\cancel{\rm CP}$)\\
$940g_\ast$ (CP)
\end{tabular}
\\
\hline\hline

Generic $\mathrm{U}(3)\times\mathrm{CP}$ &
Current &
$0.64\sqrt{g_\ast}$ &
$2.9\sqrt{g_\ast}$ &
$0.9$ (CP in the UV)
\\
\cline{2-5}
&
Future &
$1.1\sqrt{g_\ast}$ &
$20\sqrt{g_\ast}$ &
$18$ (CP in the UV)
\\
\hline\hline

Aligned $\mathrm{U}(3)\times\mathrm{CP}$ &
Current &
$0.17\sqrt{g_\ast}$ &
$0.78\sqrt{g_\ast}$ &
$0.03$
\\
\cline{2-5}
&
Future &
$0.30\sqrt{g_\ast}$ &
$5.3\sqrt{g_\ast}$ &
$0.62$
\\
\hline\hline

Generic $\mathrm{U}(2)\times\mathrm{U}(1)\times\mathrm{CP}$ &
Current &
$1.7g_\ast$ &
$2.9\sqrt{g_\ast}$ &
$3.0g_\ast$
\\
\cline{2-5}
&
Future &
$3.0g_\ast$ &
$20\sqrt{g_\ast}$ &
$61g_\ast$
\\
\hline\hline

Aligned $\mathrm{U}(2)\times\mathrm{U}(1)\times\mathrm{CP}$ &
Current &
$1.7g_\ast$ &
$0.78\sqrt{g_\ast}$ &
$3.0g_\ast$
\\
\cline{2-5}
&
Future &
$3.0g_\ast$ &
$5.3\sqrt{g_\ast}$ &
$61g_\ast$
\\
\hline\hline

$\mathrm{U}(1)^3\times\mathrm{CP}$ &
Current &
$6.8g_\ast$ &
$0.78\sqrt{g_\ast}$ &
$12g_\ast$
\\
\cline{2-5}
&
Future &
$12g_\ast$ &
$5.3\sqrt{g_\ast}$ &
$250g_\ast$
\\
\hline\hline

pRU and pLU &
Current &
$5.8\sqrt{a_{\ell,e}}\,g_\ast$ &
$2.7\sqrt{a_{\ell,e}}\,\sqrt{g_\ast}$ &
$-$
\\
\cline{2-5}
&
Future &
$10\sqrt{a_{\ell,e}}\,g_\ast$ &
$19\sqrt{a_{\ell,e}}\,\sqrt{g_\ast}$ &
$-$
\\
\hline
\end{tabular}
}
\caption{
Current bounds and future reach on the compositeness scale $m_*$, in TeV, for the different scenarios considered.
The numbers shown correspond to the left--right symmetric
case, $r_i=1$. All constraints should be understood as estimates with
$\mathcal{O}(1)$ uncertainty. The current dipole constraints are dominated by
$\mu\to e\gamma$, while in the future $\mu\to e\gamma$,
$\mu N\to eN$, and $\mu\to3e$ all give comparable constraints, with $\mu\to3e$ ultimately giving the slightly higher reach which we quote. The Higgs-current constraints are currently dominated by $\mu N\to eN$, while in the future both $\mu\to3e$ and $\mu N\to eN$ will significantly increase the reach, with $\mu N\to eN$ ultimately dominating.
}
\label{tab:model-comparison}
\end{table}

\subsection{Comparison of all scenarios}

\begin{figure}[t]
    \centering
\includegraphics[width=0.99\linewidth]{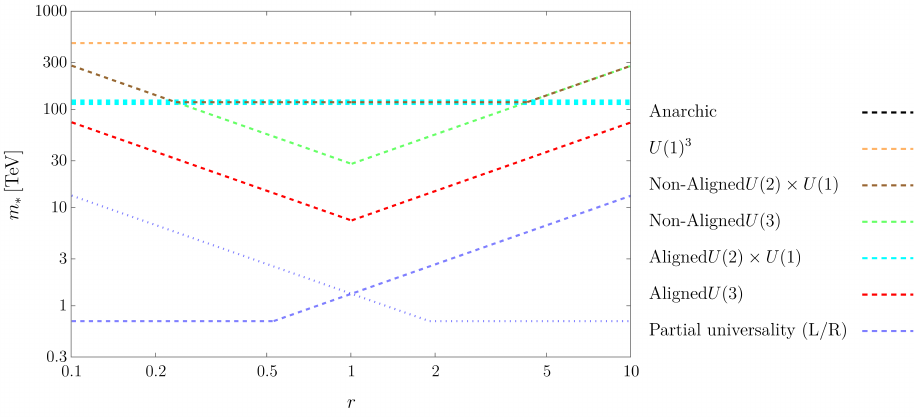}
    \caption{Comparison of the projected reach in the compositeness scale $m_*$ from lepton-flavor-violating observables and the electron EDM in the different flavor scenarios considered. For the scenarios of partial universality we set $r_\tau=r_e=r$, while in other scenarios we show the projections as a function of $r=r_e=r_\mu=r_\tau$. All projections should be understood as estimates with $\mathcal{O}(1)$ uncertainty.}
\label{fig:comparison_models_projection}
\end{figure}

We summarize the main constraints from CP and flavor violation in the lepton sector on the scenarios considered in \cref{tab:model-comparison}. For simplicity of presentation, these are shown only for $r_i=1$. We also show a comparison of the different scenarios in \cref{fig:comparison_models}. Overall, we highlight the following results:
\begin{itemize}
    \item Only scenarios with universal or partially universal mixings (MFV, pLU or pRU) and the scenarios we studied in \cref{sec:flavorprotection} i.e. based on the strong sector symmetry $\U{3}_\mathcal{O}\times \text{CP}$ or $\U{2}_\mathcal{O}\times \U{1}_\mathcal{O}\times \text{CP}$, are compatible with a compositeness scale $m_*$ below $\sim 10$~TeV.
    \item Only in the scenario based on $\U{3}_\mathcal{O}\times\text{CP}$ is the current bound on the compositeness scale already dominated by $\mu N\to eN$. In this case, the projected four-order-of-magnitude improvement in sensitivity to $\mu N\to eN$ and $\mu\to eee$ will extend the reach in the compositeness scale by an order of magnitude.
    \item  In the $\U{2}_\mathcal{O}\times \U{1}_\mathcal{O}\times \text{CP}$ scenario, the bound on the compositeness scale is currently dominated by $\mu \to e \gamma$ and the electron EDM. Nevertheless, the improvement in the sensitivity to $\mu N\to e N$ and $\mu \to eee$ probes a significant part of parameter space.
    \item The scenarios where $\tau$ decays provide a non-negligible probe are the scenarios of $\U{2}_\mathcal{O}\times \U{1}_\mathcal{O}\times \text{CP}$ and aligned $\U{3}_\mathcal{O}\times\text{CP}$. While for $r_i=1$ these are not leading, for $r_\tau \ne 1$ these can become the leading probe, as seen e.g. in \cref{fig:U2alignedtau}.
    \item Although the case of partial universality is already tightly constrained with bounds on $m_*$ in the ten TeV range, the improvement in sensitivity to $\mu N\to e N$ and $\mu \to eee$ probes a compositeness scale in the range of $\sim 30$~TeV.

\end{itemize}

\section{Conclusions}
\label{sec:conclusion}

\begingroup
\color{black}

The forthcoming improvements in searches for charged lepton flavor violation and the electron EDM offer an opportunity to test structures of new physics associated with the Higgs hierarchy problem. We have studied their implications in the framework of a Strongly Interacting Light Higgs with partial compositeness. In the anarchic scenario, present bounds from $\mu\to e\gamma$ and the electron EDM push the compositeness scale far beyond the reach of the LHC and envisioned future colliders, while the upcoming $\mu\to eee$ and $\mu N\to eN$ searches provide barely comparable bounds. As we show in this work, accidental flavor and CP symmetries of the strong sector can substantially change this conclusion.

We have studied a scenario with an accidental
$\mathrm{U}(3)_{\mathcal O}\times\mathrm{CP}$
symmetry of the strongly coupled sector, generically broken by the elementary--composite mixings. At leading order in these mixings, the symmetry restricts the Yukawa and dipole couplings to a single flavor structure, which results in the suppression of $\mu\to e\gamma$ and the electron EDM, allowing a compositeness scale in the few-TeV range. Flavor-changing $Z$ couplings generally survive, leaving $\mu\to eee$ and $\mu N\to eN$ conversion as leading probes of these models. Where these processes set the bounds, the projected improvement
in rate sensitivity extends the reach in the compositeness scale by about an order of magnitude. Improved EDM searches also test the CP violation that survives at higher order in the mixings.

We have also considered a $\mathrm{U}(2)_{\mathcal O}\times \mathrm{U}(1)_{\mathcal O}\times\mathrm{CP}$ scenario, in which the strong-sector flavor symmetry distinguishes
the third family. The $\mu\to e\gamma$ and electron EDM bounds remain important, while $\mu\to eee$, $\mu N\to eN$ conversion and flavor-violating $\tau$ decays provide complementary tests. In particular, $\tau$ decays can become leading probes when the left- or right-handed $\tau$ is fully composite.

These symmetry protections can coexist with a dynamical origin of the charged-lepton hierarchies. The flavor symmetry need only emerge in the infrared. Non-universal operator dimensions can generate hierarchical mixings during an earlier stage of RG evolution, while an approximately symmetric strong sector near the compositeness scale suppresses dangerous flavor and
CP violation. Therefore, the dynamics responsible for the flavor hierarchies do not spoil the symmetry protection emerging at low energies.

These scenarios provide concrete targets for upcoming
experiments and show how flavor protection can bring new
physics associated with the Higgs hierarchy problem within collider reach. Even if the forthcoming searches find no signal, some of the symmetry and mixing configurations considered here can remain viable at the TeV scale. Direct resonance searches and precision measurements at current and future colliders therefore complement probes of charged lepton flavor violation and the electron EDM. Together, they can test both the scale of the new dynamics and the flavor and CP structure that makes it viable.

We will pursue two further directions in future work.
The first concerns neutrino masses and leptogenesis in
composite Higgs models~\cite{Agashe:2024uvp}. Extending the flavor symmetry to the neutral resonances could
provide the near degeneracy among different families that
enhances the asymmetry in inverse-seesaw
leptogenesis~\cite{Agashe:2018cuf}. We will investigate whether this framework can accommodate successful leptogenesis from decays of TeV-scale resonances while preserving charged-lepton flavor protection. The second direction is to study realizations of the flavor
structure studied here beyond composite Higgs models.
A single Yukawa-like spurion could control the leading
chirality-changing interactions, while additional sources
of flavor breaking enter the left- and right-handed currents. We will characterize this structure and its power counting within an effective field theory containing only the SM degrees of freedom, extending the experimental targets identified here to a broader class of theories.

\endgroup
\acknowledgments

We thank Dario Buttazzo, Alfredo Glioti, Marco Nardecchia, Riccardo Rattazzi,
Marc Riembau and Andrea Wulzer for useful discussions. A.D.L. is partially funded by the European Union - Next Generation EU and MUR through the grant PRIN 202289JEW4, and by INFN.
M.E.\ is supported by the ERC (NOTIMEFORCOSMO, 101126304), by Scuola Normale Superiore, and by INFN (IS GSS-Pi).
M.P.\ was supported by the Swiss National Science Foundation (SNSF) through grant TMSGI2-225951 until October 2026, and is currently supported by an SNSF postdoctoral Mobility fellowship. S.S. acknowledges financial support from the Spanish Ministry of Science and Innovation (MICINN) through the Spanish State Research Agency, under Severo Ochoa Centres of Excellence Programme 2025-2029 (CEX2024001442-S). This work is also part of the R\&D\&i project PID2023-146686NB-C31, funded by MICIU/AEI/10.13039/501100011033/ and by ERDF/EU. IFAE is partially funded by the CERCA program of the Generalitat de Catalunya.

\newpage
\appendix
\section{$P_{LR}$ symmetry \label{sec:PLR}}
In this section, we investigate the implications of $P_{LR}$ for charged-lepton flavor-violating observables. 
In particular, we will review how the assumption of the $P_{LR}$ symmetry  forbids  modification of certain couplings of the $Z$ boson to fermionic currents at zero momentum \cite{Agashe_2006}. For such fermions, we then estimate the leading correction at non-zero momentum. 
To define $P_{LR}$ symmetry, it is convenient to first recall the custodial symmetry 
of the Higgs sector, $\SO{4}\simeq \SU{2}_L\times \SU{2}_R$, which is spontaneously broken by the Higgs vacuum expectation value to $ \SO{3} \simeq \SU{2}_V$.
$P_{LR}$ is defined as the discrete transformation exchanging the two $\SU{2}$ factors, $T_L^a \leftrightarrow T_R^a$.  The $P_{LR}$ can be seen as a parity transformation acting on $\mathbf{4}$ of $\SO{4}$ as the matrix $P_{LR}=\text{diag} ( 1,-1,-1,-1 )$. Note also that $P_{LR}$ commutes with the unbroken $ \SO{3} \simeq \SU{2}_V$. 

This symmetry is  broken by gauging only a subgroup of $\SU{2}_L\times \SU{2}_R$.
To systematically keep track of the symmetry breaking, 
it is useful to introduce
spurionic sources $W_L^{\mu,a}$ and $W_R^{\mu,a}$ gauging the full $\SU{2}_L\times \SU{2}_R$ symmetry. 
Under $P_{LR}$, these sources transform then as $W_{L\mu}^{a}
\leftrightarrow
W_{R\mu}^{a}$.
At the end of the
analysis, the sources are set to their physical values,
\begin{equation}
\begin{array}{ccc}
W_L^{\mu,a}=g W^{\mu,a}
& &
W_R^{\mu,a}=g'B^\mu\delta^{a3}.
\end{array}
\end{equation}
We would like to study the couplings of the $Z$ boson to the fermionic currents. To do so, we construct invariants under $\SU{2}_V$, which survives electroweak symmetry breaking. Under $\SU{2}_V$ the gauge sources transform as
\begin{equation}
W_{L\mu}
\longrightarrow
g_{V}
\left(
W_{L\mu}
+
i\partial_{\mu}
\right)
g_{V}^{\dagger},
\qquad
W_{R\mu}
\longrightarrow
g_{V}
\left(
W_{R\mu}
+
i\partial_{\mu}
\right)
g_{V}^{\dagger}.
\end{equation}
It is convenient to first define the combinations
\begin{equation}
W_{\pm \mu}^{a}
\equiv
\frac{1}{2}
\left(
W_{L\mu}^{a}
\pm
W_{R\mu}^{a}
\right),
\label{eq: PLR Wplusminus}
\end{equation}
which transform under $\SU{2}_V$ as
\begin{equation}
\begin{aligned}
W_{\mu}^{+}
&\longrightarrow
g_{V}
\left(
W_{\mu}^{+}
+
i\partial_{\mu}
\right)
g_{V}^{\dagger},
\\
W_{\mu}^{-}
&\longrightarrow
g_{V}
W_{\mu}^{-}
g_{V}^{\dagger}.
\end{aligned}
\end{equation}
Note also that under $P_{LR}$, they transform as $W^{\pm,a}_{\mu} \longrightarrow
\pm W^{\pm, a}_{ \mu}$.
Considering only the neutral component and setting the gauge spurions to their physical values gives
\begin{equation}
W_{\mu}^{-\;3}
=
\frac{g}{2c_{W}} Z_{\mu},
\qquad
W_{\mu}^{+\;3}
=
\frac{g}{2c_{W}}
\left(
c_{W}^{2}-s_{W}^{2}
\right)
Z_{\mu}
+
e A_{\mu}.
\label{eq:Wplusminus_physical}
\end{equation}
Following ref.~\cite{Panico:2015jxa}, we review how certain zero-momentum coupling corrections to the $Z$ are forbidden by imposing the $P_{LR}$ symmetry.

\paragraph{Zero-momentum $Z$ couplings:}
Let $\psi$ denote one chiral component of an elementary SM multiplet, such as $\nu_L$, $e_L$, or $e_R$, and let $\mathcal O_\psi$ denote the component of a strong-sector operator linearly mixing with $\psi$. Suppose that $\mathcal O_\psi$ is a $P_{LR}$ eigenstate. For definite custodial quantum numbers, this requires
\begin{equation}
 T_L=T_R,\qquad T_L^3=T_R^3,
 \label{eq: PLR eigenstate condition}
\end{equation}
where these labels refer to $\mathcal O_\psi$. Assigning $\psi$ the same parity then makes this mixing invariant.

To constrain the neutral vertex, it suffices to restrict the $\SU{2}_V$ source transformations to the subgroup generated by $T_V^3=T_L^3+T_R^3$. Writing invariants under this subgroup, the terms determining the zero-momentum neutral vertex can be written as
\begin{equation}
 \mathcal L_{\mathrm{neutral}}
 =i\bar\psi\gamma^\mu
 \left(\partial_\mu-2iT_L^3W_{+\mu}^3-iX X_\mu\right)\psi
 +2\delta g_\psi W_{-\mu}^3\bar\psi\gamma^\mu\psi,
 \label{eq: PLR zero momentum}
\end{equation}
where $X_\mu$ is the source for the conserved $\U{1}_X$ charge denoted by $X$. 
At the physical source values, $X_\mu=g'B_\mu$ and $Y=T_R^3+X=T_L^3+X$, the covariant derivative reproduces the SM neutral couplings. Since $2W_{-\mu}^3=(g/c_W)Z_\mu$, $\delta g_\psi$ is the shift in the $Z$ coupling. Under $P_{LR}$, the current $\bar\psi\gamma^\mu\psi$ is even and $W_-^3$ is odd, forcing $\delta g_\psi=0$.

For flavors with the same $T_L^3$ and $X$, $P_{LR}\psi_i=\eta_i\psi_i$ implies that $\bar\psi_i\gamma^\mu\psi_j$ has parity $\eta_i\eta_j$. The argument therefore also forbids flavor-changing vertices between states of equal parity. 

The criterion applies separately to $\nu_L$, $e_L$, and $e_R$. For a single custodial representation of the operator coupled to $\ell_L=(\nu_L,e_L)^T$, the two components share $T_R^3$ but have opposite $T_L^3$. They therefore cannot both satisfy eq.~\eqref{eq: PLR eigenstate condition}. With $Y=T_R^3+X$, protection of $e_L$ requires the operator to transform as $(\mathbf2,\mathbf2)_0$, whereas protection of $\nu_L$ requires $(\mathbf2,\mathbf2)_{-1}$. Protection of $e_R$ by this criterion requires $(\mathbf1,\mathbf1)_{-1}$. For mixings to generate charged-lepton Yukawa couplings, conservation of $X$ requires the operators that mix with $e_L$ and $e_R$ to carry the same $X$. Therefore, the couplings of $Z$ to $e_L$ and $e_R$ cannot be simultaneously protected. The assignments protecting $\nu_L$ and $e_R$, however, are compatible. 

\paragraph{Nonzero momentum:}
The zero-momentum protection does not eliminate momentum-dependent LFV interactions. For protected fields of equal $P_{LR}$ parity and common custodial charges, define $j_{ij}^{\mu}=\bar\psi_i\gamma^\mu\psi_j$. At linear order in the neutral sources, $\partial_\nu W_+^{\nu\mu,3}$ and $\partial_\nu X^{\nu\mu}$ are invariant under the neutral source transformations and even under $P_{LR}$. The symmetries therefore allow
\begin{equation}
 \mathcal L\supset
 \frac{\epsilon_{\psi_i}\epsilon_{\psi_j}}{m_*^2}
 j_{ij\mu}
 \left(
 c_+^{ij}\partial_\nu W_+^{\nu\mu,3}
 +c_X^{ij}\partial_\nu X^{\nu\mu}
 \right),
 \label{eq:PLR_derivative_sources}
\end{equation}
where flavor indices are summed. The coupling to $\partial_\nu W_-^{\nu\mu,3}$ is instead forbidden by $P_{LR}$. For a protected $e_R$ coupled to $(\mathbf1,\mathbf1)_{-1}$, custodial $\SU{2}_V$ invariance further requires $c_+^{ij}=0$: its singlet current cannot couple to a triplet derivative source. The $X$-source interaction remains allowed.

Setting the sources to their physical values gives
\begin{equation}
 \mathcal L\supset
 j_{ij\mu}
 \left(
 d_Z^{ij}\partial_\nu Z^{\nu\mu}
 +d_\gamma^{ij}\partial_\nu F^{\nu\mu}
 \right),
 \label{eq: PLR  nderivative interaction}
\end{equation}
with
\begin{equation}
 \begin{aligned}
 d_Z^{ij}
 &=
 \frac{g_Z\epsilon_{\psi_i}\epsilon_{\psi_j}}{m_*^2}
 \left[
 \frac{c_W^2-s_W^2}{2}c_+^{ij}
 -s_W^2c_X^{ij}
 \right],
 \\[3pt]
 d_\gamma^{ij}
 &=
 \frac{e\epsilon_{\psi_i}\epsilon_{\psi_j}}{m_*^2}
 \left(c_+^{ij}+c_X^{ij}\right),
 \end{aligned}
 \label{eq: PLR gZ and g gamma coupling}
\end{equation}
where $g_Z=g/c_W$ and we used $X_\mu=g'B_\mu=eA_\mu-g_Zs_W^2Z_\mu$.

At low energies, $|q^2|\ll m_Z^2$, the derivative $Z$ vertex supplies a factor of $q^2$, while the propagator supplies $1/m_Z^2$. The transverse vector contribution is consequently suppressed by $|q^2|/m_Z^2$. Relative to an unprotected Higgs-current contribution with coefficient $C_H^{ij}\sim g_*^2\epsilon_{\psi_i}\epsilon_{\psi_j}/m_*^2$, its amplitude satisfies
\begin{equation}
 \left|
 \frac{\mathcal A_{\partial Z}}{\mathcal A_H}
 \right|
 \sim
 \frac{g_Z^2}{g_*^2}\frac{|q^2|}{m_Z^2}
 \sim
 \frac{|q^2|}{g_*^2v^2} \sim \epsilon_{\ell_j}^2\epsilon_{e_j}^2\,,
 \label{eq:PLR_Z_suppression}
\end{equation}
where we used that for the decay of a charged-lepton with flavor $j$, $|q^2|\sim m_{\ell_j}^2$. Note, however, that considering the $Z$ pole observables, this low-energy suppression is absent: $\delta g_\psi^{ij}\sim\epsilon_{\psi_i}\epsilon_{\psi_j}m_Z^2/m_*^2$, corresponding to a suppression $\sim g_Z^2/g_*^2$ relative to an unprotected Higgs-current correction. 

For photon exchange, the factor of $q^2$ cancels the massless propagator, leaving a contact interaction without the above low-energy suppression. To compare it with direct strong-sector exchange, consider the operator
\[
 (\bar\psi_i\gamma^\mu\psi_j)(\bar f\gamma_\mu f),
\]
where $f$ is a charged fermion of specified flavor and chirality. The photon and direct contributions to its coefficient generically scale as
\begin{equation}
 C_\gamma^{ij;f}
 \sim
 \frac{e^2\epsilon_{\psi_i}\epsilon_{\psi_j}}{m_*^2},
 \qquad
 C_{\mathrm{direct}}^{ij;f}
 \sim
 \frac{g_*^2\epsilon_{\psi_i}\epsilon_{\psi_j}\epsilon_f^2}{m_*^2},
\end{equation}
with electric charges and order-one coefficients suppressed. Thus
\begin{equation}
 \left|
 \frac{C_\gamma^{ij;f}}{C_{\mathrm{direct}}^{ij;f}}
 \right|
 \sim
 \frac{e^2}{g_*^2\epsilon_f^2}.
 \label{eq:PLR_photon_vs_fourfermion}
\end{equation}
Photon exchange can therefore dominate for sufficiently elementary $f$, with $g_*\epsilon_f\lesssim e$, and must be retained in that regime. 

\paragraph{Custodial $\SO{4}$ protection:}
Custodial $\SO{4}$ can also protect the $Z$ couplings to right-handed charged leptons \cite{Agashe_2006}. In the custodial limit, the leading correction encoded in $\mathcal{O}_{He}$ is proportional to the $T_R^3$ eigenvalue of the lepton embedding and therefore vanishes for $T_R^3=0$.
This condition is satisfied, for example, when $e_R$ is embedded in a $(\mathbf1,\mathbf1)_{-1}$ or in the $T_R^3=0$ component of a $(\mathbf1,\mathbf3)_{-1}$ of $\SU{2}_L\times \SU{2}_R \times \U1_X$. This protection cannot be combined with $P_{LR}$ protection of the left-handed charged lepton. 
Indeed, protecting $e_L$ through $P_{LR}$ requires $T_R^3=T_L^3=-1/2$, which fixes $X_\ell=0$, whereas protecting $e_R$ requires $T_R^3=0$, and hence $X_e=-1$.
The protection applies at zero momentum while momentum-dependent corrections remain allowed through derivative interactions.

\providecommand{\tauadd}[1]{{\color{brown}#1}}
\providecommand{\taudel}[1]{\sout{#1}}
\providecommand{\taudelmath}[1]{\cancel{#1}}
\providecommand{\tauReview}[1]{#1}
\providecommand{\tauvalue}[2]{%
  \makecell[c]{\(#1\)\\[-1pt]\(\tauadd{#2}\)}}

\providecommand{\tauadd}[1]{{\color{brown}#1}}
\providecommand{\taudel}[1]{\sout{#1}}
\providecommand{\taudelmath}[1]{\cancel{#1}}
\providecommand{\tauReview}[1]{#1}

\providecommand{\CLFVCurrent}{\textsc{Current}}
\providecommand{\CLFVMid}{\textsc{Mid}}
\providecommand{\CLFVFuture}{\textsc{Future}}
\providecommand{\CLFVNoData}{\textemdash}
\providecommand{\adl}[1]{\textcolor{green!50!black}{#1}}
\providecommand{\me}[1]{\textcolor{blue}{#1}}
\providecommand{\MPcol}[1]{\textcolor{Orange}{#1}}
\providecommand{\MP}[1]{\textsf{\MPcol{[MP: #1]}}}

\section{Experimental prospects for LFV in the $\tau$ sector}
\label{app:taudecays}
We summarize the current status and prospects of LFV searches in the
$\tau$ sector. We focus on $\tau\to\mu$ transitions, since the
corresponding $\tau\to e$ rates are parametrically suppressed in our
scenarios, while the experimental sensitivities are comparable
\cite{banerjee2022snowmass2021whitepaper}.
\Cref{tab:tau-clfv-sensitivities} collects the relevant radiative,
three-lepton and semileptonic searches, including channels beyond
those discussed in the main text. Belle~II can study all these final
states, while LHCb, ATLAS and CMS provide complementary sensitivity
to $\tau\to3\mu$. Several nonradiative searches are projected to reach
branching ratios of a few $10^{-10}$
\cite{banerjee2022snowmass2021whitepaper,theatlascollaboration2025projectionskeymeasurementsheavy}.
\Cref{tab:tau-operator-bounds} gives the corresponding bounds and
reach on the dipole and Higgs-current coefficients.

{\small
\setlength{\tabcolsep}{5pt}
\setlength{\LTcapwidth}{\textwidth}
\renewcommand{\arraystretch}{1.05}

\begin{longtable}{
  >{\raggedright\arraybackslash}p{0.14\textwidth}
  |>{\raggedright\arraybackslash}p{0.50\textwidth}
  >{\centering\arraybackslash}p{0.25\textwidth}}
\caption{Experimental upper limits and projected sensitivities at
90\% confidence or credibility level, following each analysis,
for selected LFV $\tau$ decays. \CLFVCurrent\ denotes observed upper limits;  \CLFVMid\ denotes sensitivities expected in the early 2030s, with
$10\,\mathrm{ab}^{-1}$ at Belle~II and $50\,\mathrm{fb}^{-1}$ at LHCb.
\CLFVFuture\ denotes sensitivities expected with the final data samples
in the early 2040s, assuming the upgrades and timelines considered in
ref.~\cite{theatlascollaboration2025projectionskeymeasurementsheavy}.
The projections are expected upper limits in the absence of a signal;
ranges reflect different background assumptions. 
Hadronic projections follow
ref.~\cite{banerjee2022snowmass2021whitepaper}.}
\label{tab:tau-clfv-sensitivities}\\

\toprule
\multicolumn{1}{c|}{\textbf{Status}} &
\multicolumn{1}{c}{\textbf{Experiment and data sample}} &
\multicolumn{1}{c}{\makecell[c]{\textbf{Upper limit /}\\\textbf{sensitivity}}}\\
\midrule
\endfirsthead

\multicolumn{3}{l}{%
  {\bfseries \tablename\ \thetable{}} --- Continued from previous page}\\
\toprule
\multicolumn{1}{c|}{\textbf{Status}} &
\multicolumn{1}{c}{\textbf{Experiment and data sample}} &
\multicolumn{1}{c}{\makecell[c]{\textbf{Upper limit /}\\\textbf{sensitivity}}}\\
\midrule
\endhead

\midrule
\multicolumn{3}{r}{Continued on next page}\\
\bottomrule
\endfoot

\bottomrule
\endlastfoot

\multicolumn{3}{l}{\(\tau\to\mu\gamma\)}\\*
\CLFVCurrent
  & Belle, \(988\,\mathrm{fb}^{-1}\) \cite{2021}
  & \(<4.2\times10^{-8}\)\\
\CLFVCurrent
  & Belle~II, \(428\,\mathrm{fb}^{-1}\) \cite{BelleII:TauMuGamma2026}
  & \(<9.5\times10^{-8}\)\\
\CLFVMid
  & Belle~II, \(10\,\mathrm{ab}^{-1}\) \cite{theatlascollaboration2025projectionskeymeasurementsheavy}
  & \(<1.5\times10^{-8}\)\\
\CLFVFuture
  & Belle~II, \(50\,\mathrm{ab}^{-1}\) \cite{theatlascollaboration2025projectionskeymeasurementsheavy}
  & \(<7.0\times10^{-9}\)\\\hline

\multicolumn{3}{l}{\(\tau^-\to\mu^-\mu^+\mu^-\)}\\*
\CLFVCurrent
  & Belle, \(782\,\mathrm{fb}^{-1}\) \cite{banerjee2022snowmass2021whitepaper}
  & \(<2.1\times10^{-8}\)\\
\CLFVCurrent
  & LHCb, \(5.4\,\mathrm{fb}^{-1} \)  \cite{LHCb:Tau3Mu2026}
  & \(<1.9 \times10^{-8}\)\\
\CLFVCurrent
  & CMS, \(131\,\mathrm{fb}^{-1}\)  \cite{CMS:Tau3Mu2024}
  & \(<2.9\times10^{-8}\)\\
\CLFVCurrent
  & ATLAS, \(137\,\mathrm{fb}^{-1}\)  \cite{ATLAS:Tau3Mu2026}
  & \(<8.7\times10^{-8}\)\\
\CLFVCurrent
  & Belle~II, \(424\,\mathrm{fb}^{-1}\)  \cite{BelleII:Tau3Mu2024}
  & \(<1.9\times10^{-8}\)\\
\CLFVMid
  & LHCb, \(50\,\mathrm{fb}^{-1}\) \cite{theatlascollaboration2025projectionskeymeasurementsheavy}
  & \(<6.4\times10^{-9}\)\\
\CLFVMid
  & Belle~II, \(10\,\mathrm{ab}^{-1}\) \cite{theatlascollaboration2025projectionskeymeasurementsheavy}
  & \(<(0.8\text{--}3.7)\times10^{-9}\)\\
\CLFVFuture
  & LHCb, \(300\,\mathrm{fb}^{-1}\) \cite{theatlascollaboration2025projectionskeymeasurementsheavy}
  & \(<2.6\times10^{-9}\)\\
\CLFVFuture
  & CMS, \(3000\,\mathrm{fb}^{-1}\) \cite{theatlascollaboration2025projectionskeymeasurementsheavy}
  & \(<3.9\times10^{-9}\)\\
\CLFVFuture
  & ATLAS, \(3000\,\mathrm{fb}^{-1}\) \cite{theatlascollaboration2025projectionskeymeasurementsheavy}
  & \(<(1.3\text{--}6.4)\times10^{-9}\)\\
\CLFVFuture
  & Belle~II, \(50\,\mathrm{ab}^{-1}\) \cite{theatlascollaboration2025projectionskeymeasurementsheavy}
  & \(<(0.2\text{--}1.7)\times10^{-9}\)\\\hline

\multicolumn{3}{l}{\(\tau^-\to\mu^-e^+e^-\)}\\*

\CLFVCurrent
  & Belle~II, \(428\,\mathrm{fb}^{-1}\) \cite{BelleII:TauThreeLeptons2025}
  & \(<1.6\times10^{-8}\)\\

\CLFVFuture
  & Belle~II, \(50\,\mathrm{ab}^{-1}\) \cite{banerjee2022snowmass2021whitepaper}
  & \(<2.9\times10^{-10}\)\\\hline
\multicolumn{3}{l}{\(\tau\to\mu\rho^0\)}\\*

\CLFVCurrent
  & Belle, \(980\,\mathrm{fb}^{-1}\)
    \cite{2023}
  & \(<1.7\times10^{-8}\)\\
\CLFVFuture
  & Belle~II, \(50\,\mathrm{ab}^{-1}\) \cite{banerjee2022snowmass2021whitepaper}
  & \(<5.5\times10^{-10}\)\\\hline

\multicolumn{3}{l}{\(\tau\to\mu K_S^0\)}\\*
\CLFVCurrent
  & Belle, \(980\,\mathrm{fb}^{-1}\), and Belle~II,
    \(428\,\mathrm{fb}^{-1}\) \cite{2025}
  & \(<1.2\times10^{-8}\)\\
\CLFVFuture
  & Belle~II, \(50\,\mathrm{ab}^{-1}\) \cite{banerjee2022snowmass2021whitepaper}
  & \(<4.0\times10^{-10}\)\\\hline

\multicolumn{3}{l}{\(\tau\to\mu\omega\)}\\*
\CLFVCurrent
  & Belle, \(980\,\mathrm{fb}^{-1}\) \cite{2023}
  & \(<3.9\times10^{-8}\)\\
\CLFVFuture
  & Belle~II, \(50\,\mathrm{ab}^{-1}\) \cite{banerjee2022snowmass2021whitepaper}
  & \(<1.4\times10^{-9}\)\\\hline

\multicolumn{3}{l}{\(\tau\to\mu\phi\)}\\*
\CLFVCurrent
  & Belle, \(980\,\mathrm{fb}^{-1}\) \cite{2023}
  & \(<2.3\times10^{-8}\)\\
\CLFVFuture
  & Belle~II, \(50\,\mathrm{ab}^{-1}\) \cite{banerjee2022snowmass2021whitepaper}
  & \(<8.4\times10^{-10}\)\\\hline

\multicolumn{3}{l}{\(\tau\to\mu\pi^0\)}\\*

\CLFVCurrent
  & BaBar, \(339\,\mathrm{fb}^{-1}\) \cite{BaBarTauPseudoscalars2007}
  & \(<1.1\times10^{-7}\)\\*
\CLFVFuture
  & Belle~II, \(50\,\mathrm{ab}^{-1}\) \cite{banerjee2022snowmass2021whitepaper}
  & \(<7.1\times10^{-10}\)\\\hline

\multicolumn{3}{l}{\(\tau\to\mu\pi^+\pi^-\)}\\*
\CLFVCurrent
  & Belle, \(854\,\mathrm{fb}^{-1}\) \cite{banerjee2022snowmass2021whitepaper}
  & \(<2.1\times10^{-8}\)\\
\CLFVFuture
  & Belle~II, \(50\,\mathrm{ab}^{-1}\) \cite{banerjee2022snowmass2021whitepaper}
  & \(<5.6\times10^{-10}\)\\\hline

\multicolumn{3}{l}{\(\tau\to\mu\eta\)}\\*
\CLFVCurrent
  & Belle~II (preliminary), \(428\,\mathrm{fb}^{-1}\)
    \cite{Liao:PhiPsi2026Tau}
  & \(<4.2\times10^{-8}\)\\
\CLFVFuture
  & Belle~II, \(50\,\mathrm{ab}^{-1}\)
    \cite{banerjee2022snowmass2021whitepaper}
  & \(<8.0\times10^{-10}\)\\\hline

\multicolumn{3}{l}{\(\tau\to\mu\eta'\)}\\*
\CLFVCurrent
  & Belle, \(401\,\mathrm{fb}^{-1}\) \cite{BellePseudoscalars2007}
  & \(<1.3\times10^{-7}\)\\
\CLFVFuture
  & Belle~II, \(50\,\mathrm{ab}^{-1}\)
    \cite{banerjee2022snowmass2021whitepaper}
  & \(<1.2\times10^{-9}\)\\\hline

\multicolumn{3}{l}{\(\tau\to\mu K^+K^-\)}\\*
\CLFVCurrent
  & Belle, \(854\,\mathrm{fb}^{-1}\)
    \cite{banerjee2022snowmass2021whitepaper}
  & \(<4.4\times10^{-8}\)\\
\CLFVFuture
  & Belle~II, \(50\,\mathrm{ab}^{-1}\)
    \cite{banerjee2022snowmass2021whitepaper}
  & \(<1.1\times10^{-9}\)\\\hline

\end{longtable}
} %

\begingroup
\setlength{\emergencystretch}{1em}
We translate the experimental sensitivities into bounds on the
dimensionful Wilson coefficients of the main text, with $[C_i]=-2$ \cite{altmannshofer2023probingleptonflavorviolation}.
The relevant Higgs-current combinations are
\begin{equation}
    C_L^{\tau\mu}=(C_{H\ell}^{(1)}+C_{H\ell}^{(3)})^{\tau\mu},
    \qquad C_R^{\tau\mu}=C_{He}^{\tau\mu}.
\end{equation}
At momenta well below $m_Z$, they give
\begin{equation}
    \mathcal M_H(\tau\to\mu X)
    \propto (C_\chi^{\tau\mu})^*\,
    \bar u_\mu\gamma_\alpha P_\chi u_\tau\,
    \langle X|J_Z^\alpha|0\rangle,
    \qquad \chi=L,R,
\end{equation}
where $P_{L,R}=(1\mp\gamma_5)/2$ and
$J_Z^\alpha=\sum_f\bar f\gamma^\alpha(g_V^f-g_A^f\gamma_5)f$,
with $g_V^f=T_3^f-2Q_fs_W^2$ and $g_A^f=T_3^f$.

The strongest dipole bound comes from $\tau\to\mu\gamma$.
Its width is $v^2m_\tau^3|C_{e\gamma}|^2/(8\pi)$ when the muon
mass is neglected.
Dipole contributions to three-lepton decays require conversion of a virtual
photon, giving
\begin{equation}
    \frac{\Gamma_D(\tau\to\mu\ell^+\ell^-)}
         {\Gamma_D(\tau\to\mu\gamma)}
    \simeq \frac{\alpha}{3\pi}
    \left[\log\!\left(\frac{m_\tau^2}{m_\ell^2}\right)-c_\ell\right],
    \qquad c_e=3,\quad c_\mu=\frac{11}{4}.
\end{equation}
These ratios are approximately $0.010$ for $\mu e^+e^-$ and
$0.0022$ for $3\mu$. The larger electron logarithm makes
$\mu e^+e^-$ more sensitive to the dipole at comparable
branching-ratio limits.

Hadronic rates depend on the meson decay constants and quark charges.
We use the decay-constant conventions of ref.~\cite{BallZwicky2005}
for vector mesons and 
\cite{FeldmannKrollStech1998} for $\eta$ and $\eta'$, with
$f_\pi\simeq130\,\mathrm{MeV}$.
In the isospin limit and neglecting $\omega$--$\phi$ mixing,
the relevant matrix elements are
\begin{equation}
\begin{aligned}
    \langle\pi^0|J_Z^\alpha|0\rangle
        &=i\frac{f_\pi}{\sqrt{2}}(g_A^u-g_A^d)p^\alpha,\\
    \langle P|J_Z^\alpha|0\rangle
        &=i g_A^s f_P^s p^\alpha,\qquad P=\eta,\eta',\\
    \langle\rho^0|J_Z^\alpha|0\rangle
        &=\frac{f_\rho}{\sqrt{2}}(g_V^u-g_V^d)m_\rho\varepsilon^{*\alpha},\\
    \langle\omega|J_Z^\alpha|0\rangle
        &=\frac{f_\omega}{\sqrt{2}}(g_V^u+g_V^d)m_\omega\varepsilon^{*\alpha},\\
    \langle\phi|J_Z^\alpha|0\rangle
        &=f_\phi g_V^s m_\phi\varepsilon^{*\alpha}.
\end{aligned}
\end{equation}
Here $\varepsilon^\alpha$ denotes the vector-meson polarization.
The non-strange axial contributions to $\eta$ and $\eta'$ cancel
because $g_A^u+g_A^d=0$ \cite{Goto_2015}.
For the electromagnetic current, the corresponding combinations are
$f_V^{\rm em}=f_\rho/\sqrt{2},\,f_\omega/(3\sqrt{2}),\,-f_\phi/3$
for $V=\rho^0,\omega,\phi$, respectively.
The dipole rates relative to $\mu\gamma$ scale as
$e^2|f_V^{\rm em}|^2/m_V^2$, so the smaller electromagnetic
charges weaken the $\omega$ and $\phi$ bounds relative to $\rho^0$.

Higgs-current operators produce both leptonic and hadronic modes
at tree level. Their widths scale as
$|C_\chi|^2m_\tau^5/(384\pi^3)$ for three leptons and
$|C_\chi|^2f_M^2m_\tau^3/(32\pi)$ for a single meson,
up to numerical factors, weak charges and kinematic corrections.
The two-body phase space compensates for much of the decay-constant
suppression, while meson masses introduce only order-one corrections.
The relevant hadronic charges are
\begin{equation}
\begin{gathered}
    g_A^u-g_A^d=1,\qquad g_A^s=-\frac12,\\
    g_V^u-g_V^d\simeq0.54,\qquad
    g_V^u+g_V^d\simeq-0.15,\qquad
    g_V^s\simeq-0.35.
\end{gathered}
\end{equation}
The weak-isospin cancellation in the $\omega$ current suppresses
its rate by roughly an order of magnitude relative to $\rho^0$
or $\phi$. In three-lepton decays, the small vector charge
$g_V^\ell\simeq-0.04$ does not suppress the rate because
$g_A^\ell=-1/2$.

Parity forbids a single photon from producing $\pi^0$, $\eta$ or
$\eta'$. Higgs-current operators produce these modes through the
axial quark current, without an additional suppression by the
muon or light-quark masses \cite{Goto_2015}. Neither operator produces $\mu K_S^0$
at tree level, since the electromagnetic and SM neutral quark
currents conserve strangeness.

For the numerical estimates we use the hadronic normalization of
ref.~\cite{altmannshofer2023probingleptonflavorviolation},
$f_\omega=195\,\mathrm{MeV}$ \cite{BallZwicky2005}, and
$f_\eta^s=-110\,\mathrm{MeV}$,
$f_{\eta'}^s=135\,\mathrm{MeV}$ \cite{Goto_2015}.
We use the decay rates given in \cite{Calibbi:2021pyh}.
The $\pi^+\pi^-$ and $K^+K^-$ entries assume $\rho$ and $\phi$
dominance.

\begin{table}[H]
\centering
\small
\caption{Current bounds and future sensitivities on
$|C_i|^{-1/2}$ in TeV, using \cref{tab:tau-clfv-sensitivities}
and one coefficient taken to be non-zero at a time. Future sensitivities correspond to Belle~II at
$50\,\mathrm{ab}^{-1}$; the $3\mu$ range reflects different
background assumptions. A dash denotes no leading-order
contribution from the operator considered.
The $\pi^+\pi^-$ and $K^+K^-$ entries are indicative estimates
assuming $\rho$ and $\phi$ dominance, respectively.
The present $\mu\eta$ bound uses the preliminary Belle~II result. The conversions assume the signal efficiencies used in the quoted
experimental analyses and projections.}
\label{tab:tau-operator-bounds}
\setlength{\tabcolsep}{4pt}
\renewcommand{\arraystretch}{1.15}
\begin{tabular*}{\textwidth}{@{\extracolsep{\fill}}lcccc@{}}
\toprule
& \multicolumn{2}{c}{Dipole}
& \multicolumn{2}{c}{Higgs current}\\
\cmidrule(lr){2-3}\cmidrule(l){4-5}
Channel & Current & Future & Current & Future\\
\midrule
$\tau\to\mu\gamma$ & $610$ & $960$
    & \textemdash & \textemdash\\
$\tau\to3\mu$ & $160$ & $300\text{--}510$
    & \makecell[c]{$9.0$}
    & \makecell[c]{$17\text{--}28$}\\
$\tau\to\mu e^+e^-$ & $250$ & $680$ & $8.5$ & $23$\\
$\tau\to\mu\rho^0$ & $170$ & $400$ & $9.4$ & $22$\\
$\tau\to\mu\phi$ & $90$ & $210$ & $8.3$ & $19$\\
$\tau\to\mu\omega$ & $77$ & $180$ & $4.0$ & $9.2$\\
$\tau\to\mu\pi^0$ & \textemdash & \textemdash & $6.5$ & $23$\\
$\tau\to\mu\pi^+\pi^-$ & $\sim160$ & $\sim400$
    & $\sim9.0$ & $\sim22$\\
$\tau\to\mu\eta$ & \textemdash & \textemdash & $6.2$ & $17$\\
$\tau\to\mu\eta'$ & \textemdash & \textemdash & $4.6$ & $15$\\
$\tau\to\mu K^+K^-$ & $\sim65$ & $\sim160$
    & $\sim5.9$ & $\sim15$\\
\bottomrule
\end{tabular*}
\end{table}
\endgroup

\clearpage

\bibliographystyle{JHEP}
\bibliography{Refs}

@article{Panico:2016ull,
    author = {Panico, Giuliano and Pomarol, Alex},
    title = {{Flavor hierarchies from dynamical scales}},
    journal = {JHEP},
    volume = {07},
    pages = {097},
    year = {2016},
    doi = {10.1007/JHEP07(2016)097},
    eprint = {1603.06609},
    archivePrefix = {arXiv},
    primaryClass = {hep-ph},
    reportNumber = {CERN-TH-2016-065}
}

@article{Frigerio:2018uwx,
    author = {Frigerio, Michele and Nardecchia, Marco and Serra, Javi and Vecchi, Luca},
    title = {{The Bearable Compositeness of Leptons}},
    journal = {JHEP},
    volume = {10},
    pages = {017},
    year = {2018},
    doi = {10.1007/JHEP10(2018)017},
    eprint = {1807.04279},
    archivePrefix = {arXiv},
    primaryClass = {hep-ph},
    reportNumber = {CERN-TH-2018-160}
}

@article{Isidori:2012ts,
    author = {Isidori, Gino and Straub, David M.},
    title = {{Minimal Flavour Violation and Beyond}},
    journal = {Eur. Phys. J. C},
    volume = {72},
    pages = {2103},
    year = {2012},
    doi = {10.1140/epjc/s10052-012-2103-1},
    eprint = {1202.0464},
    archivePrefix = {arXiv},
    primaryClass = {hep-ph}
}

@inproceedings{deBlas:2022ofj,
    author = {de Blas, Jorge and Du, Yong and Grojean, Christophe and Gu, Jiayin and Miralles, Victor and Peskin, Michael E. and Tian, Junping and Vos, Marcel and Vryonidou, Eleni},
    title = {{Global SMEFT Fits at Future Colliders}},
    booktitle = {{Snowmass 2021}},
    year = {2022},
    eprint = {2206.08326},
    archivePrefix = {arXiv},
    primaryClass = {hep-ph}
}

@article{Contino:2013kra,
    author = {Contino, Roberto and Ghezzi, Margherita and Grojean, Christophe and Muhlleitner, Margarete and Spira, Michael},
    title = {{Effective Lagrangian for a light Higgs-like scalar}},
    journal = {JHEP},
    volume = {07},
    pages = {035},
    year = {2013},
    doi = {10.1007/JHEP07(2013)035},
    eprint = {1303.3876},
    archivePrefix = {arXiv},
    primaryClass = {hep-ph},
    reportNumber = {CERN-PH-TH-2013-047, KA-TP-06-2013, PSI-PR-13-04}
}

@article{Barbieri:2012bh,
    author = "Barbieri, Riccardo and Buttazzo, Dario and Sala, Filippo and Straub, David M.",
    title = "{Less Minimal Flavour Violation}",
    eprint = "1206.1327",
    archivePrefix = "arXiv",
    primaryClass = "hep-ph",
    doi = "10.1007/JHEP10(2012)040",
    journal = "JHEP",
    volume = "10",
    pages = "040",
    year = "2012"
}

@article{Barbieri:2012uh,
    author = "Barbieri, Riccardo and Buttazzo, Dario and Sala, Filippo and Straub, David M.",
    title = "{Flavour physics from an approximate $U(2)^3$ symmetry}",
    eprint = "1203.4218",
    archivePrefix = "arXiv",
    primaryClass = "hep-ph",
    doi = "10.1007/JHEP07(2012)181",
    journal = "JHEP",
    volume = "07",
    pages = "181",
    year = "2012"
}

@article{Cacciapaglia:2007fw,
    author = "Cacciapaglia, Giacomo and Csaki, Csaba and Galloway, Jamison and Marandella, Guido and Terning, John and Weiler, Andreas",
    title = "{A GIM Mechanism from Extra Dimensions}",
    eprint = "0709.1714",
    archivePrefix = "arXiv",
    primaryClass = "hep-ph",
    doi = "10.1088/1126-6708/2008/04/006",
    journal = "JHEP",
    volume = "04",
    pages = "006",
    year = "2008"
}

@article{Fitzpatrick:2007sa,
    author = "Fitzpatrick, A. Liam and Perez, Gilad and Randall, Lisa",
    title = "{Flavor anarchy in a Randall-Sundrum model with 5D minimal flavor violation and a low Kaluza-Klein scale}",
    eprint = "0710.1869",
    archivePrefix = "arXiv",
    primaryClass = "hep-ph",
    doi = "10.1103/PhysRevLett.100.171604",
    journal = "Phys. Rev. Lett.",
    volume = "100",
    pages = "171604",
    year = "2008"
}

@Article{Barbieri2012,
  author        = {Barbieri, Riccardo and Buttazzo, Dario and Sala, Filippo and Straub, David M. and Tesi, Andrea},
  journal       = {Journal of High Energy Physics},
  title         = {A 125 GeV composite Higgs boson versus flavour and electroweak precision tests},
  year          = {2012},
  issn          = {1029-8479},
  month         = may,
  number        = {5},
  volume        = {2013},
  archiveprefix = {arXiv},
  copyright     = {arXiv.org perpetual, non-exclusive license},
  date          = {2012-11-21},
  doi           = {10.1007/jhep05(2013)069},
  eprint        = {1211.5085},
  primaryclass  = {hep-ph},
  publisher     = {Springer Science and Business Media LLC},
}

@article{Redi:2011zi,
    author = {Redi, Michele and Weiler, Andreas},
    title = {{Flavor and CP Invariant Composite Higgs Models}},
    journal = {JHEP},
    volume = {11},
    pages = {108},
    year = {2011},
    doi = {10.1007/JHEP11(2011)108},
    eprint = {1106.6357},
    archivePrefix = {arXiv},
    primaryClass = {hep-ph},
    reportNumber = {DESY-11-109, CERN-PH-TH-154}
}

@article{Redi:2012uj,
    author = "Redi, Michele",
    title = "{Composite MFV and Beyond}",
    eprint = "1203.4220",
    archivePrefix = "arXiv",
    primaryClass = "hep-ph",
    reportNumber = "CERN-PH-TH-2012-067",
    doi = "10.1140/epjc/s10052-012-2030-1",
    journal = "Eur. Phys. J. C",
    volume = "72",
    pages = "2030",
    year = "2012"
}

@article{Santiago:2008vq,
    author = "Santiago, Jose",
    title = "{Minimal Flavor Protection: A New Flavor Paradigm in Warped Models}",
    eprint = "0806.1230",
    archivePrefix = "arXiv",
    primaryClass = "hep-ph",
    doi = "10.1088/1126-6708/2008/12/046",
    journal = "JHEP",
    volume = "12",
    pages = "046",
    year = "2008"
}

@article{Delaunay:2010dw,
    author = "Delaunay, Cedric and Gedalia, Oram and Lee, Seung J. and Perez, Gilad and Ponton, Eduardo",
    title = "{Ultra Visible Warped Model from Flavor Triviality and Improved Naturalness}",
    eprint = "1007.0243",
    archivePrefix = "arXiv",
    primaryClass = "hep-ph",
    doi = "10.1103/PhysRevD.83.115003",
    journal = "Phys. Rev. D",
    volume = "83",
    pages = "115003",
    year = "2011"
}

@article{Csaki:2009wc,
    author = "Csaki, Csaba and Perez, Gilad and Surujon, Ze'ev and Weiler, Andreas",
    title = "{Flavor Alignment via Shining in RS}",
    eprint = "0907.0474",
    archivePrefix = "arXiv",
    primaryClass = "hep-ph",
    reportNumber = "CERN-PH-TH-2009-109, UCSD-PTH-09-06",
    doi = "10.1103/PhysRevD.81.075025",
    journal = "Phys. Rev. D",
    volume = "81",
    pages = "075025",
    year = "2010"
}

@article{Calibbi:2021pyh,
    author = "Calibbi, Lorenzo and Marcano, Xabier and Roy, Joydeep",
    title = "{Z lepton flavour violation as a probe for new physics at future $e^+e^-$ colliders}",
    eprint = "2107.10273",
    archivePrefix = "arXiv",
    primaryClass = "hep-ph",
    reportNumber = "TUM-HEP 1352/21",
    doi = "10.1140/epjc/s10052-021-09777-3",
    journal = "Eur. Phys. J. C",
    volume = "81",
    number = "12",
    pages = "1054",
    year = "2021"
}

@article{Csaki:2008eh,
    author = "Csaki, Csaba and Falkowski, Adam and Weiler, Andreas",
    title = "{A Simple Flavor Protection for RS}",
    eprint = "0806.3757",
    archivePrefix = "arXiv",
    primaryClass = "hep-ph",
    doi = "10.1103/PhysRevD.80.016001",
    journal = "Phys. Rev. D",
    volume = "80",
    pages = "016001",
    year = "2009"
}

@article{Huber:2003sf,
    author = "Huber, Stephan J. and Shafi, Qaisar",
    title = "{Seesaw mechanism in warped geometry}",
    eprint = "hep-ph/0309252",
    archivePrefix = "arXiv",
    reportNumber = "DESY-03-110, BA-03-11",
    doi = "10.1016/j.physletb.2003.12.012",
    journal = "Phys. Lett. B",
    volume = "583",
    pages = "293--303",
    year = "2004"
}

@article{Agashe:2025tge,
    author = {Agashe, Kaustubh and Ricci, Lorenzo and Sundrum, Raman},
    title = {{Partial compositeness: From anarchy to symmetry}},
    journal = {Phys. Rev. D},
    volume = {112},
    number = {7},
    pages = {075025},
    year = {2025},
    doi = {10.1103/hv22-69yz},
    eprint = {2507.05332},
    archivePrefix = {arXiv},
    primaryClass = {hep-ph},
    reportNumber = {UMD-PP-025-04}
}

@article{Glioti:2024hye,
    author = {Glioti, Alfredo and Rattazzi, Riccardo and Ricci, Lorenzo and Vecchi, Luca},
    title = {{Exploring the flavor symmetry landscape}},
    journal = {SciPost Phys.},
    volume = {18},
    number = {6},
    pages = {201},
    year = {2025},
    doi = {10.21468/SciPostPhys.18.6.201},
    eprint = {2402.09503},
    archivePrefix = {arXiv},
    primaryClass = {hep-ph}
}

@article{MEGII:2025new,
    author = {Afanaciev, K. and others},
    collaboration = {MEG II},
    title = {{New limit on the $\mu^+ \to e^+\gamma$ decay with the MEG II experiment}},
    year = {2025},
    eprint = {2504.15711},
    archivePrefix = {arXiv},
    primaryClass = {hep-ex}
}

@article{Liu:2016idz,
    author = {Liu, Da and Pomarol, Alex and Rattazzi, Riccardo and Riva, Francesco},
    title = {{Patterns of Strong Coupling for LHC Searches}},
    journal = {JHEP},
    volume = {11},
    pages = {141},
    year = {2016},
    doi = {10.1007/JHEP11(2016)141},
    eprint = {1603.03064},
    archivePrefix = {arXiv},
    primaryClass = {hep-ph},
    reportNumber = {CERN-TH-2016-048}
}

@article{MEGII:2018design,
    author = {Baldini, A. M. and others},
    collaboration = {MEG II},
    title = {{The design of the MEG II experiment}},
    journal = {Eur. Phys. J. C},
    volume = {78},
    number = {5},
    pages = {380},
    year = {2018},
    doi = {10.1140/epjc/s10052-018-5845-6},
    eprint = {1801.04688},
    archivePrefix = {arXiv},
    primaryClass = {physics.ins-det}
}

@article{SINDRUM:1988mueee,
    author = {Bellgardt, U. and others},
    collaboration = {SINDRUM},
    title = {{Search for the Decay $\mu^+ \to e^+ e^+ e^-$}},
    journal = {Nucl. Phys. B},
    volume = {299},
    pages = {1--6},
    year = {1988},
    doi = {10.1016/0550-3213(88)90462-2}
}

@article{Mu3e:2013proposal,
    author = {Blondel, A. and others},
    title = {{Research Proposal for an Experiment to Search for the Decay $\mu \to eee$}},
    year = {2013},
    eprint = {1301.6113},
    archivePrefix = {arXiv},
    primaryClass = {physics.ins-det}
}

@article{Mu3e:2020TDR,
    author = {Arndt, K. and others},
    collaboration = {Mu3e},
    title = {{Technical design of the phase I Mu3e experiment}},
    journal = {Nucl. Instrum. Meth. A},
    volume = {1014},
    pages = {165679},
    year = {2021},
    doi = {10.1016/j.nima.2021.165679},
    eprint = {2009.11690},
    archivePrefix = {arXiv},
    primaryClass = {physics.ins-det}
}

@article{SINDRUMII:2006mueconv,
    author = {Bertl, W. H. and others},
    collaboration = {SINDRUM II},
    title = {{A Search for muon to electron conversion in muonic gold}},
    journal = {Eur. Phys. J. C},
    volume = {47},
    pages = {337--346},
    year = {2006},
    doi = {10.1140/epjc/s2006-02582-x}
}

@article{Mu2e:2019,
    author = {Bernstein, Robert H.},
    title = {{The Mu2e Experiment}},
    journal = {Front. Phys.},
    volume = {7},
    pages = {1},
    year = {2019},
    doi = {10.3389/fphy.2019.00001}
}

@article{COMET:2020TDR,
    author = {Abramishvili, R. and others},
    collaboration = {COMET},
    title = {{COMET Phase-I technical design report}},
    journal = {Prog. Theor. Exp. Phys.},
    volume = {2020},
    number = {3},
    pages = {033C01},
    year = {2020},
    doi = {10.1093/ptep/ptz125}
}

@article{Roussy:2022cmp,
    author = {Roussy, Tanya S. and Caldwell, Luke and Wright, Trevor and Cairncross, William B. and Shagam, Yuval and Ng, Kia Boon and Schlossberger, Noah and Park, Sun Yool and Wang, Anzhou and Ye, Jun and Cornell, Eric A.},
    title = {{An improved bound on the electron's electric dipole moment}},
    journal = {Science},
    volume = {381},
    number = {6653},
    pages = {46--50},
    year = {2023},
    doi = {10.1126/science.adg4084},
    eprint = {2212.11841},
    archivePrefix = {arXiv},
    primaryClass = {physics.atom-ph}
}

@article{Alarcon:2022ero,
    author = {Alarcon, R. and others},
    title = {{Electric dipole moments and the search for new physics}},
    year = {2022},
    eprint = {2203.08103},
    archivePrefix = {arXiv},
    primaryClass = {hep-ph}
}

@article{Kuno:1999jp,
    author = {Kuno, Yoshitaka and Okada, Yasuhiro},
    title = {{Muon decay and physics beyond the standard model}},
    journal = {Rev. Mod. Phys.},
    volume = {73},
    pages = {151--202},
    year = {2001},
    doi = {10.1103/RevModPhys.73.151},
    eprint = {hep-ph/9909265},
    archivePrefix = {arXiv},
    reportNumber = {KEK-PREPRINT-99-69, KEK-TH-639}
}

@article{DAmbrosio:2002vsn,
    author = {D'Ambrosio, G. and Giudice, G. F. and Isidori, G. and Strumia, A.},
    title = {{Minimal flavor violation: An Effective field theory approach}},
    journal = {Nucl. Phys. B},
    volume = {645},
    pages = {155--187},
    year = {2002},
    doi = {10.1016/S0550-3213(02)00836-2},
    eprint = {hep-ph/0207036},
    archivePrefix = {arXiv},
    reportNumber = {CERN-TH-2002-147, IFUP-TH-2002-17}
}

@article{Redi:2013pga,
    author = {Redi, Michele},
    title = {{Leptons in Composite MFV}},
    journal = {JHEP},
    volume = {09},
    pages = {060},
    year = {2013},
    doi = {10.1007/JHEP09(2013)060},
    eprint = {1306.1525},
    archivePrefix = {arXiv},
    primaryClass = {hep-ph}
}

@article{Agashe:2004rs,
    author = {Agashe, Kaustubh and Contino, Roberto and Pomarol, Alex},
    title = {{The Minimal composite Higgs model}},
    journal = {Nucl. Phys. B},
    volume = {719},
    pages = {165--187},
    year = {2005},
    doi = {10.1016/j.nuclphysb.2005.04.035},
    eprint = {hep-ph/0412089},
    archivePrefix = {arXiv},
    reportNumber = {UAB-FT-567}
}

@article{Covone:2024elw,
    author = {Covone, Sebastiano and Davighi, Joe and Isidori, Gino and Pesut, Marko},
    title = {{Flavour deconstructing the composite Higgs}},
    journal = {JHEP},
    volume = {01},
    pages = {041},
    year = {2025},
    doi = {10.1007/JHEP01(2025)041},
    eprint = {2407.10950},
    archivePrefix = {arXiv},
    primaryClass = {hep-ph},
    reportNumber = {CERN-TH-2024-112}
}

@article{Kaplan:1991dc,
    author = {Kaplan, David B.},
    title = {{Flavor at SSC energies: A New mechanism for dynamically generated fermion masses}},
    journal = {Nucl. Phys. B},
    volume = {365},
    pages = {259--278},
    year = {1991},
    doi = {10.1016/S0550-3213(05)80021-5},
    reportNumber = {UCSD-PTH-91-04}
}

@article{Giudice:2007fh,
    author = {Giudice, G. F. and Grojean, C. and Pomarol, A. and Rattazzi, R.},
    title = {{The Strongly-Interacting Light Higgs}},
    journal = {JHEP},
    volume = {06},
    pages = {045},
    year = {2007},
    doi = {10.1088/1126-6708/2007/06/045},
    eprint = {hep-ph/0703164},
    archivePrefix = {arXiv},
    reportNumber = {CERN-PH-TH-2007-47}
}

@book{Panico:2015jxa,
    author = {Panico, Giuliano and Wulzer, Andrea},
    title = {{The Composite Nambu-Goldstone Higgs}},
    series = {Lecture Notes in Physics},
    volume = {913},
    publisher = {Springer},
    year = {2016},
    doi = {10.1007/978-3-319-22617-0},
    eprint = {1506.01961},
    archivePrefix = {arXiv},
    primaryClass = {hep-ph},
    reportNumber = {DFPD-2015TH9}
}

@article{Davighi:2025cqx,
    author = {Davighi, Joe and Isidori, Gino},
    title = {{A Composite Theory of Higgs and Flavour}},
    year = {2025},
    eprint = {2512.19650},
    archivePrefix = {arXiv},
    primaryClass = {hep-ph},
    reportNumber = {CERN-TH-2025-267, ZU-TH 88/25}
}

@inproceedings{Pesut:2025ook,
    author = {Pesut, Marko},
    title = {{Flavour Non-Universality and Higgs Compositeness}},
    booktitle = {{35th Rencontres de Blois}: {Particle Physics and Cosmology}},
    year = {2025},
    eprint = {2501.11376},
    archivePrefix = {arXiv},
    primaryClass = {hep-ph}
}

@article{Frigerio2018,
    author = {Frigerio, Michele and Nardecchia, Marco and Serra, Javi and Vecchi, Luca},
    title = {{The Bearable Compositeness of Leptons}},
    journal = {JHEP},
    volume = {10},
    pages = {017},
    year = {2018},
    doi = {10.1007/JHEP10(2018)017},
    eprint = {1807.04279},
    archivePrefix = {arXiv},
    primaryClass = {hep-ph},
    reportNumber = {CERN-TH-2018-160}
}

@article{banerjee2022snowmass2021whitepaper,
    author = {Swagato Banerjee and Vincenzo Cirigliano and Mogens Dam and Abhay Deshpande and Luca Fiorini and Kaori Fuyuto and Ciprian Gal and Tom{\'{a}}{\v{s}} Husek and Emanuele Mereghetti and Kevin Mons{\'{a}}lvez-Pozo and Haiping Peng and Francesco Polci and Jorge Portol{\'{e}}s and Armine Rostomyan and Michel Hern{\'{a}}ndez Villanueva and Bin Yan and Jinlong Zhang and Xiaorong Zhou},
    title = {{Snowmass 2021 White Paper: Charged lepton flavor violation in the tau sector}},
    year = {2022},
    eprint = {2203.14919},
    archivePrefix = {arXiv},
    primaryClass = {hep-ph}
}

@article{EliasMiro:2021jgu,
    author = {Elias Miro, Joan and Fernandez, Clara and Gumus, Mehmet Asim and Pomarol, Alex},
    title = {{Gearing up for the next generation of LFV experiments, via on-shell methods}},
    journal = {JHEP},
    volume = {06},
    pages = {126},
    year = {2022},
    doi = {10.1007/JHEP06(2022)126},
    eprint = {2112.12131},
    archivePrefix = {arXiv},
    primaryClass = {hep-ph}
}

@article{Agashe_2006,
    author = {Agashe, Kaustubh and Contino, Roberto and Da Rold, Leandro and Pomarol, Alex},
    title = {{A custodial symmetry for $Zb\bar b$}},
    journal = {Phys. Lett. B},
    volume = {641},
    number = {1},
    pages = {62--66},
    publisher = {Elsevier BV},
    year = {2006},
    doi = {10.1016/j.physletb.2006.08.005},
    eprint = {hep-ph/0605341},
    archivePrefix = {arXiv}
}

@article{2023,
    author = {Tsuzuki, N. and others},
    collaboration = {Belle},
    title = {{Search for lepton-flavor-violating $\tau$ decays into a lepton and a vector meson using the full Belle data sample}},
    journal = {JHEP},
    volume = {06},
    pages = {118},
    publisher = {Springer Science and Business Media LLC},
    year = {2023},
    doi = {10.1007/JHEP06(2023)118},
    eprint = {2301.03768},
    archivePrefix = {arXiv},
    primaryClass = {hep-ex}
}

@article{2025,
    author = {Adachi, I. and others},
    collaboration = {Belle, Belle II},
    title = {{Search for lepton-flavor-violating $\tau^-\to\ell^-K_S^0$ decays at Belle and Belle II}},
    journal = {JHEP},
    volume = {08},
    pages = {092},
    publisher = {Springer Science and Business Media LLC},
    year = {2025},
    doi = {10.1007/JHEP08(2025)092},
    eprint = {2504.15745},
    archivePrefix = {arXiv},
    primaryClass = {hep-ex}
}

@article{2021,
    author = {Uno, K. and others},
    collaboration = {Belle},
    title = {{Search for lepton-flavor-violating tau-lepton decays to $\ell\gamma$ at Belle}},
    journal = {JHEP},
    volume = {10},
    pages = {019},
    publisher = {Springer Science and Business Media LLC},
    year = {2021},
    doi = {10.1007/JHEP10(2021)019},
    eprint = {2103.12994},
    archivePrefix = {arXiv},
    primaryClass = {hep-ex}
}

@article{theatlascollaboration2025projectionskeymeasurementsheavy,
    author = {{ATLAS Collaboration} and {Belle II Collaboration} and {CMS Collaboration} and {LHCb Collaboration}},
    title = {{Projections for Key Measurements in Heavy Flavour Physics}},
    year = {2025},
    eprint = {2503.24346},
    archivePrefix = {arXiv},
    primaryClass = {hep-ex}
}

@article{altmannshofer2023probingleptonflavorviolation,
    author = {Altmannshofer, Wolfgang and Munbodh, Pankaj and Oh, Talise},
    title = {{Probing lepton flavor violation at Circular Electron-Positron Colliders}},
    journal = {JHEP},
    volume = {08},
    pages = {026},
    year = {2023},
    doi = {10.1007/JHEP08(2023)026},
    eprint = {2305.03869},
    archivePrefix = {arXiv},
    primaryClass = {hep-ph}
}

@article{Goto_2015,
    author = {Goto, Toru and Kitano, Ryuichiro and Mori, Shingo},
    title = {{Lepton flavor violating $Z$-boson couplings from nonstandard Higgs interactions}},
    journal = {Phys. Rev. D},
    volume = {92},
    number = {7},
    pages = {075021},
    publisher = {American Physical Society (APS)},
    year = {2015},
    doi = {10.1103/PhysRevD.92.075021},
    eprint = {1507.03234},
    archivePrefix = {arXiv},
    primaryClass = {hep-ph}
}

@article{Keren-Zur:2012buf,
    author = {Keren-Zur, Boaz and Lodone, Paolo and Nardecchia, Marco and Pappadopulo, Duccio and Rattazzi, Riccardo and Vecchi, Luca},
    title = {{On Partial Compositeness and the CP asymmetry in charm decays}},
    journal = {Nucl. Phys. B},
    volume = {867},
    pages = {394--428},
    year = {2013},
    doi = {10.1016/j.nuclphysb.2012.10.012},
    eprint = {1205.5803},
    archivePrefix = {arXiv},
    primaryClass = {hep-ph}
}

@article{Vecchi:2012fv,
    author = {Vecchi, Luca},
    title = {{A Flavor Sector for the Composite Higgs}},
    journal = {Phys. Lett. B},
    volume = {727},
    pages = {130--135},
    year = {2013},
    doi = {10.1016/j.physletb.2013.08.006},
    eprint = {1206.4701},
    archivePrefix = {arXiv},
    primaryClass = {hep-ph},
    reportNumber = {LA-UR-12-22220}
}

@article{Agashe:2015izu,
    author = {Agashe, Kaustubh and Hong, Sungwoo and Vecchi, Luca},
    title = {{Warped seesaw mechanism is physically inverted}},
    journal = {Phys. Rev. D},
    volume = {94},
    number = {1},
    pages = {013001},
    year = {2016},
    doi = {10.1103/PhysRevD.94.013001},
    eprint = {1512.06742},
    archivePrefix = {arXiv},
    primaryClass = {hep-ph}
}

@article{Cirigliano:2005ck,
    author = "Cirigliano, Vincenzo and Grinstein, Benjamin and Isidori, Gino and Wise, Mark B.",
    title = "{Minimal flavor violation in the lepton sector}",
    eprint = "hep-ph/0507001",
    archivePrefix = "arXiv",
    reportNumber = "UCSD-PTH-05-11, CALT-68-2566",
    doi = "10.1016/j.nuclphysb.2005.08.037",
    journal = "Nucl. Phys. B",
    volume = "728",
    pages = "121--134",
    year = "2005"
}

@article{Davighi:2023iks,
    author = "Davighi, Joe and Isidori, Gino",
    title = "{Non-universal gauge interactions addressing the inescapable link between Higgs and flavour}",
    eprint = "2303.01520",
    archivePrefix = "arXiv",
    primaryClass = "hep-ph",
    doi = "10.1007/JHEP07(2023)147",
    journal = "JHEP",
    volume = "07",
    pages = "147",
    year = "2023"
}

@article{Agashe:2018cuf,
    author = {Agashe, Kaustubh and Du, Peizhi and Ekhterachian, Majid and Fong, Chee Sheng and Hong, Sungwoo and Vecchi, Luca},
    title = {{Natural Seesaw and Leptogenesis from Hybrid of High-Scale Type I and TeV-Scale Inverse}},
    journal = {JHEP},
    volume = {04},
    pages = {029},
    year = {2019},
    doi = {10.1007/JHEP04(2019)029},
    eprint = {1812.08204},
    archivePrefix = {arXiv},
    primaryClass = {hep-ph}
}

@article{Agashe:2024uvp,
    author = {Agashe, Kaustubh and Du, Peizhi and Ekhterachian, Majid and Fong, Chee Sheng and Hong, Sungwoo and Vecchi, Luca},
    title = {{Conformal leptogenesis in composite Higgs models}},
    journal = {JHEP},
    volume = {02},
    pages = {132},
    year = {2025},
    doi = {10.1007/JHEP02(2025)132},
    eprint = {2410.00960},
    archivePrefix = {arXiv},
    primaryClass = {hep-ph},
    reportNumber = {UMD-PP-024-10}
}

@article{LHCb:Tau3Mu2026,
    author = {Aaij, R. and others},
    collaboration = {LHCb},
    title = {{Search for $\tau^-\to\mu^-\mu^+\mu^-$ decays at the LHCb experiment with Run 2 data}},
    journal = {JHEP},
    volume = {09},
    pages = {069},
    year = {2026},
    doi = {10.1007/JHEP09(2026)069},
    eprint = {2601.20785},
    archivePrefix = {arXiv},
    primaryClass = {hep-ex}
}

@article{ATLAS:Tau3Mu2026,
    author = {Aad, G. and others},
    collaboration = {ATLAS},
    title = {{A search for lepton-flavour violating $\tau\to3\mu$ decays with the ATLAS detector}},
    journal = {Phys. Lett. B},
    volume = {880},
    pages = {140748},
    year = {2026},
    doi = {10.1016/j.physletb.2026.140748},
    eprint = {2603.18099},
    archivePrefix = {arXiv},
    primaryClass = {hep-ex}
}

@article{BelleII:TauThreeLeptons2025,
    author = {Adachi, I. and others},
    collaboration = {Belle II},
    title = {{Search for the lepton-flavor-violating $\tau^-\to e^\mp\ell^\pm\ell^-$ decays at Belle II}},
    journal = {JHEP},
    volume = {12},
    pages = {169},
    year = {2025},
    doi = {10.1007/JHEP12(2025)169},
    eprint = {2507.18236},
    archivePrefix = {arXiv},
    primaryClass = {hep-ex}
}

@article{BelleII:TauMuGamma2026,
    author = {Abumusabh, M. and others},
    collaboration = {Belle II},
    title = {{Search for the lepton-flavor-violating decay $\tau^\pm\to\mu^\pm\gamma$ at Belle II}},
    year = {2026},
    eprint = {2608.23009},
    archivePrefix = {arXiv},
    primaryClass = {hep-ex}
}

@article{BelleII:Tau3Mu2024,
    author = {Adachi, I. and others},
    collaboration = {Belle II},
    title = {{Search for lepton-flavor-violating $\tau^-\to\mu^-\mu^+\mu^-$ decays at Belle II}},
    journal = {JHEP},
    volume = {09},
    pages = {062},
    year = {2024},
    doi = {10.1007/JHEP09(2024)062},
    eprint = {2405.07386},
    archivePrefix = {arXiv},
    primaryClass = {hep-ex}
}

@article{CMS:Tau3Mu2024,
    author = {Hayrapetyan, A. and others},
    collaboration = {CMS},
    title = {{Search for the lepton flavor violating $\tau\to3\mu$ decay in proton-proton collisions at $\sqrt{s}=13$ TeV}},
    journal = {Phys. Lett. B},
    volume = {853},
    pages = {138633},
    year = {2024},
    doi = {10.1016/j.physletb.2024.138633},
    eprint = {2312.02371},
    archivePrefix = {arXiv},
    primaryClass = {hep-ex}
}

@article{BaBarTauPseudoscalars2007,
    author = {Aubert, B. and others},
    collaboration = {BaBar},
    title = {{Search for Lepton Flavor Violating Decays $\tau^\pm\to\ell^\pm\pi^0,\ell^\pm\eta,\ell^\pm\eta'$}},
    journal = {Phys. Rev. Lett.},
    volume = {98},
    pages = {061803},
    year = {2007},
    doi = {10.1103/PhysRevLett.98.061803},
    eprint = {hep-ex/0610067},
    archivePrefix = {arXiv}
}

@article{BallZwicky2005,
    author        = {Ball, Patricia and Zwicky, Roman},
    title         = {{$B_{d,s}\to\rho,\omega,K^*,\phi$ Decay Form Factors
                      from Light-Cone Sum Rules Revisited}},
    journal       = {Phys. Rev. D},
    volume        = {71},
    pages         = {014029},
    year          = {2005},
    doi           = {10.1103/PhysRevD.71.014029},
    eprint        = {hep-ph/0412079},
    archivePrefix = {arXiv}
}

@article{FeldmannKrollStech1998,
    author        = {Feldmann, Thorsten and Kroll, Peter and Stech, Berthold},
    title         = {{Mixing and Decay Constants of Pseudoscalar Mesons}},
    journal       = {Phys. Rev. D},
    volume        = {58},
    pages         = {114006},
    year          = {1998},
    doi           = {10.1103/PhysRevD.58.114006},
    eprint        = {hep-ph/9802409},
    archivePrefix = {arXiv}
}

@article{BellePseudoscalars2007,
    author        = {Miyazaki, Y. and others},
    collaboration = {Belle},
    title         = {{Search for Lepton Flavor Violating $\tau^-$ Decays
                      into $\ell^-\eta$, $\ell^-\eta'$ and $\ell^-\pi^0$}},
    journal       = {Phys. Lett. B},
    volume        = {648},
    pages         = {341--350},
    year          = {2007},
    doi           = {10.1016/j.physletb.2007.03.027},
    eprint        = {hep-ex/0703009},
    archivePrefix = {arXiv}
}

@misc{Liao:PhiPsi2026Tau,
    author = {Liao, Yipu},
    title  = {{$\tau$ lepton physics at Belle and Belle II}},
    year   = {2026},
    note   = {Talk at the 14th International Workshop on $e^+e^-$
              Collisions from Phi to Psi, Pisa, 8 June 2026,
              slides 22--23; Belle II preliminary},
    url    = {https://indico.sns.it/event/140/contributions/1262/attachments/397/1174/PhiPsi26_tau_B1B2_YipuLiao.pdf}
}

@misc{2collaboration2026improveddirectlimitmuon,
      title={An improved direct limit on the muon electric dipole moment}, 
      author={2 Collaboration and D. P. Aguillard and T. Albahri and D. Allspach and J. Annala and K. Badgley and S. Baeßler and L. Bailey and E. Barlas-Yucel and T. Barrett and E. Barzi and F. Bedeschi and M. Berz and M. Bhattacharya and H. P. Binney and P. Bloom and J. Bono and E. Bottalico and T. Bowcock and S. Braun and M. Bressler and G. Cantatore and R. M. Carey and B. C. K. Casey and D. Cauz and R. Chakraborty and A. Chapelain and S. Chappa and S. Charity and C. Chen and M. Cheng and R. Chislett and Z. Chu and T. E. Chupp and C. Claessens and F. Confortini and M. E. Convery and S. Corrodi and L. Cotrozzi and J. D. Crnkovic and S. Dabagov and P. T. Debevec and S. Di Falco and G. Di Sciascio and S. Donati and B. Drendel and A. Driutti and M. Eads and A. Edmonds and J. Esquivel and M. Farooq and R. Fatemi and K. Ferraby and C. Ferrari and M. Fertl and A. T. Fienberg and A. Fioretti and D. Flay and S. B. Foster and H. Friedsam and N. S. Froemming and C. Gabbanini and I. Gaines and S. Ganguly and J. George and L. K. Gibbons and A. Gioiosa and K. L. Giovanetti and P. Girotti and W. Gohn and L. Goodenough and T. Gorringe and J. Grange and S. Grant and F. Gray and S. Haciomeroglu and T. Halewood-Leagas and D. Hampai and F. Han and J. Hempstead and D. W. Hertzog and G. Hesketh and E. Hess and A. Hibbert and Z. Hodge and S. Y. Hoh and K. W. Hong and R. Hong and T. Hu and Y. Hu and M. Iacovacci and M. Incagli and S. Israel and P. Kammel and M. Kargiantoulakis and M. Karuza and J. Kaspar and D. Kawall and L. Kelton and A. Keshavarzi and D. S. Kessler and K. S. Khaw and Z. Khechadoorian and B. Kiburg and M. Kiburg and O. Kim and N. Kinnaird and E. Kraegeloh and J. LaBounty and K. R. Labe and M. Lancaster and S. Lee and B. Li and D. Li and L. Li and I. Logashenko and A. Lorente Campos and Z. Lu and A. Lucà and G. Lukicov and A. Lusiani and A. L. Lyon and B. MacCoy and R. Madrak and K. Makino and S. Mastroianni and R. McCarthy and J. P. Miller and S. Miozzi and B. Mitra and J. P. Morgan and W. M. Morse and J. Mott and A. Nath and J. K. Ng and H. Nguyen and Y. Oksuzian and Z. Omarov and W. Osar and R. Osofsky and S. Park and G. Pauletta and J. Peck and G. M. Piacentino and R. N. Pilato and K. T. Pitts and B. Plaster and N. Pohlman and C. C. Polly and D. Počanić and J. Price and B. Quinn and M. U. H. Qureshi and G. Rakness and S. Ramachandran and E. Ramberg and R. Reimann and B. L. Roberts and D. L. Rubin and M. Sakurai and L. Santi and C. Schlesier and A. Schreckenberger and Y. K. Semertzidis and A. K. Soha and M. Sorbara and J. Stapleton and D. Still and C. Stoughton and D. Stratakis and D. Stöckinger and H. E. Swanson and G. Sweetmore and D. A. Sweigart and M. J. Syphers and Y. Takeuchi and D. A. Tarazona and T. Teubner and A. E. Tewsley-Booth and V. Tishchenko and N. H. Tran and W. Turner and E. Valetov and D. Vasilkova and G. Venanzoni and T. Walton and A. Weisskopf and L. Welty-Rieger and P. Winter and Y. Wu and B. Yu and M. Yucel and E. Zaid and Y. Zeng and C. Zhang},
      year={2026},
      eprint={2608.11124},
      archivePrefix={arXiv},
      primaryClass={hep-ex},
      url={https://arxiv.org/abs/2608.11124}, 
}

@article{Feruglio2015LFV,
  author        = {Feruglio, Ferruccio and Paradisi, Paride
                   and Pattori, Andrea},
  title         = {{Lepton Flavour Violation in Composite Higgs Models}},
  journal       = {Eur. Phys. J. C},
  volume        = {75},
  pages         = {579},
  year          = {2015},
  doi           = {10.1140/epjc/s10052-015-3807-9},
  eprint        = {1509.03241},
  archivePrefix = {arXiv},
  primaryClass  = {hep-ph}
}
\end{document}